\documentclass[a4paper,11pt]{article}
\pdfoutput=1 

\usepackage{jcappub} 

\usepackage[T1]{fontenc} 
\usepackage{natbib}

\usepackage{array}
\usepackage{makecell}

\DeclareMathVersion{mymath}
\usepackage{feynmp} 
\usepackage{float}
\usepackage{graphicx}
\usepackage{caption}
\usepackage{subcaption}
\usepackage[toc,page]{appendix}
\usepackage{wrapfig}
\usepackage{etoolbox}
\usepackage{xcolor}
\usepackage[labelfont=it]{caption}
\usepackage{soul}
\usepackage{cleveref}
\usepackage{titlesec}
\titlespacing*{\section}{0pt}{1.1\baselineskip}{\baselineskip}

\title{\boldmath Cubic Halo Bias in Eulerian and Lagrangian Space}

\author[a]{Muntazir Mehdi Abidi}
\author[a]{and Tobias Baldauf}

\affiliation[a]{Center for Theoretical Cosmology, \\
DAMTP, University of Cambridge, CB3 0WA, United Kingdom}

\emailAdd{sma74@cam.ac.uk}

\abstract{Predictions of the next-to-leading order, i.e. one-loop, halo power spectra, depend on local and non-local bias parameters up to cubic order. The linear bias parameter can be estimated from the large scale limit of the halo-matter power spectrum, and the second order bias parameters from the large scale, tree-level bispectrum. Cubic operators would naturally be quantified using the tree-level trispectrum. As the latter is computationally expensive, we extend the quadratic field method proposed in Schmittfull et al. 2014 to cubic fields, in order to estimate cubic bias parameters.\\
We cross-correlate a basis set of cubic bias operators with the halo field and express the result in terms of the cross-spectra of these operators, in order to cancel cosmic variance. We obtain significant detections of local and non-local cubic bias parameters, which are partially in tension with predictions based on local Lagrangian bias schemes. We directly measure the Lagrangian bias parameters of the protohaloes associated with our halo sample and clearly detect a non-local quadratic term in Lagrangian space. We do not find a clear detection of non-local cubic Lagrangian terms for low mass bins, but there is some mild evidence for their presence for the highest mass bin.
While the method presented here focuses on cubic bias parameters, the approach could also be applied to quantifications of cubic primordial non-Gaussianity.
}

\newcommand{\de}{\delta}

\newcommand{\be}{\begin{equation}}
\newcommand{\ee}{\end{equation}}

\newcommand{\bk}{\boldsymbol{k}}
\newcommand{\bq}{\boldsymbol{q}}
\newcommand{\bx}{\boldsymbol{x}}
 \color{black}
 \color{black}
 \color{black}
 \color{black}
   
\newcommand{\bp}{\boldsymbol{p}}

\newcommand{\ba}{\begin{aligned}}
\newcommand{\ea}{\end{aligned}}

\newcommand{\correc}[1]{{\color{black}#1}}

\newcommand{\newcorrec}[1]{{\color{black}#1}}

\begin{document}
\maketitle
\flushbottom

\section{Introduction}

The Large Scale Structure (LSS) of the Universe contains a wealth of information about the origin, composition, and evolution of the Universe. In order to extract this information from on-going and future LSS surveys, we have to understand various sources of non-linearities present in the late-time LSS observables. In general, there are three main sources of non-linearities: 
\begin{itemize}
\item non-linear matter clustering due to gravity
\item non-linear biasing: the relation between the distribution of tracers and dark matter
\item primordial non-Gaussianity (PNG), which induces non-linearities on the initial conditions
\end{itemize}
Recently, the  powerful framework of the Effective Field Theory of Large Scale Structure (EFTofLSS) \cite{Baumann:2010tm, Carrasco:2012cv, Porto:2013qua, Senatore:2014via, Baldauf:2014qfa, Angulo:2014tfa,Baldauf:2015aha, Bertolini:2016bmt, Pajer:2013jj}, which is an extension and correction of Standard Perturbation Theory \cite{Bernardeau:2001qr}, has provided a valuable insight into the non-linear nature of  matter clustering due to gravity. It has been shown that the EFTofLSS prediction for the two-loop dark matter power spectrum agrees to $1\%$ precision both with the $N$-body simulations up to $k_{\text{max}}\approx 0.3 h$ Mpc$^{-1}$ \cite{Carrasco:2013sva,Baldauf:2015aha} and with the one-loop bispectrum up to $k_{\text{max}}\approx 0.22 h$ Mpc$^{-1}$ at redshift $z=0$ \cite{Angulo:2014tfa,Baldauf:2014qfa}. However, to provide a consistent model for the statistics of biased tracers using the framework of the EFTofLSS, we need to understand the non-linearities due to biasing between the distribution of the tracers (halos or galaxies) and the matter distribution. 

The predictions of the next-to-leading order, that is one-loop halo power spectra and halo-matter cross spectra, depend on the bias parameters up to cubic order \cite{Assassi:2014fva, McDonald:2009dh, Senatore:2014eva}. The one-loop halo bispectra, on the other hand, depend on the bias parameters up to quartic order \cite{Assassi:2014fva}. Therefore, quantifying the higher order bias parameters precisely is a crucial step towards the modeling of the statistics of biased tracers. As we will describe in more detail in Sec.~\ref{sec:phm}, the halo-matter cross power spectrum depends on a particular combination of two cubic bias parameters. Study \cite{Saito:2014qha} attempted to measure this combination of bias parameters by fitting the scale dependence of the halo-matter power spectrum. However, the authors neglected the presence of derivative (or $k^2$) bias parameters, which are degenerate with the effect of the cubic bias operators. Their constraints are likely to be biased.

In this paper, we focus on the biasing problem and the measurements of halo bias parameters up to cubic order. There are two ways to study the halo bias: one is called the \emph{Eulerian bias model} and the other is known as the \emph{Lagrangian bias model}. In the  Eulerian bias model, the halo overdensity field $\de_{\text{h}}(\bx,\tau)$ is described in terms of co-moving coordinates $\bx$ as 
\begin{equation}
\de_{\text{h}}(\bx,\tau) = \sum_{\mathcal{O}} b_{\mathcal{O}}\mathcal{O}(\bx,\tau)\, ,
\label{EulBias}
\end{equation}
where $b_{\mathcal{O}}$ are the bias parameters and $\mathcal{O}(\bx)$ are bias operators that are functionals of matter density $\de(\bx)$. Eulerian biasing beyond linear order was first studied by \cite{Fry:1992vr} who introduced the local Eulerian bias model, where $\mathcal{O}(\bx)$ are local functions of $\de(\bx)$ expanded into a Taylor series. However, based on symmetry arguments, it was shown in \cite{McDonald:2009dh, Assassi:2014fva, Chan:2012jj} that the local Eulerian model is incomplete, making it important to include non-local terms at quadratic and cubic order. Numerical evidence for the presence of a quadratic non-local term in the Eulerian bias model in $N$-body simulations was given by \citep{Chan:2012jj, Baldauf:2012hs}. In principle, the halo field contains a typical scale, for instance the Eulerian or Lagrangian extent of a halo. For this and for numerical reasons, we will evaluate the operators in the right hand side of Eq.~\eqref{EulBias} smoothed on $R_\text{h}$. Physical bias models, based on the notion of halos being formed from a patch of size $R_\text{h}\propto M$ in Lagrangian space which exceeds the critical collapse density, have a physical scale built in. This scale can be fitted from the actual halos, as in \cite{Baldauf:2014fza}, as a function of mass, but we will rather pick a fixed value independent of mass and account for the residual uncertainty.

On the other hand, in a \emph{Lagrangian bias model} we identify protohalos, the regions in the initial density field that collapse and form halos at late-time, and describe the relation of the protohalo density field $\de_{\text{h}}(\bq)$ and  the initial density field $\de_{\text{G}}(\bq)$ in Lagrangian coordinates $\bq$. Writing the biasing relation in Lagrangian space is very useful, because it separates the non-linearities due to biasing from the non-linearities generated from gravitational instabilities. The most studied Lagrangian bias model so far is the local Lagrangian Bias (LLB) model. However, some evidence for the presence of a non-local tidal term in the Lagrangian model has been found recently in \cite{Modi:2016dah}. The time evolution of Lagrangian protohalos can be studied in the framework of co-evolution of a halo fluid coupled to the dynamically dominant dark matter component through its gravitational potential, as we will describe in the next section.

One of the aims of this paper is to constrain the bias parameters up to cubic order in Eulerian and Lagrangian spaces. The linear bias parameter can be estimated from the large scale halo-matter cross power spectrum and the second order bias parameters from the large scale, tree-level, bispectrum. Furthermore, the natural statistic to constrain cubic bias parameters is the large-scale, tree-level trispectrum. \correc{We summarize the N-point functions and relevant bias parameters in Tab.~\ref{tab:par}}. Estimating the bispectrum and trispectrum is computationally expensive, so we use the quadratic field method proposed in \cite{Schmittfull:2014tca} to estimate the quadratic bias and extend the method to cubic fields to estimate cubic bias parameters.

The key idea is to cross-correlate a basis of cubic bias operators (i.e. a weighted sum of three smoothed Gaussian fields) with the protohalo field and the late-time halo field, and to express the results in terms of the cross-spectra of cubic operators with themselves. The smoothing on the scale $R_\text{f}$ serves as a high-$k$ cutoff in our analysis. In perturbation theory (PT), the cross correlation of cubic fields with themselves can be expressed in terms of two-loop power spectrum diagrams. These diagrams contain one UV-sensitive reducible two-loop diagram and one two-loop irreducible diagram. Because of our ignorance of the exact scale of halos, the UV-sensitive diagrams might affect the measurements, of the bias parameters depending on which fiducial halo smoothing scale (cutoff) we choose. In our approach, we remove the strongly UV-sensitive diagrams by removing the part of the field that correlates with the linear density field. We will refer to this procedure as \emph{orthogonalization}. The quadratic correlators do not contain this sort of UV-sensitive diagrams, so there is no need to orthogonalize them.

The two-loop irreducible diagrams contain two cut-off scales. One scale is the artificially induced $R_{\text{f}}$, which we choose to be $20 h^{-1}$ Mpc, and the other is the fiducial halo smoothing scale $R_{\text{h}}$. The smoothing scale $R_\text{f}$ corresponds to the $1/k_{\text{max}}$ in a bispectrum or trispectrum analysis. Even though the irreducible diagrams at quadratic field and cubic field level are not highly UV-sensitive, they are still affected by the choice of $R_{\text{h}}$ and this dependency can affect the bias measurements. To make our measurements of the bias parameters independent of the halo smoothing scale, we Taylor-expand the cross-spectra around $R_{\text{h}}=4 h^{-1}$ Mpc and introduce a one parameter counterterm $\text{d}R$ for both quadratic and cubic statistics. This pragmatic approach is introduced in order to avoid dealing with a large number $\mathcal{O}(20)$ of EFT counterterms.

We find clear detection of the presence of cubic local and non-local terms in Eulerian space. On the other hand, we find clear evidence of a non-local Lagrangian tidal field. In addition, we do not find a clear detection of Lagrangian cubic non-local terms for low mass bins; however, for the highest mass bin we do find some presence of cubic non-local Lagrangian terms. Furthermore, we find that the mass dependence of the Eulerian cubic non-local bias terms prefer a co-evolution prediction of the Lagrangian bias model with a non-zero tidal field and no cubic fields. We also find that the presence of the Lagrangian tidal field does not induce new cubic bias operators at late-times; rather it merely changes the amplitude of cubic bias operators, which has been previously discussed in \cite{Chan:2012jj,Mirbabayi:2014zca,Desjacques:2016bnm}.  

This paper is organized as follows. In Section \ref{section2}, we discuss the bias models in Eulerian and Lagrangian space, as well as the co-evolution of the dark matter halos and dark matter. We present both the general definitions of the cubic operators and the co-evolution predictions for cubic bias parameters in the presence of the Lagrangian tidal field. In Section \ref{section3}, we discuss the quadratic and cubic field methods, and discuss how to remove UV-sensitive diagrams. In Section \ref{section4}, we describe our methodology to measure bias parameter from $N$-body simulations. In Section \ref{section5}, we present our results. We conclude in Section \ref{conclusion}.

\begin{table}[h]
\small
\begin{center}
\begin{tabular}{|c|c|c|}
\hline
   & \thead{Tree-level}& \thead{One-loop} \\  [0.5ex] 
\hline
Power Spectrum $P_{\text{hm}}$ & $b_1$ &  \makecell{$b_1$, \color{purple}$b_2$, $b_{s^2}$, \color{blue} $b_{\Gamma_3}$\\
\color{brown} derivative bias ($b_{\nabla^2\delta}\dots$)\color{black}} \\ 
 \hline
Bispectrum $B_{\text{hmm}}$ &  \makecell{$b_1$, \color{purple}$b_{2}$, $b_{s^2}$ \color{black}} &    \makecell{$b_1$, \color{purple} $b_{2}$, $b_{s^2}$, \\
\color{blue}$b_{\Gamma_3}$, $b_{\mathcal{G}_3}$, $b_{\mathcal{G}_2\delta}$,  $b_{3}$, \\ 
\color{orange} $b_{\Gamma_4}$, $b_{\Delta_4}$, $b_{\Gamma_3 \delta}$, $b_{\bar{\Gamma}_4}$ \color{black},
\color{brown} derivative bias ($b_{\nabla^2\delta}\dots$)\color{black} } \\
 \hline
Trispectrum $T_{\text{hmmm}}$ &  \makecell{$b_1$, \color{purple} $b_{2}$, $b_{s^2}$, \\ \color{blue}$b_{\Gamma_3}$, $b_{\mathcal{G}_3}$, $b_{\mathcal{G}_2\delta}$,  $b_{3}$\color{teal}}  &  \color{teal}many bias terms... \color{black} \\
 \hline
\end{tabular}
\end{center}
\caption{Bias parameter estimation from $N$-point functions. The tree-level power spectrum, bispectrum and trispectrum are natural statistics to obtain cleanest (and non-degenerate) constraints on the linear, quadratic and cubic bias parameters respectively. On the other hand the constraints on the derivative bias can be obtained from the loop statistics once the other bias parameters have been fixed from the tree-level statistics. \correc{The terms in Orange are quartic bias parameters which are beyond the scope of this paper}. A similar table is also given in \cite{Assassi:2014fva}.}
\label{tab:par}
\end{table}


\section{Halo Bias}

\label{section2} 

There are two ways to write down the halo bias relation: (1) in evolved Eulerian space and (2) in initial Lagrangian space. We will discuss both viewpoints in this Section.

\subsection{Eulerian Bias Model}
Following \cite{McDonald:2009dh,Assassi:2014fva,Senatore:2014eva}, without loss of generality the bias relation in Eulerian space given in Eq. \eqref{EulBias} can be written up to cubic order as
\begin{equation}
\ba
\de_{\text{h}}(\bx) &= b_{1}\Big(\de^{(1)}(\bx)+\de^{(2)}(\bx)+\de^{(3)}(\bx)\Big) + \frac{b_2}{2!}\Big(\de^2(\bx) - \langle\de^2(\bx)\rangle\Big) + b_2\Big(\delta^{(1)}(\bx)\delta^{(2)}(\bx)  - \langle\de^{(1)}\de^{(2)}\rangle\Big) \\
&+ b_{s^2}\Big(s^2(\bx) - \langle s^2(\bx)\rangle\Big)+2 b_{s^2}\Big(s^{(1)}_{ij}(\bx)s^{(2)}_{ij}(\bx) -\langle s_{ij}^{(1)}s^{(2)}_{ij}\rangle\Big) +b_{\delta^3}\Big(\de^3(\bx)-3\de(\bx) \langle\de^2(\bx)\rangle\Big)\\
&+ b_{\mathcal{G}_3}\mathcal{G}_3(\bx)+ b_{\mathcal{G}_2\de}\Big(\mathcal{G}_2\delta(\bx)- \langle\mathcal{G}_2\de(\bx)\rangle\Big)+ b_{\Gamma_3}\Big(\Gamma_3(\bx)-\langle\Gamma_3(\bx)\rangle\Big)\\
&+ b_{\nabla^2\de}
\nabla^2\de(\bx) + \dots 
\ea
\label{E1}
\end{equation}
Here, $s^{(1)}_{ij}(\bx)s^{(2)}_{ij}(\bx)$ describes the tidal bias contribution propagated to cubic order and $\delta^{(1)}(\bx)\delta^{(2)}(\bx)$ quadratic bias contribution propagated to third order. \correc{The quadratic tidal field $s^2(\bx)=s_{ij}(\bx)s_{ij}(\bx)$ is given as the trace of the square of the tidal tensor
\begin{equation}
s_{ij}(\bx) = \Bigg(\frac{\nabla_i\nabla_j}{\nabla^2}-\frac{1}{3}\de^{(\text{K})}_{ij}\Bigg)\de(\bx),
\end{equation}
where $\de^{(\text{K})}_{ij}$ is the Kronecker delta function}. For simplicity, from now onwards we adopt the notation $s^{(3)}(\bx)$ for $s^{(1)}_{ij}(\bx)s^{(2)}_{ij}(\bx)$. The remaining operators will be introduced in detail later. 
We remove the variance of the quadratic fields in order to ensure a mean zero halo overdensity $\langle \delta_\text{h}\rangle=0$ and subtract contributions proportional to $\sigma^2 \delta$ from the cubic terms, to avoid a renormalization \cite{McDonald:2006mx} of the low-$k$ limit of the halo-matter power spectrum. The bias parameters appearing in Eq.~\eqref{E1} are thus renormalized and physical bias parameters. Evaluating the one point moments $\sigma^2 = \langle \de^2_{\text{G}}\rangle = \int_{\bq}P_{\text{lin}}(\bq)$ in Eq.~\eqref{E1} yields
\begin{equation}
\ba
\de_{\text{h}}(\bx) &= b_{1}\Big(\de^{(1)}(\bx)+\de^{(2)}(\bx)+\de^{(3)}(\bx)\Big) + \frac{b_2}{2!}\Big(\de^2(\bx) - \sigma^2\Big) + b_2\Big(\delta^{(1)}\delta^{(2)}(\bx)  - \frac{68}{21} \sigma^2\de(\bx)\Big) \\
&+ b_{s^2}\Big(s^2(\bx) - \frac{2}{3}\sigma^2 \Big)+2 b_{s^2}\Big(s^{(3)}(\bx) -\frac{136}{63}\de(\bx)\sigma^2\Big) +b_{\delta^3}\Big(\de^3(\bx)-3\de(\bx)\sigma^2\Big)\\
&+ b_{\mathcal{G}_3}\mathcal{G}_3(\bx)+ b_{\mathcal{G}_2\de}\Big(\mathcal{G}_2\delta(\bx)+4\de(\bx)\sigma^2\Big)+ b_{\Gamma_3}\Big(\Gamma_3(\bx)+\frac{32}{35}\de(\bx)\sigma^2\Big)\\
&+ b_{\nabla^2\de}
\nabla^2\de(\bx) + \dots 
\ea
\label{E2}
\end{equation}
The second order terms $\de^{(2)}$, $\de^2$ and $s^2$ are the second order density field, the density-squared, and the square of the tidal tensor terms respectively, and form a basis of the quadratic bias operators \correc{$\mathcal{O}_2$}. There are seven distinct bias operators at cubic order, corresponding to seven bias parameters in general. Among seven bias operators at cubic order, four correspond to four new bias parameters: the coefficients of $\de^3$, $\mathcal{G}_3$, $\mathcal{G}_2\de$ and $\Gamma_3$. These are the most general operators made up of the second derivatives of the gravitational and velocity potentials, $\Phi_{\text{g},\text{v}}$, which are invariant under the symmetries of the equations of motion. At second order, there is no distinction between the gravitational and velocity potentials, because the contributions arise from squares of the linear potentials and at this order $\de^{(1)} = -\theta^{(1)}$. However, the velocity potential becomes an independent degree of freedom at cubic order \cite{Assassi:2014fva}. In fact, $\Gamma_3$ depends on the gravitational as well as velocity potentials explicitly. The Galileon operators and $\Gamma_3$ are defined as follows \cite{Chan:2012jj, Assassi:2014fva}:
\begin{equation}
\ba
\mathcal{G}_{1}(\Phi_{\text{g}}) &= \nabla^2\Phi_g=\delta\\
\mathcal{G}_{2}(\Phi_{\text{g}}) &= (\nabla_{i}\nabla_{j}\Phi_{\text{g}})^2-(\nabla^2\Phi_{\text{g}})^2\\
\mathcal{G}_{3}(\Phi_{\text{g}}) &= -\frac{1}{2}\Bigg[(\nabla^2\Phi_{\text{g}})^3 +2\nabla_{i}\nabla_{j}\Phi_{\text{g}}\nabla^{j}\nabla^{k}\Phi_{\text{g}}\nabla_{k}\nabla_{i}\Phi_{\text{g}} - 3(\nabla_{i}\nabla_{j}\Phi_{\text{g}})^2\nabla^2\Phi_{\text{g}}\Bigg]
\ea
\end{equation}
and
\begin{equation}
\Gamma_3(\Phi_{\text{v}},\Phi_{\text{g}}) = \mathcal{G}_{2}^{(3)}(\Phi_{\text{g}}) - \mathcal{G}^{(3)}_{2}(\Phi_{\text{v}})
\end{equation}
where $\mathcal{G}_{2}^{(3)}$ is given by
\begin{equation}
\mathcal{G}_{2}^{(3)}(\Phi) = 2 \Big(\nabla_{i}\nabla_{j}\Phi^{(1)}\nabla^{i}\nabla^{j}\Phi^{(2)} - \nabla^2\Phi^{(1)}\nabla^2\Phi^{(2)} \Big).
\end{equation}
The second order potential $\Phi^{(2)}_{\text{g,v}}$ depends on the second order density $\de^{(2)}$ or velocity divergence $\theta^{(2)}$. We define our basis of quadratic bias operators $\mathcal{O}_2$ and cubic bias operators $\mathcal{O}_3$ in Eulerian space as:
\begin{equation}
\mathcal{O}_2 \in \Big\{\de^{(2)}(\bx), \de^2(\bx), s^2(\bx)\Big\}
\label{Ebasis2}
\end{equation}
\begin{equation}
\mathcal{O}_3 \in \Big\{\de^{(3)}(\bx), \de^{(1)}\de^{(2)}(\bx), s^{(3)}(\bx),\de^3(\bx), \mathcal{G}_3(\bx)+\frac{1}{9}\de^3(\bx), \mathcal{G}_2\de(\bx)+\frac{2}{3}\de^3(\bx), \Gamma_3(\bx)+\frac{16}{63}\de^3(\bx)\Big\}.
\label{Ebasisnew}
\end{equation}
In order to reduce degeneracies in the fitting and to make the results more aligned with the usual notion of local cubic bias $b_3$, we removed $\delta^3$ contributions from the non-local bias operators $\Gamma_3,\mathcal{G}_2\delta$ and $\mathcal{G}_3$.
In Appendix A, we show how this basis can be mapped to the basis employed in \cite{Angulo:2015eqa,Fujita:2016dne}.
Our full basis of Eulerian bias parameters is given by\footnote{\correc{Note that corresponding to the cubic bias operators defined in eq.\eqref{Ebasisnew}, the local cubic bias parameter has changed to $b_3$, which is the coefficient of $\delta^3$ as predicted by the spherical collapse model. However, in a naive expansion in terms of the cubic bias operators, the coefficient becomes
\be
b_{\delta^3}=b_3+\frac{1}{9}b_{\mathcal{G}_3}+\frac{2}{3}b_{\mathcal{G}_2 \delta}+\frac{16}{63}b_{\Gamma_3}\, .
\ee
as shown in Eqs~\eqref{E1} and ~\eqref{E2}.
}}
\begin{equation}
\mathcal{B}\in \Big\{b_1,b_2,b_{s^2},b_3, b_{\mathcal{G}_3},b_{\mathcal{G}_2\delta},b_{\Gamma_3}\Big\}.
\end{equation}
\subsection{Lagrangian Bias Model}
In Lagrangian space, all gravitational coupling kernels $F_n$ \correc{(for $n>1$)} are zero, so matter field equals the linear Gaussian field. We write the Lagrangian bias model with local and non-local terms up to cubic order as
\begin{equation}
\ba
\de_{\text{h}}(\bq) &= b^{\text{L}}_{1}\de_{\text{G}}(\bq)+\frac{b^{\text{L}}_2}{2!}\Big(\de^2_{\text{G}}(\bq) -\sigma^2 \Big)+ \frac{b^{\text{L}}_3}{3!} \Big(\de^3_{\text{G}}(\bq)-3\sigma^2\de_{\text{G}}(\bq)\Big) + b^{\text{L}}_{s^2}\Big(s^{2}(\bq)- \frac{2}{3} \sigma^2\Big) \\
&+ b^{\text{L}}_{\mathcal{G}_3}\mathcal{G}_3(\bq)+b^{\text{L}}_{\mathcal{G}_2\delta}\Big(\mathcal{G}_2\delta(\bq) +4\sigma^2\de_{\text{G}}(\bq)\Big) + b^{\text{L}}_{\Gamma_3}\Big(\Gamma_3(\bq) + \frac{32}{35}\sigma^2\de_{\text{G}}(\bq)\Big)+ b_{\nabla^2\delta}\nabla^2\delta_{\text{G}}(\bq)+\dots
\ea
\label{Lbias2a}
\end{equation}
where $\bq$ is the Lagrangian coordinate of protohalos, $b^{\text{L}}_{i}$ are the Lagrangian bias parameters, and $\de_{\text{h}}(\bq)$ is the protohalo density field. This expansion in Hermite polynomials ensures that there is no renormalization of the bias parameters in the correlators \cite{Szalay:1988,Ferraro:2013}. Thus, the bias parameters in the model are the physical bias parameters occuring in the low-$k$ limit of $n$-point functions. 
Our basis of quadratic bias operator $\mathcal{O}_2^{\text{L}}$ and cubic bias operators $\mathcal{O}_3^{\text{L}}$ in Lagrangian space are defined as:
\begin{equation}
\mathcal{O}_2^{\text{L}} \in \Big\{\de^2_{\text{G}}(\bq), s^2(\bq)\Big\} \quad \text{and}\quad \mathcal{O}_3^{\text{L}} \in \Big\{\de^3_{\text{G}}(\bq), \mathcal{G}_3(\bq)+\frac{1}{9}\de^3_{\text{G}}(\bq), \mathcal{G}_2\de(\bq)+\frac{2}{3}\de^3_{\text{G}}(\bq), \Gamma_3(\bq)+\frac{16}{63}\de^3_{\text{G}}(\bq)\Big\}.
\label{Lbasis2b}
\end{equation}
\correc{Similar to Eulerian bias parameters, we define a basis of the cubic Lagrangian bias parameters as
\begin{equation}
\mathcal{B}^{\text{L}}\in \Big\{b_1^{\text{L}},b_2^{\text{L}},b_{s^2}^{\text{L}},b_3^{\text{L}}, b_{\mathcal{G}_3}^{\text{L}},b_{\mathcal{G}_2\delta}^{\text{L}},b_{\Gamma_3}^{\text{L}}\Big\}.
\end{equation}
}


\subsection{Time evolution of the Lagrangian bias}
For simplicity, we consider the local Lagrangian bias model and study its time evolution. At some initial time $\tau_i$ we define the protohalo density field with only local terms as
\be
\de_\text{h} (\bq, \tau_i) = b^{\text{L}}_{1}(\tau_i)\de_{\text{G}}(\bq, \tau_i) + \frac{b^{\text{L}}_{2}(\tau_i)}{2!} \Big( \de^{2}_{\text{G}}(\bq,\tau_i)-\sigma^2\Big) + \frac{b^{\text{L}}_{3}(\tau_i)}{3!} \Big( \de^{3}_{\text{G}}(\bq,\tau_i)-3\de_{\text{G}}(\bq,\tau_i)\sigma^2\Big)  + \dots
\label{Lagdel}
\ee
The time dependence of bias parameters $b^{\text{L}}_{n}(\tau)$ and the linear density field $\de_{\text{G}}(\bq,\tau)$ are defined as
\be
 b^{\text{L}}_{n}(\tau) = \Bigg(\frac{D(\tau_i)}{D(\tau)}\Bigg)^{n}b^{\text{L}}_{n}(\tau_i)\quad \text{and}\quad \de_{\text{G}}(\bq,\tau) = \frac{D(\tau)}{D(\tau_i)}\de_{\text{G}}(\bq,\tau_i)\, ,
\ee
where $\tau$ is the conformal time and $D(\tau)$ is the linear growth factor. \correc{From now onwards, we will choose $\tau_i=0$}.\footnote{\correc{To avoid confusion, we use $\de_{\text{h}}(\bq)\equiv\de_{\text{h}}(\bq,\tau_i=0)$ to represent the halo density field in the Lagrangian coordinates at the initial time $\tau_i=0$ and $\de_{\text{h}}(\bx,\tau)$ as the evolved halo density field in the Eulerian coordinates}} We now transform the fields from Lagrangian to Eulerian coordinates. The Eulerian comoving coordinates $\bx$ and Lagrangian coordinates $\bq$ are related through the displacement field vector $\Psi(\bq,\tau)$ as
\be
\bx(\bq,\tau) = \bq + \Psi(\bq,\tau).
\label{eq:trans}
\ee
We can use this relation and expand the Lagrangian density field up to third order \correc{by expressing the Lagrangian coordinates in the Eulerian coordinates as}
\be
\de_{\text{G}}(\bq,\tau) = \de_{\text{G}}(\bx,\tau) - \underbrace{\Psi(\bq,\tau)\cdot\nabla\de_{\text{G}}(\bq,\tau) + \frac{1}{2}\Psi_{i}(\bq,\tau)\Psi_{j}(\bq,\tau)\nabla_{i}\nabla_{j}\de_{\text{G}}(\bq,\tau)}_{\text{shift terms}} + \dots 
\label{del}
\ee
The second and third term in Eq.~\eqref{del} describe the shift terms. 
We define the second and third order shift terms \correc{in the Eulerian coordinates as  $H^{(2)}(\bx,\tau)$ and $H^{(3)}(\bx,\tau)$  respectively as below}:
\begin{equation}
\ba
H^{(2)}(\bx,\tau) &= \Psi^{(1)}(\bx,\tau)\cdot\nabla\de_{\text{G}}(\bx,\tau),\\
H^{(3)}(\bx,\tau) &= \Psi^{(2)}(\bx,\tau)\cdot\nabla\de_{\text{G}}(\bx,\tau)- \frac{1}{2}\Psi^{(1)}_{i}(\bx,\tau)\Psi^{(1)}_{j}(\bx,\tau)\nabla_{i}\nabla_{j}\de_{\text{G}}(\bx,\tau) \\
&- \Psi^{(1)}_{i}(\bx,\tau)\Big(\nabla_i \Psi^{(1)}_{j}(\bx,\tau)\Big)\nabla_{j}\de_{\text{G}}(\bx,\tau),
\label{shifts}
\ea
\end{equation}
\correc{where $\Psi^{(1)}$ and $\Psi^{(2)}$ are the first and second order displacement fields in Lagrangian Perturbation Theory (LPT) (see \cite{Matsubara:2007wj} for more details)}. In order to transform the halo density field in eq.~\eqref{Lagdel} from Lagrangian to Eulerian coordinates, we have to use the continuity equations for halos and dark matter given by
\be
[1+\delta_{\text{h}}(\bx,\tau)]\text{d}^3\bx = [1+\de_{\text{h}}(\bq)]\text{d}^3\bq\quad\text{and} \quad [1+\de(\bx,\tau)]\text{d}^3\bx = \text{d}^3\bq,
\ee
which lead to
\begin{equation}
\de_{\text{h}}(\bx,\tau) = \de_{\text{h}}(\bq) + \de(\bx,\tau) + \de_{\text{h}}(\bq)\de(\bx,\tau).
\label{delh}
\end{equation}

\correc{In Eq.~\eqref{delh},  $\de(\bx,\tau)$ represents the fully evolved non-linear matter field at late time $\tau$}. Using Eqs. \eqref{Lagdel}, \eqref{eq:trans}, \eqref{shifts}, and \eqref{delh} the second order halo field in Eulerian space is written as
\begin{equation}
\ba
\de^{(2)}_{\text{h}}(\bx,\tau) &= \Big( 1+b^{\text{L}}_{1}(\tau)\Big)\de^{(2)}(\bx,\tau) + \Big(\frac{1}{2}b_{2}^{\text{L}}(\tau)+\frac{4}{21}b_{1}^{\text{L}}(\tau)\Big)\de^{2}(\bx,\tau)
-\frac{2}{7}b_{1}^{\text{L}}s^{2}(\bx,\tau)
\ea
\label{ourpred2}
\end{equation}
In deriving the above expression, we use the relation
\be
\de^{(2)}(\bx,\tau) = \frac{17}{21} \de^2(\bx,\tau) - H^{(2)}(\bx,\tau)  + \frac{2}{7}s^{2}(\bx,\tau).
\ee
The third order solution contains many terms, and it is not convenient to write the full expression here. Instead we compare the final expression with our Eulerian cubic basis defined in Eq.~\eqref{Ebasisnew}. The full expression of the cubic halo density field can then be obtained by multiplying the basis with the coefficient vector
\begin{equation}
 \left\{1,\frac{4}{21},-\frac{2}{7},0,-\frac{22}{63},0,\frac{23}{42}\right\}b_{1}^{\text{L}}+
 \left\{0,\frac{1}{2},0,-\frac{1}{2},0,-\frac{2}{7},0\right\}b_{2}^{\text{L}}
+ \left\{0,0,0,\frac{1}{6},0,0,0\right\}b_{3}^{\text{L}}\, ,
\label{eq:ourpred2}
\end{equation}
which gives
\begin{equation}
\ba
\de^{(3)}_{\text{h}}(\bx,\tau) &= \left(1+b_1^{\text{L}}(\tau)\right) \de^{(3)}(\bx,\tau) + \left(\frac{b^{\text{L}}_{3}(\tau)}{6}-\frac{b^{\text{L}}_{2}(\tau)}{2}\right)\de^{3}(\bx,\tau) -\frac{22}{63}b_{1}^{\text{L}}(\tau)\mathcal{G}_3(\bx,\tau) \\
& -\frac{2}{7}b_{2}^{\text{L}}(\tau)\de\mathcal{G}_2(\bx,\tau) +\frac{23}{42}b_{1}^{\text{L}}(\tau)\Gamma_3(\bx,\tau) + \Big(\frac{8}{21}b_{1}^{\text{L}}(\tau)+b_{2}^{\text{L}}(\tau)\Big)\de\de^{(2)}(\bx,\tau)\\
&-\frac{4}{7}b_{1}^{\text{L}}s^{(3)}(\bx,\tau). 
\ea
\label{ourpred3}
\end{equation}
This time evolution of quadratic and cubic bias parameters in the Lagrangian framework has already been previously discussed in \cite{Mirbabayi:2014zca,Desjacques:2016bnm}. If we carry out the same calculations assuming a non-zero tidal field in the initial Lagrangian bias model we get the following second and third order solutions:
\begin{equation}
\ba
\widetilde{\de}^{(2)}_{\text{h}}(\bx,\tau) &= \Big( 1+b^{\text{L}}_{1}(\tau)\Big)\de^{(2)}(\bx,\tau) + \Big(\frac{1}{2}b_{2}^{\text{L}}(\tau)+\frac{4}{21}b_{1}^{\text{L}}(\tau)\Big)\de^{2}(\bx,\tau)
-\Big(\frac{2}{7}b_{1}^{\text{L}}(\tau)-b^{\text{L}}_{s^2}(\tau)\Big)s^{2}(\bx,\tau)
\ea
\label{ourpred2s2}
\end{equation}
and
\begin{equation}
\ba
\widetilde{\de}^{(3)}_{\text{h}}(\bx,\tau) &= \left(1+b_1^{\text{L}}(\tau)\right) \de^{(3)}(\bx,\tau) + \left(\frac{b^{\text{L}}_{3}(\tau)}{6}-\frac{b^{\text{L}}_{2}(\tau)}{2}-\frac{2}{3}b^{\text{L}}_{s^2}(\tau)\right)\de^{3}(\bx,\tau)\\
&-\Big(\frac{22}{63}b_{1}^{\text{L}}(\tau)-2 b^{\text{L}}_{s^2}(\tau)\Big)\mathcal{G}_3(\bx,\tau) -\Big(\frac{2}{7}b_{2}^{\text{L}}(\tau)+\frac{8}{21}b^{\text{L}}_{s^2}(\tau)\Big)\de\mathcal{G}_2(\bx,\tau) \\
&+\Big(\frac{23}{42}b_{1}^{\text{L}}(\tau)-\frac{5}{2}b^{\text{L}}_{s^2}(\tau)\Big)\Gamma_3(\bx,\tau) 
+ \Big(\frac{8}{21}b_{1}^{\text{L}}(\tau)+b_{2}^{\text{L}}(\tau)\Big)\de\de^{(2)}(\bx,\tau) \\
&-\Big(\frac{4}{7}b_{1}^{\text{L}}(\tau) - 2b^{\text{L}}_{s^2}(\tau)\Big)s^{(3)}(\bx,\tau). 
\ea
\label{ourpred3s2}
\end{equation}
The Lagrangian tidal term leaks into the cubic bias parameters, but does not change the Eulerian basis. 

\subsection{Co-evolution of dark matter and halos}
Gravity naturally introduces non-local terms in the bias relation. To see this we do the following. Under the assumptions of no velocity bias (that is the velocity of halos traces the velocity of dark matter) and the conservation of halos, one can solve the coupled equations of motion for dark matter and dark matter halos. The continuity and Euler equations are given by
\begin{equation}
\begin{split}
\de'_{\text{m}}(\bk, \tau) + \theta_{\text{m}}(\bk, \tau) =& -\int_{\bq}\alpha(\bq,\bk-\bq)\theta_{\text{m}}(\bq, \tau)\de_{\text{m}}(\bk-\bq, \tau),\\
\theta'_{\text{m}}(\bk, \tau) +\mathcal{H}\theta_{\text{m}}(\bk, \tau)+ \frac{3}{2}\mathcal{H}^2\Omega_{\text{m}}\de_{\text{m}}(\bk, \tau) =& -\int_{\bq}\beta(\bq,\bk-\bq)\theta_{\text{m}}(\bq, \tau)\theta_{\text{m}}(\bk-\bq, \tau),\\
\de'_{\text{h}}(\bk, \tau) + \theta_{\text{h}}(\bk, \tau) =& -\int_{\bq}\alpha(\bq,\bk-\bq)\theta_{\text{h}}(\bq, \tau)\de_{\text{h}}(\bk-\bq, \tau),\\
\theta'_{\text{h}}(\bk, \tau) +\mathcal{H}\theta_{\text{h}}(\bk, \tau)+ \frac{3}{2}\mathcal{H}^2\Omega_{\text{m}}\de_{\text{h}}(\bk, \tau) =& -\int_{\bq}\beta(\bq,\bk-\bq)\theta_{\text{h}}(\bq, \tau)\theta_{\text{h}}(\bk-\bq, \tau),
\end{split}
\end{equation}
where $\int_{\bq}=\int{\text{d}^3\bq}/{(2\pi)^3}$.
If we assume that the Eulerian bias model was completely local at some initial time $\tau_i$, then the second and third order solutions of the coupled system of equations described above can tell us how much non-locality is induced by gravity in the late-time bias relation. For a detailed discussion/calculations of the co-evolution of dark matter and dark matter halos we refer to \cite{Chan:2012jj,Baldauf:2012hs,Saito:2014qha}. The second order solution is
\begin{equation}
\ba
\de^{(2)}_\text{h}(\bk,\tau) &= \de^{(2)}_\text{h}(\bk,\tau_i) + (b^{\text{L}}_1(\tau)+1)\int_{\bq}F_2(\bq,\bk-\bq)\de_{\text{G}}(\bq,\tau)\de_{\text{G}}(\bk-\bq, \tau) \\
&+  \frac{4}{21}b^{\text{L}}_1(\tau)\int_{\bq}\de_{\text{G}}(\bq,\tau)\de_{\text{G}}(\bk-\bq,\tau) - \frac{2}{7}b^{\text{L}}_1(\tau)\int_{\bq}S_2(\bq,\bk-\bq)\de_{\text{G}}(\bq,\tau)\de_{\text{G}}(\bk-\bq,\tau)
\ea
\end{equation}
and the third order solution is given in \cite{Saito:2014qha} and previously in different notation in \cite{Chan:2012jj}:
\begin{equation}
\de^{(3)}_\text{h}(\bk,\tau) = \int_{\bq_1}\int_{\bq_2}\mathcal{K}_{3}(\bq_1,\bq_2,\bk-\bq_1-\bq_2; \tau)\de_{\text{G}}(\bq_{1},\tau)\de_{\text{G}}(\bq_{2},\tau)\de_{\text{G}}(\bk-\bq_{1}-\bq_{2},\tau)\, ,
\label{eq:coev2}
\end{equation}
where
\begin{equation}
\ba
\mathcal{K}_{3} &= \frac{1}{3}b^{\text{L}}_{3}(\tau) + \frac{1}{3}G_{3}(\bq_1,\bq_2,\bk-\bq_1-\bq_2) + \Big(\frac{1}{2}b^{\text{L}}_{1}(\tau)+\frac{1}{3}\Big)\alpha(\bq_1,\bq_2+\bq_3)F_2(\bq_2,\bq_3)\\
& 
+\Big(\frac{1}{2}b^{\text{L}}_{2}(\tau)+\frac{2}{21}b_{1}^{\text{L}}(\tau)\Big)\alpha(\bq_1,\bq_2+\bq_3) + 
\frac{1}{14}b_{1}^{\text{L}}(\tau)\alpha(\bq_1,\bq_2+\bq_3)S_2(\bq_2,\bq_3)\\
&+\Big(\frac{1}{2}b^{\text{L}}_{1}(\tau)+\frac{1}{3}\Big)\alpha(\bq_1,\bq_2+\bq_3)G_2(\bq_2,\bq_3)\, .
\ea
\label{eq:coev3}
\end{equation}
Eqs.~\eqref{eq:coev2} and ~\eqref{eq:coev3} agree with Eqs.~\eqref{eq:ourpred2} and ~\eqref{ourpred3} respectively.
\begin{figure}[t!]
\begin{center}
\includegraphics[width=10cm]{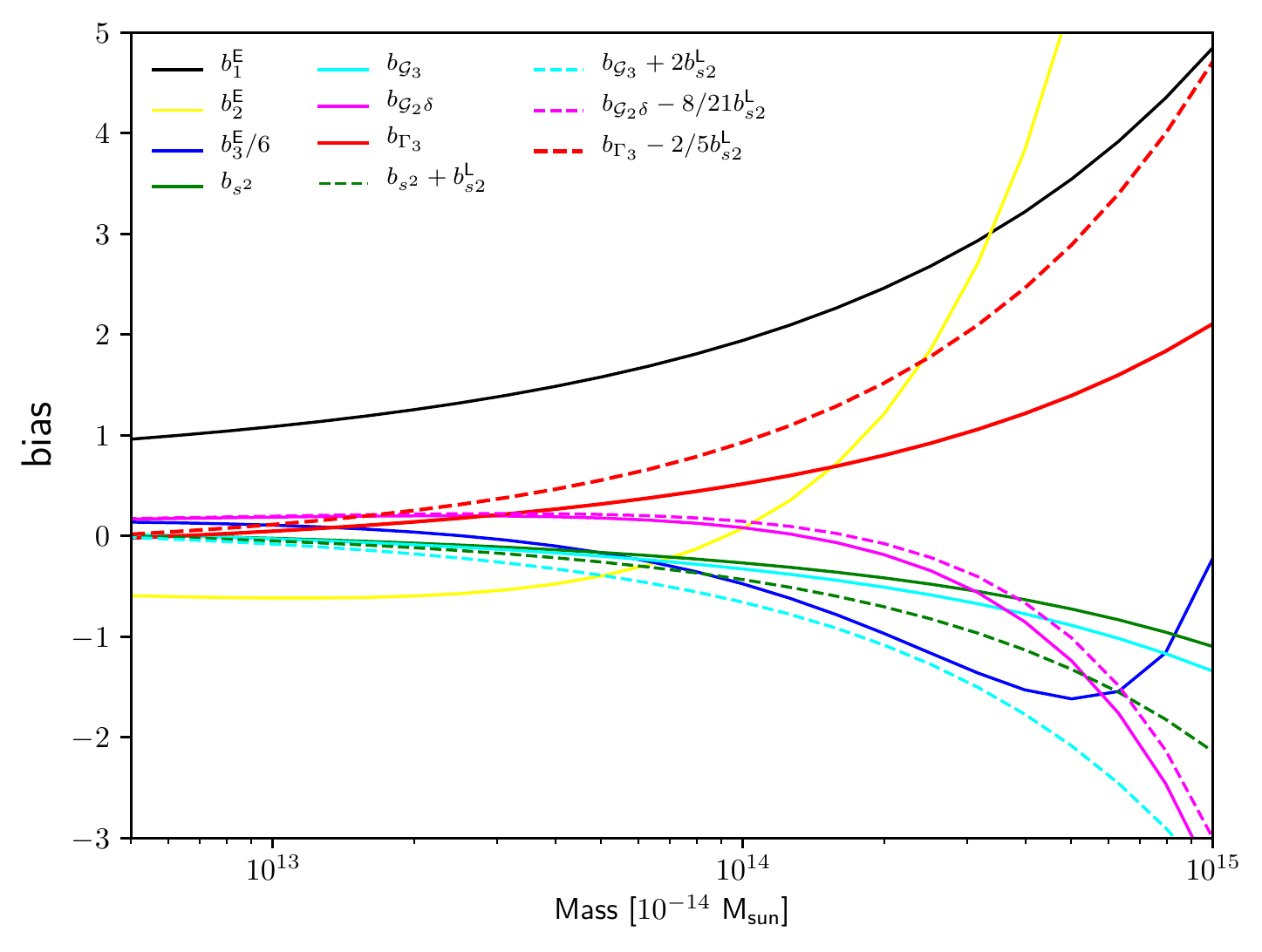}
\end{center}
\caption{Theoretical predictions of bias parameters are obtained from the co-evolution of the local Lagrangian bias model (LLB) and the co-evolution of the Lagrangian bias model with the non-zero tidal term $b^{\text{L}}_{s2}$ at second order. The mass dependence of the initial Lagrangian bias is defined in Eq. \eqref{Ltidal}.}
\label{coevfig}
\end{figure}
\subsection{Bias predictions}
Let us summarize the predictions for the coefficients of our basis Eq.~\eqref{Ebasisnew}.
We study two cases:
\begin{itemize}
\item \textbf{Local Lagrangian Bias Model:} Under the assumption of a local Lagrangian bias model all non-local terms in the late-time bias model are generated from gravitational instability. The late-time bias parameters are then given by:
\begin{equation}
\ba
&b_{1} = b^{\text{L}}_1 + 1;\qquad &b_{2} = \frac{4}{21}b^{\text{L}}_1 + \frac{1}{2}b^{\text{L}}_2\\
&b_{s^2} = -\frac{2}{7}b^{\text{L}}_{1};\qquad &b_{\delta^3} = -\frac{1}{2}b^{\text{L}}_2 + \frac{1}{6}b^{\text{L}}_3\\
&b_{\mathcal{G}_3} = -\frac{22}{63}b^{\text{L}}_{1} ;\qquad &b_{\mathcal{G}_2\delta} = -\frac{2}{7}b^{\text{L}}_{2};\\
&b_{\Gamma_3} = \frac{23}{42}b^{\text{L}}_{1}; \qquad &b_3 = -\frac{398}{3969}b_{1}^{\text{L}}-\frac{13}{42}b_{2}^{\text{L}}+\frac{1}{6}b_{3}^{\text{L}}.
\label{predictions1}
\ea
\end{equation}

\item \textbf{Local Lagrangian Bias Model + $b^{\text{L}}_{s^2}$:}
We extend the local Lagrangian model and include a non-local tidal term at second order. Such a term would arise in ellipsoidal collapse models \cite{Castorina:2016tuc,Modi:2016dah}. Propagating the additional contribution through the co-evolution calculation, we obtain the following prediction for late-time bias parameters:
\begin{equation}
\ba
&b_{1} = b^{\text{L}}_1 + 1;\qquad &b_{2} = \frac{4}{21}b^{\text{L}}_1 + \frac{1}{2}b^{\text{L}}_2\\
&b_{s^2} = -\frac{2}{7}b^{\text{L}}_{1} + b^{\text{L}}_{s^2};\qquad &b_{\delta^3} = -\frac{1}{2}b^{\text{L}}_2 + \frac{1}{6}b^{\text{L}}_3-\frac{2}{3}b^{\text{L}}_{s^2}\\
&b_{\mathcal{G}_3} = -\frac{22}{63}b^{\text{L}}_{1} +2b^{\text{L}}_{s^2};\qquad &b_{\mathcal{G}_2\delta} = -\frac{2}{7}b^{\text{L}}_{2} - \frac{8}{21}b^{\text{L}}_{s^2};\\
&b_{\Gamma_3} = \frac{23}{42}b^{\text{L}}_{1} -\frac{5}{2}b^{\text{L}}_{s^2};\qquad &b_3 = -\frac{398}{3969}b_{1}^{\text{L}}-\frac{13}{42}b_{2}^{\text{L}}+\frac{1}{6}b_{3}^{\text{L}}.
\ea
\label{predictions2}
\end{equation}
\end{itemize}
Fig. \ref{coevfig} shows the co-evolution bias predictions with and without the Lagrangian tidal field $b^{\text{L}}_{s^2}$ based on local bias parameters derived from a Sheth-Tormen (ST) mass function \cite{Sheth:1999mn}. The initial Lagrangian tidal bias used in this plot is motivated by our observations and given in Eq. \eqref{Ltidal}.

\section{Quadratic and Cubic Fields}
\label{section3}

In this section we discuss the quadratic field method proposed in \cite{Schmittfull:2014tca} and extend it to cubic fields. First, we discuss the quadratic fields and describe the PT expressions for the cross-correlation of quadratic fields with the density fields. Then, we describe our full basis of cubic fields and the cross correlation with the non-linear matter field and halo field. The cross-spectra with cubic fields contain diagrams that are highly UV-sensitive. To remove these diagrams from our model, we describe a procedure that we denote \emph{orthogonalization}. Finally, we discuss why including the counter terms is essential for making the bias measurements insensitive to our ignorance of the halo smoothing scale $R_{\text{h}}$.  
\subsection{Quadratic Fields}
As proposed in \cite{Schmittfull:2014tca} we consider three quadratic fields: the density-squared $\de^2(\bx)$, the shift $\Psi(\bx)\cdot\nabla\delta(\bx)$, and the square of the tidal tensor $s^2(\bx)$. In Fourier space, these fields are defined as
\begin{equation}
\mathcal{D}_2[\de](\bk) = \int_{\bq}\de_{\text{G}}(\bq)\de_{\text{G}}(\bk-\bq)\mathcal{K}_{D_2}(\bq,\bk-\bq)W_{\text{R}_\text{f}}(\bq)W_{\text{R}_\text{f}}(\bk-\bq) \, ,
\label{quadfield}
\end{equation}
where $\mathcal{D}_2[\de]\in\{\de^2, -\Psi\cdot\nabla\de,s^2\}$ and $\mathcal{K}_{D_2}\in \{1,H_2,S_2\}$ with $H_2$ and $S_2$ defined as
\begin{equation}
H_2(\bq_1,\bq_2) = -\frac{1}{2}(\bq_1\cdot\bq_2)\Big(\frac{q_1}{q_2} + \frac{q_2}{q_1}\Big)\quad \text{and}\quad S_{2}(\bq_1,\bq_2) = \frac{(\bq_1\cdot\bq_2)^2}{q_1^2q_2^2} -\frac{1}{3}.
\label{s2}
\end{equation}
The density squared and $s^2$ correspond to the bias parameters $b_{2}$ and $b_{s^2}$ respectively. Due to the equivalence principle, there is no separate bias parameter corresponding to the shift term.

\begin{figure}[H]
\begin{center}
\includegraphics[width=5cm]{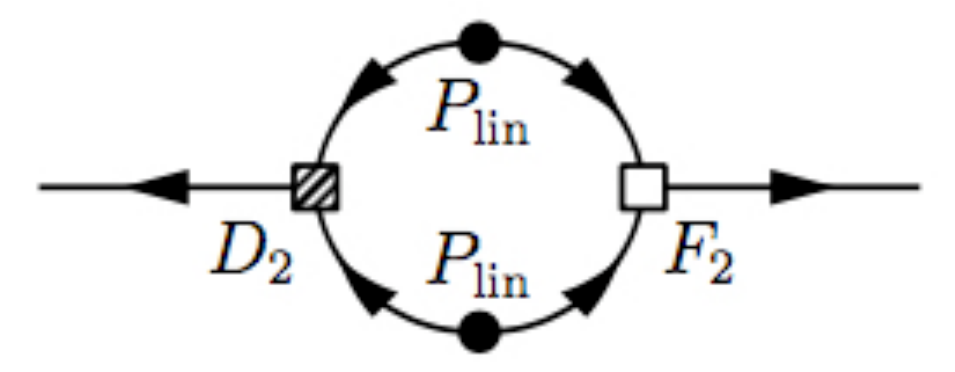} 
\end{center}
\caption{Diagrammatic representation of $\langle \mathcal{D}_{2}(\bk)|\de^{(2)}(\bk')\rangle'$.}
\label{D2f2a}
\end{figure}

Due to the convolution integrals, Eq.~\eqref{quadfield} receives contributions from all modes. To restrict to large scale modes, we implement a cut-off by smoothing the density field: $\de_{\text{G}}(\bk) \rightarrow W_{\text{R}_{\text{f}}}(\bk)\de_{\text{G}}(\bk)$. For definiteness, we choose a Gaussian filter $W_{R_{\text{f}}}(k) = \exp(-k^2 R_{\text{f}}^2/2)$ with a fiducial $R_{\text{f}}=20 h^{-1}$ Mpc, corresponding to a $k_{\text{max}}\approx 1/R_\text{f}$ cutoff. This cutoff $k_{\text{max}} \approx 0.05 h$ Mpc$^{-1}$ is the maximum wavenumber contributing to the quadratic and cubic field integrals. This choice of smoothing can be motivated from the fact that one-loop PT is typically valid for wavenumbers $k<0.1 h$ Mpc$^{-1}$ \cite{Baldauf:2015aha}. The cross-correlation of the smoothed quadratic fields with the halo density $\delta_{\text{h}}$ can be expanded as
\begin{equation}
\ba
\langle \mathcal{D}_{2}[\de](\bk)|\de_{\text{h}}(\bk')\rangle =  b_1\langle \mathcal{D}_{2}[\de](\bk)|\de^{(2)}(\bk')\rangle +\frac{b_2}{2}\langle \mathcal{D}_{2}[\de](\bk)|\delta^2(\bk')\rangle +b_{s^2}\langle \mathcal{D}_{2}[\de](\bk)|s^2(\bk')\rangle.
\label{quadraticmodel}
\ea
\end{equation}
In the above equation, each term can be expressed as a one-loop PT integral as\footnote{The prime on the correlator signifies that the Fourier space expectation value is equal to the power spectrum, i.e., $\langle\delta(\bk)\delta(\bk')\rangle'=P(k)$ as opposed to $\langle\delta(\bk)\delta(\bk')\rangle=(2\pi)^3\delta_\text{(D)}^3(\bk+\bk')P(k)$.}
\begin{equation}
\ba
\langle \mathcal{D}_{2}[\de](\bk)|\mathcal{O}_{2}(\bk')\rangle' =2\int_{\bq} W_{\text{R}_{\text{f}}}(\bq)W_{\text{R}_{\text{f}}}(\bk-\bq)W_{\text{R}_{\text{h}}}(\bq)&W_{\text{R}_{\text{h}}}(\bk-\bq)P_{\text{lin}}(q)P_{\text{lin}}(|\bk-\bq|) \\
&\times \mathcal{K}_{D_{2}}(\bq,\bk-\bq)\mathcal{K}_{\mathcal{O}_2}(\bq,\bk-\bq),
\ea
\label{a}
\end{equation}
where $\mathcal{O}_2(\bk)$ is defined in Eq. \eqref{Ebasis2} and $\mathcal{K}_{\mathcal{O}_2}(\bq_1,\bq_2) \in \{F_2(\bq_1,\bq_2), 1, S_2(\bq_1,\bq_2)\}$ respectively. The diagrammatic representation of \eqref{a} is shown in Fig.~\ref{D2f2a}, for $\mathcal{K}_{\mathcal{O}_2} = F_2$. 

There are two different smoothing scales in Eq.~\eqref{a}. The halo smoothing scale $R_{\text{h}}$ has a physical meaning and corresponds to the size of the Lagrangian patch collapsing into the halo. However, since we don't know $R_{\text{h}}$ a priori, we will have to take into account our ignorance of this fact while constraining the physical bias parameters. In general, the cut-offs are not physical and should not appear in the model. One has to add appropriate counter terms to remove the cut-off dependence. We will discuss our choice of the counter term in a later section. The external smoothing scale $R_{\text{f}}$ is an analysis cutoff avoiding high-$k$ contributions to the estimator. As long as this requirement is satisfied the results will not depend on the choice of $R_{\text{f}}$ since it is consistently implemented.
\subsection{Cubic Fields}
The definition of cubic fields follows straightforwardly from the above considerations concerning quadratic fields. We define a smoothed cubic field in Fourier space as
\begin{equation}
\ba
\mathcal{D}_{3}[\de](\bk) = \int_{\bq_1}\int_{\bq_2}\de_{\text{G}}(\bq_1)\de_{\text{G}}(\bq_2)&\de_{\text{G}}(\bk-\bq_1-\bq_2)W_{\text{R}_{\text{f}}}(\bq_1)W_{\text{R}_{\text{f}}}(\bq_2) \\
&\times W_{\text{R}_{\text{f}}}(\bk-\bq_1-\bq_2)\mathcal{K}_{D_{3}}(\bq_1,\bq_2,\bk-\bq_1-\bq_2)\, 
\label{cubic}
\ea
\end{equation}
where $\mathcal{K}_{D_{3}}$ is cubic kernels after symmetrization. In our model, there are seven cubic bias fields as described in Eq.~\eqref{Ebasisnew}. The unsymmetrized cubic kernels are defined as
\begin{equation}
\ba
\mathcal{K}_{\mathcal{G}_2\de}(\bq_1,\bq_2,\bq_3) =\Bigg(\frac{(\bq_1\cdot\bq_2)^2}{q_1^2q_2^2}-1\Bigg)\, ,
\ea
\end{equation}
\begin{equation}
\ba
\mathcal{K}_{\mathcal{G}_3}(\bq_1,\bq_2,\bq_3) = -\frac{1}{2}\Bigg[\frac{(\bq_1\cdot\bq_2) (\bq_1\cdot\bq_3) (\bq_2\cdot\bq_3) }{q_1^2q_2^2q_3^2}-\frac{(\bq_1\cdot\bq_2)^2}{q_1^2q_2^2}-\frac{(\bq_1\cdot\bq_3)^2}{q_1^2
   q_3^2}-\frac{(\bq_2\cdot\bq_3)^2}{q_2^2 q_3^2}\Bigg]\, ,
\ea
\end{equation}
\begin{equation}
\mathcal{K}_{\Gamma_3}(\bq_1,\bq_2,\bq_3) = 2 \Bigg(\frac{\big(\bq_1\cdot(\bq_2+\bq_3)\big)^2}{q_1^2(\bq_2+\bq_3)^2}-1\Bigg)\Bigg(F_2(\bq_2,\bq_3)-G_2(\bq_2,\bq_3)\Bigg)\, ,
\end{equation}
\begin{equation}
\mathcal{K}_{s^{(3)}}(\bq_1,\bq_2,\bq_3) = 2S_{2}(\bq_1,\bq_2+\bq_3)F_2(\bq_2,\bq_3)\, ,
\end{equation}
\begin{equation}
\mathcal{K}_{\de^{(1)}\de^{(2)}}(\bq_1,\bq_2,\bq_3) = 2 F_2(\bq_2,\bq_3)\, ,
\end{equation}
\begin{equation}
\mathcal{K}_{\de^{(3)}}(\bq_1,\bq_2,\bq_3) = F_3(\bq_1,\bq_2,\bq_3)\, .
\end{equation}
We cross correlate these cubic fields with the halo density field. The cubic fields are of order $\mathcal{O}(\de^3)$ and they correlate only with the linear and cubic operators in $\de_{\text{h}}$. The cross-correlation of the cubic fields with the quadratic fields are five-point functions which vanish in an infinite volume universe and hence do not contribute to the signal. However, for finite ensembles, these five-point functions do contribute to the noise. In order to make more precise measurements, we remove them to reduce the noise.

\begin{figure}[h]
\centering
\includegraphics[width=13cm]{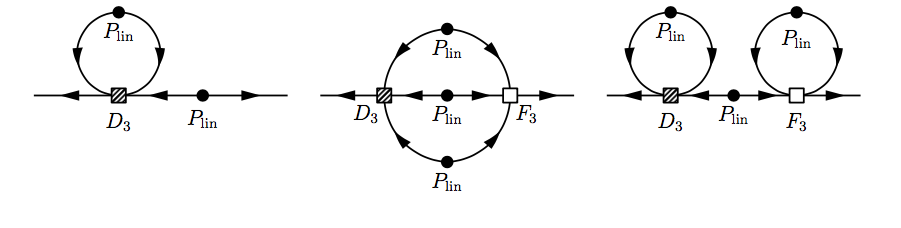}
\caption{Perturbative expressions for one-loop, and two-loop irreducible and two-loop reducible terms of $\langle \mathcal{D}_{3}(\bk)|\de_{\text{NL}}(\bk') \rangle$  are shown in diagrammatic form.  The propagators are represented by linear power spectra $P_{\text{lin}}$, the cubic field kernel $\mathcal{D}_{3}$ is represented by the hatched square. Finally, empty squares correspond to the gravitational kernel $F_{3}$. Loops correspond to integrals over all wavenumbers $\bq$ or $\bp$ and arrows represent the flow of momentum.}
\label{D3NL}
\end{figure}

\begin{figure}[h]
\begin{center}
\includegraphics[width=13cm]{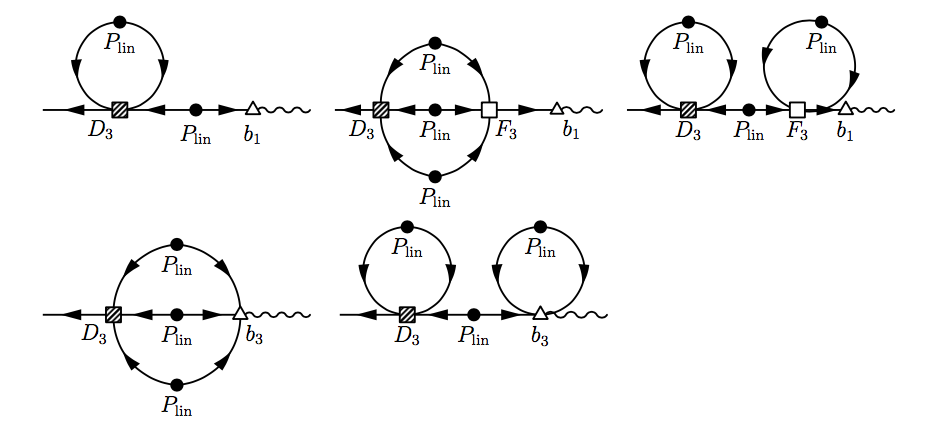}
\end{center}
\caption{Diagrams contributing to the correlation of cubic fields with the halo field in Eq.~\eqref{D3dh}. The triangles represent the linear, quadratic and cubic bias kernels. The straight lines are used to describe the density field, whereas halo fields are described by wiggly lines. The Feynman rules are discussed in detail in \cite{Baldauf:2010vn}.}
\label{D3diagram}
\end{figure}

We provide a step by step explanation as things get more elaborate at cubic level. First, we describe the cubic correlations with the non-linear density field $\de_{\text{NL}}$  up to order $\mathcal{O}(\de^6)$: 
\begin{equation}
\langle \mathcal{D}_3[\de](\bk)|\de_{\text{NL}}(\bk')\rangle' = \langle \mathcal{D}_3[\de](\bk)|\de^{(1)}(\bk')\rangle' + \langle \mathcal{D}_3[\de](\bk)|\de^{(3)}(\bk')\rangle'\, .
\end{equation}
The first term is the one-loop term and the second is the two-loop term. The two-loop term consists of an irreducible part and a reducible part, where the latter can be written as the product of two one-loop diagrams. The diagrammatic representation of these terms is shown in Fig.~\ref{D3NL} and the PT expressions are given as follows:
\begin{equation}
\text{One-Loop}\Rightarrow 3W_{\text{R}_{\text{f}}}(\bk)P_{\text{lin}}(k)\int_{\bq}P_{\text{lin}}(q) \mathcal{K}_{D_3}(\bk,\bq,-\bq)W_{\text{R}_{\text{f}}}(\bq)^2
\label{D3oneloop}
\end{equation}
\begin{equation}
\ba
\text{Two-Loop Reducible}\Rightarrow 9W_{\text{R}_{\text{f}}}(\bk)P_{\text{lin}}(k)&\int_{\bq_1}P_{\text{lin}}(q_1)W_{\text{R}_{\text{f}}}(\bq_1)\mathcal{K}_{D_3}(\bk,\bq_1,-\bq_1) \times\\
&\int_{\bq_2} P_{\text{lin}}(q_2) W_{\text{R}_{\text{f}}}(\bq_2)F_{3}(\bk,\bq_2,-\bq_2)
\ea
\label{D32loopR}
\end{equation}
\begin{equation}
\ba
\text{Two-Loop Irreducible}\Rightarrow 6\int_{\bq_1}&\int_{\bq_2}P_{\text{lin}}(|\bk-\bq_1-\bq_2|)P_{\text{lin}}(q_1)P_{\text{lin}}(q_2)W_{\text{R}_{\text{f}}}(\bq_1)W_{\text{R}_{\text{f}}}(\bq_2)\\
& \times W_{\text{R}_{\text{f}}}(\bk-\bq_1-\bq_2) \mathcal{K}_{D_3}(\bk-\bq_1-\bq_2,\bq_1,\bq_2)F_{3}(\bk-\bq_1-\bq_2,\bq_1,\bq_2)
\ea
\label{D32loopIR}
\end{equation}
Basically, the PT expressions and diagrams are similar in all cubic correlations up to order $\mathcal{O}(\de^6)$. We can write the cubic correlations with the halo density field as
\begin{equation}
\ba
\langle \mathcal{D}_{3}[\de](\bk)|\de_{\text{h}}(\bk') \rangle' &= b_1\langle \mathcal{D}_{3}[\de](\bk)|\de^{(1)}(\bk') \rangle' + b_1\langle \mathcal{D}_{3}[\de](\bk)|\de^{(3)}(\bk') \rangle' + \frac{b_3}{3!}\langle \mathcal{D}_{3}[\de](\bk)|\de^3(\bk') \rangle'  \\
&+ b_{\mathcal{G}_2\de}\langle \mathcal{D}_{3}[\de](\bk)|\mathcal{G}_2\de(\bk') \rangle' 
+ b_{\mathcal{G}_3}\langle \mathcal{D}_{3}[\de](\bk)|\mathcal{G}_3(\bk') \rangle' +  b_{\Gamma_3}\langle \mathcal{D}_{3}[\de](\bk)|\Gamma_3(\bk') \rangle'\\
&+ b_{2}\langle \mathcal{D}_{3}[\de](\bk)|\de\de^{(2)}(\bk') \rangle' +  2b_{s^2}\langle \mathcal{D}_{3}[\de](\bk)|s^{(3)}(\bk') \rangle'+ \dots
\ea
\label{D3dh}
\end{equation}
The first two terms are the same as Eqs. \eqref{D3oneloop}, \eqref{D32loopR}, and \eqref{D32loopIR}, except here they are multiplied by the linear bias $b_1$. The other terms in Eq.~\eqref{D3dh} are two-loop terms which again consist of a reducible and an irreducible diagram. The PT expressions are the same as Eqs.~\eqref{D32loopR} and \eqref{D32loopIR} except the $F_3$ kernel is replaced by the cubic kernels from $\mathcal{O}_3$ in Eq.~\eqref{Ebasisnew} and we have to add extra smoothing functions corresponding to the intrinsic halo smoothing scale $R_{\text{h}}$. The diagrams are shown in Fig.~\ref{D3diagram}.

\subsection{Bispectrum and Trispectrum Estimators}
\label{sec:bitri}
The cross-spectra between quadratic fields and the halo field are nothing but the integrated bispectra:
\be
\langle \mathcal{D}_2[\de](\bk)|\de_{\text{h}}(\bk')\rangle' = \int_{\bq} \mathcal{K}_{D_2}(\bq,\bk-\bq)B_{\text{hmm}}(\bk',\bq,\bk-\bq)W_{\text{R}_{\text{f}}}(\bq)W_{\text{R}_{\text{f}}}(\bk-\bq)\, ,
\label{bispectrum}
\ee
where all of the matter fields in the bispectrum are Gaussian fields.
Similarly, the cross power spectra between a cubic fields and the halo field can be written as integrated trispectra
\be
\ba
\langle \mathcal{D}_{3}[\de](\bk)|\de_{\text{h}}(\bk')\rangle' = \int_{\bq} \int_{\bp} \mathcal{K}_{D_3}(\bk-\bq-\bp,\bq,\bp)&T_{\text{hmmm}}(\bk',\bp,\bk-\bq-\bp,\bq)\\
&\times W_{\text{R}_{\text{f}}}(\bq)W_{\text{R}_{\text{f}}}(\bp)W_{\text{R}_{\text{f}}}(\bk-\bq-\bp).
\label{trispectrum}
\ea
\ee
The estimators defined in Eqs.~\eqref{bispectrum} and \eqref{trispectrum} contain bispectrum and trispectrum information in an optimal way. We will use these estimators to constrain quadratic and cubic bias parameters. Note that the matter fields are the Gaussian fields.
An alternative way to estimate cubic bias parameters is to calculate the bispectrum of the Gaussian field, a squared field operator and the orthogonalized halo field $\langle \de_{\text{G}}\mathcal{D}_2\de_{\text{h}}\rangle$. This measurement probes the trispectrum in terms of a one-loop bispectrum rather than a two-loop power spectrum. This measurement retains additional external configuration dependence, but a detailed exploration of the performance of this estimator exceeds the scope of this paper. 
\subsection{Removing the UV sensitive diagrams}
\begin{figure}[h]
  \centering
\includegraphics[width=10cm]{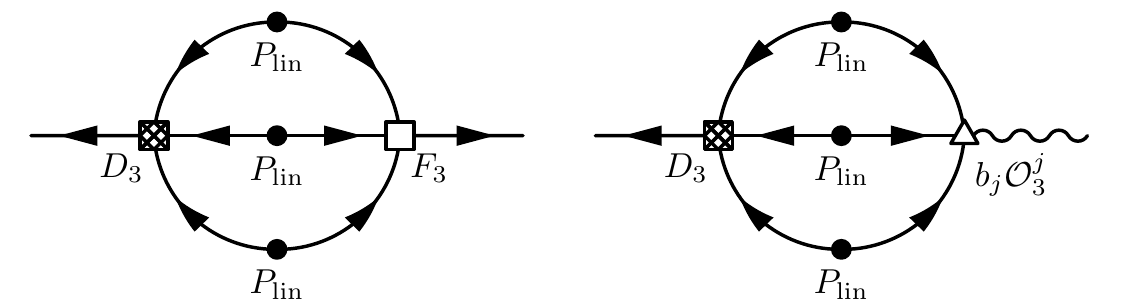} 
\caption{Cross-correlations of cubic fields with the orthogonalized fields contain only the two-loop irreducible diagram. The figure shows PT diagrams for $\langle D_3|\widetilde{F}_3 \rangle$ (left) and $\langle D_3|\widetilde{\mathcal{O}}_{3}^{j} \rangle$(right). }
\label{D2f2}
\end{figure}

The reducible diagrams introduced above contain a loop with two counteraligned momenta entering into a cubic kernel. These diagrams are highly cutoff or smoothing dependent. For instance, if we consider $\mathcal{O}_3=\delta^3$, the integral yields the variance of the field smoothed on scale $R_\text{h}$. As we describe in more detail in Appendix~\ref{app:UV}, a change of halo smoothing from $R_\text{h}=4h^{-1} \text{Mpc}$ to $R_\text{h}=6h^{-1} \text{Mpc}$ can lead to a order unity relative change in the amplitude of these contributions. Such a massive change in the template would lead to an equally significant change in the prefactors and thus bias the constraint on the bias parameters. In contrast the amplitude of the irreducible diagrams changes by a much smaller magnitude on the order of a few percent. We thus consider it important to remove the highly UV-sensitive contributions from our bias estimator. Fortunately, the reducible diagrams can be identified with the part of the halo field that correlates with the linear field.
The remaining parts of $\de_{\text{NL}}$ and $\delta_{\text{h}}$ orthogonal to the linear field are defined as
\begin{equation}
\widetilde{\de}_{A}(\bk) = \de_{A}(\bk) - \frac{\langle\de_{\text{G}}|\de_{A}\rangle}{\langle\de_{\text{G}}|\de_{\text{G}}\rangle}\de_{\text{G}}(\bk)\, ,
\label{Projection}
\end{equation}
where $A\in \{\text{NL},\text{h}\}$. These residual contributions only contribute to the irreducible diagrams.
The above definition can be extended to arbitrary operators $\mathcal{O}$. The cross-correlation of cubic operators with the orthogonal part of the halo field now only depends on the two-loop irreducible diagram. The irreducible diagram also depends on the smoothing scale; however, this dependence is less severe and if necessary can be taken into account by adding a counter term $\text{d}R$ as discussed in the next subsection. The final expression for the cross-correlation of the cubic fields with the projected halo density field that we use to constrain cubic bias is given by
\begin{equation}
\ba
\langle \mathcal{D}_3[\de](\bk)|\widetilde{\de}_{\text{h}}(\bk')\rangle' &= \sum_{j=1}^{j=7}b_{j}\langle \mathcal{D}_3[\de](\bk)|\widetilde{\mathcal{O}}_{3, j}(\bk') \rangle'.
\label{cubicmodel}
\ea
\end{equation}
\newcorrec{We diagrammatically describe the correlations of cubic fields with the orthogonalized fields in Fig. \ref{D2f2}.}
\subsection{Counter Term: Taylor expansion around $R_{\text{h}}$}


The models for the cross-correlations of quadratic and cubic fields with the halo fields in Eqs.~\eqref{quadraticmodel} and \eqref{cubicmodel} have some residual dependence on the halo smoothing scale $R_\text{h}$. For the cubic fields this is still the case after orthogonalization, but the dependence is less severe for the orthogonalized fields. In principle this cutoff or smoothing dependence would call for the inclusion of EFT inspired counterterms. \correc{The dependence of quadratic and cubic correlations on the halo smoothing scale $R_{\text{h}}$ is explained in more detail in Appendix.~\ref{app:UV}, where we explicitly quantify this effect and discuss the possibility of including EFT counter terms to remove this effect. In particular, we show that despite removing reducible diagrams and the fiducial scale being much larger than the halo scale, there are residual dependencies of the correlators on the halo smoothing scale at the several percent level.}

\correc{However,} the large number of necessary counterterms arising at the field level required to absorb the dependency of the results on the unknown halo smoothing scale $R_\text{h}$ motivates a more pragmatic approach. In particular, we are considering a Taylor expansion in the dependence on $R_\text{h}$ around the fiducial value $R_\text{h}=4h^{-1}\ \text{Mpc}$. The Taylor expansions of the quadratic and cubic correlations are thus given by

\begin{equation}
\langle \mathcal{D}_{2,i} |\mathcal{O}_{2, j}\rangle= \langle \mathcal{D}_{2,i}|\mathcal{O}_{2, j}\rangle\Big|_{R_{\text{h}}=4}  + \frac{\text{d}R}{2}\Bigg(\langle \mathcal{D}_{2,i}|\mathcal{O}_{2, j}\rangle\Big|_{R_{\text{h}}=4}-\langle \mathcal{D}_{2,i}|\mathcal{O}_{2, j}\rangle\Big|_{R_{\text{h}}=6}\Bigg)
\label{eq:d2o2}
\end{equation}
and
\begin{equation}
\langle \mathcal{\correc{D}}_{3,i}|\widetilde{\mathcal{O}}_{3,j}\rangle = \langle \mathcal{\correc{D}}_{3,i}|\widetilde{\mathcal{O}}_{3,j}\rangle\Big|_{R_{\text{h}}=4}  + \frac{\text{d}R}{2}\Bigg(\langle \mathcal{\correc{D}}_{3,i}|\widetilde{\mathcal{O}}_{3,j}\rangle\Big|_{R_{\text{h}}=4}-\langle \mathcal{\correc{D}}_{3,i}|\widetilde{\mathcal{O}}_{3,j}\rangle\Big|_{R_{\text{h}}=6}\Bigg)
\label{eq:d3o3}
\end{equation}
respectively.
\correc{The l.h.s. of Eqs.~\eqref{eq:d2o2} and \eqref{eq:d3o3} are now functions of d$R$. This dependency quantifies the effect of the deviation of the halo smoothing scale from its fiducial value $R_{\text{h}}=4 h^{-1}$ Mpc.  In the subsequent analysis, we will constrain this parameter d$R$ along with other bias parameters}.

\section{Methodology}
\label{section4}

\subsection{Numerical Simulations}
We use a suite of 15 realisations of a cosmological $N$-body simulation. The initial conditions are generated with the second order Lagrangian Perturbation Theory (\textbf{2-LPT}) code \cite{Scoccimarro:2011pz} at the initial redshift $z_
\text{i}=99$ and are subsequently evolved using \textbf{Gadget-2} \cite{Springel:2005mi}. The simulations are performed with $N_\text{p} = 1024^{3}$ dark matter particles in a cubic box of length $L=1500 h^{-1}$ Mpc with periodic boundary conditions. We assume a flat $\Lambda$CDM cosmology with the cosmological parameters $\Omega_\text{m}=0.272$, $\Omega_\Lambda=0.728$, $h=0.704$, $n_\text{s}=0.967$.

Dark matter halos in the final $z=0$ density field are identified using the Friends-of-Friends (FoF) algorithm with linking length $l=0.2$ times the mean inter particle distance. We also trace back the halo particles to the initial conditions to define the protohalos as progenitors of gravitational collapse. We will be using these protohalos to study the evolution of bias from Lagrangian to Eulerian space. The halos are binned in mass, with each bin spanning a factor of three in mass. The mass and number density of the five halo mass bins are given in Table~\ref{halotable}. Particles and halos are assigned to a regular grid using the Cloud-in-Cell (CIC) scheme. We Fourier transform the matter and halo density fields using the publicly available \textbf{FFTW} library\footnote{\url{http://www.fftw.org}}.

From the initial conditions we also extract the underlying Gaussian density field from which we generate the quadratic and cubic field using a sequence of multiplications with powers of the wavenumber in Fourier space, Fourier transform and multiplications of fields in configuration space. 

\begin{table}[H]
\small
\begin{center}
\begin{tabular}{|c|c|c|}
\hline
  \thead{Mass Bin} & \thead{Halo Mass $[10^{13} h^{-1}M_{\odot}]$}& \thead{Number Density} $[10^{-6}h^{3}$ Mpc$^{-3}$] \\  [0.2ex] 
\hline
I  & 0.773 & 627 \\ 
II  & 2.33 & 216 \\ 
 III  & 6.93 & 66.5 \\ 
 IV  & 20.1 & 16.5 \\ 
 V  & 56.8 & 2.48 \\
\hline
\end{tabular}
\end{center}
\caption{Halo mass bins employed in this study. We quote the mean mass of the sample and the number density of halos.}
\label{halotable}
\end{table}
\subsection{Parameter Estimation}
As described before, the natural statistics for estimating $b_1$ is the tree-level halo-matter power spectrum. To estimate $b_1$ we minimize $\chi^2_{\text{lin}}$ defined below
\begin{equation}
\chi^2_{\text{lin}} = \sum_{k_i}^{k_{\text{max}}}\Bigg(\frac{\hat{P}_{\text{hm}}(k_i)/\hat{P}_{\text{mm}}(k_i) - b_1}{\sigma(\hat{P}_{\text{hm}}(k_i)/\hat{P}_{\text{mm}}(k_i))}\Bigg)^2.
\label{b1chi2}
\end{equation}
Taking the ratio of two power spectra obtained from the same initial conditions cancels out the random fluctuations, resulting in the reduction of cosmic variance and improved constraints on $b_1$. The maximum wavenumber is chosen to be $k_{\text{max}} = 0.026 h$ Mpc$^{-1}$ to ensure that we are in the regime where linear theory and scale independent bias are applicable. 

To estimate the quadratic and cubic bias parameters we cross correlate three quadratic fields defined in Eq.~\eqref{quadfield} and a basis of cubic bias operators \eqref{cubic} with the orthogonalized halo density field. To do cosmic variance cancellation, we obtain the cross-spectra terms in Eq. \eqref{quadraticmodel} and Eq. \eqref{cubicmodel} from N-body simulations, rather than using the PT result. The motivation is again cosmic variance cancellation. 
At the field level $|P_{\text{sim}} - P_{\text{model}}|$ can be written as
\begin{equation}
\langle  \mathcal{D}_{2,i}|\Delta \de^{\text{quad}}_{\text{h}}  \rangle  = \sum_{j=1}^{j=3}\Big\langle  \mathcal{D}_{2,i}|\Big(\delta_{\text{h}} -b_{1}\delta_{\text{G}} -b_{j}\mathcal{O}_{2,j}\Big)\Big\rangle 
\label{4.2}
\end{equation}
and 
\begin{equation}
\langle \mathcal{\correc{D}}_{3,i}|\Delta \widetilde{\de}^{\text{cubic}}_{\text{h}} \rangle =  \sum_{j=1}^{j=7}\Big\langle \mathcal{\correc{D}}_{3,i}\Big|\Big((\widetilde{\delta}_{\text{h}}-b_{j}\widetilde{\mathcal{O}}_{3,j} - b_1\widetilde{\de}^{(2)} - b_2\widetilde{\de}^2 - b_{s^2}\widetilde{s}^2\Big)\Big\rangle
\label{4.3}
\end{equation}
for quadratic and cubic statistics respectively\footnote{\correc{Here $\mathcal{D}_2$ and $\mathcal{D}_3$ describe the quadratic and cubic fields smoothed with $R_{\text{f}}=20 h^{-1}$ Mpc. On the other hand, $\mathcal{O}_2$ and $\mathcal{O}_3$ describe the quadratic and cubic basis operators smoothed with a halo smoothing scale $R_{\text{h}}$}}. The tilde stands for orthogonalized fields. Note that in Eq.~\eqref{quadraticmodel} and Eq.~\eqref{cubicmodel} we omitted odd-correlators, that is the cross-correlations of the quadratic fields with the linear density field or cubic fields with the quadratic fields. These cross-spectra are zero in an infinite volume limit. However, in a finite simulation volume these correlations contribute to the covariance matrix. In fact, the odd cross-correlations are the leading source of noise, which  can be reduced by removing these contributions at the field level in Eq.~\eqref{4.2} and Eq.~\eqref{4.3}.

We define the $\chi^2$ for the quadratic and cubic statistics as
\begin{equation}
\chi^2_{\text{quad}} = \sum_{k_j=k_{\text{min}}}^{k_{\text{max}}}\sum_{i=1}^{i=3}\Bigg(\frac{\langle  \mathcal{D}_{2,i}(\bk_j)|\Delta \delta^{\text{quad}}_{\text{h}}(\bk'_j)\rangle'}{\sigma(\langle  \mathcal{D}_{2,i}(\bk_j)|\Delta \delta^{\text{quad}}_{\text{h}}(\bk'_j)\rangle')} \Bigg)^2
\label{chi2quad}
\end{equation}
and
\begin{equation}
\chi^2_{\text{cubic}} =\sum_{k_j=k_\text{min}}^{k_{\text{max}}} \sum_{i=1}^{i=7}\Bigg(\frac{\langle \mathcal{\correc{D}}_{3,i}(\bk_j)|\Delta \widetilde{\delta}^{\text{cubic}}_{\text{h}}(\bk'_j)\rangle'}{\sigma(\langle  \mathcal{\correc{D}}_{3,i}(\bk_j)|\Delta \widetilde{\delta}^{\text{cubic}}_{\text{h}}(\bk'_j)\rangle')} \Bigg)^2.
\label{chi2cubic}
\end{equation}
The maximum wavenumber we use in our analysis is $k_\text{max} = 0.056 h$ Mpc$^{-1}$. In Eq. \eqref{chi2quad} and Eq. \eqref{chi2cubic} we sum over quadratic and cubic fields. After defining the chi-squared for the linear, quadratic and cubic statistics                                 we run the MCMC chains to get the best-fit bias parameters that minimize the joint chi-squared (or joint likelihood function), which is defined as
\begin{equation}
\chi^2 = \chi^2_{\text{lin}}+ \chi^2_{\text{quad}}+\chi^2_{\text{cubic}}\, .
\label{eq:chitot}
\end{equation}


\section{Results} 

\label{section5} 

\subsection{Some Preliminary Checks}
\begin{figure}[t]
\begin{center}
\includegraphics[width=\textwidth]{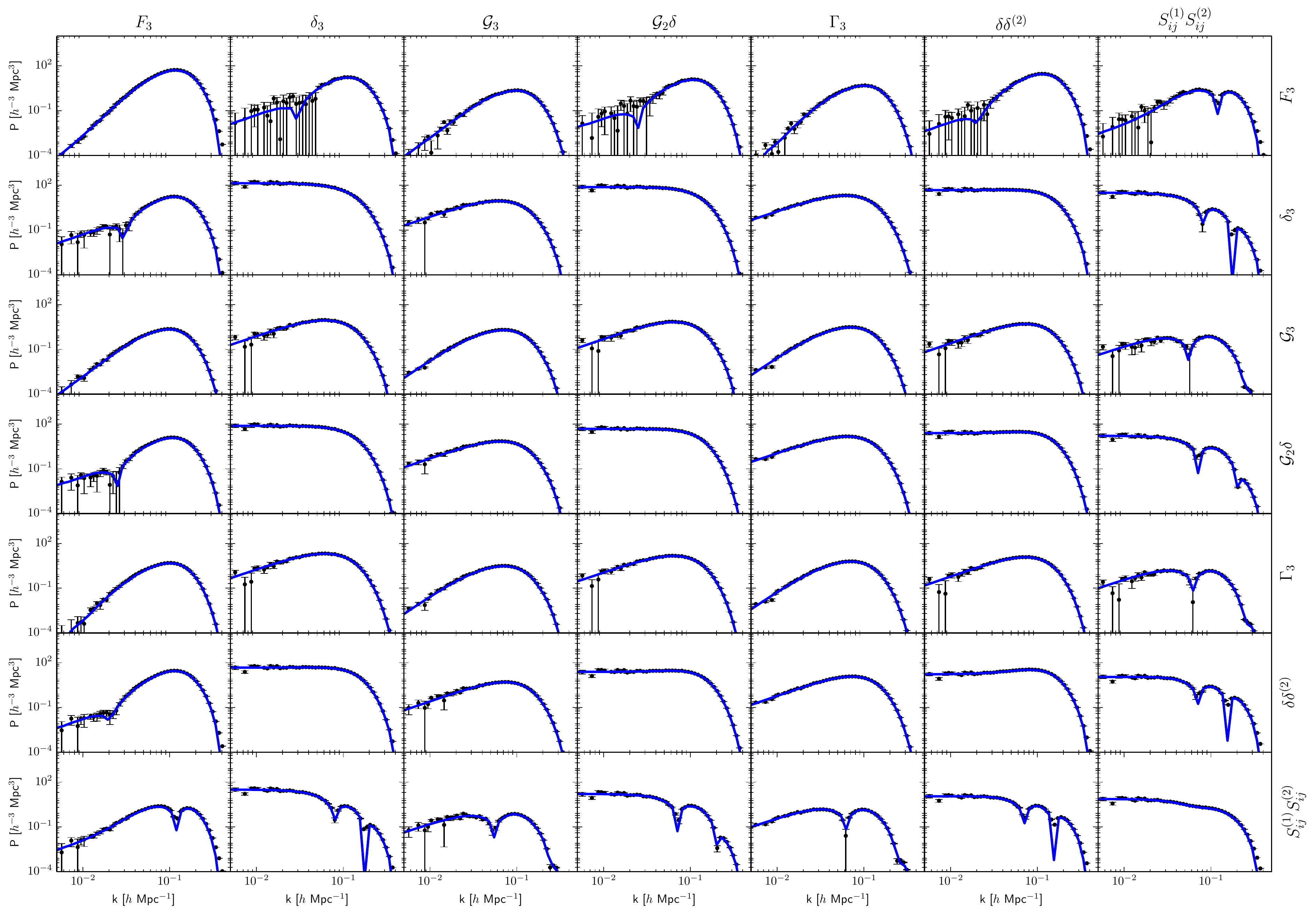}
\end{center}
\caption{Irreducible parts of the cross-spectra of cubic fields. The solid lines are the numerical evaluation of the perturbation theory loop integrals in Eq.~\eqref{D32loopIR}, while the dots with errorbars show simulation data. The cubic fields from left to right are smoothed with $R_{\text{h}}=4 h^{-1}$ Mpc, while fields from top to bottom are smoothed with the fiducial halo smoothing scale $R_{\text{f}}=20 h^{-1}$ Mpc.}
\label{7by7}
\end{figure}

Before discussing our main results, we describe some preliminary checks as follows: 
\begin{itemize}

\begin{figure}[h]
\centering
\includegraphics[width=12cm]{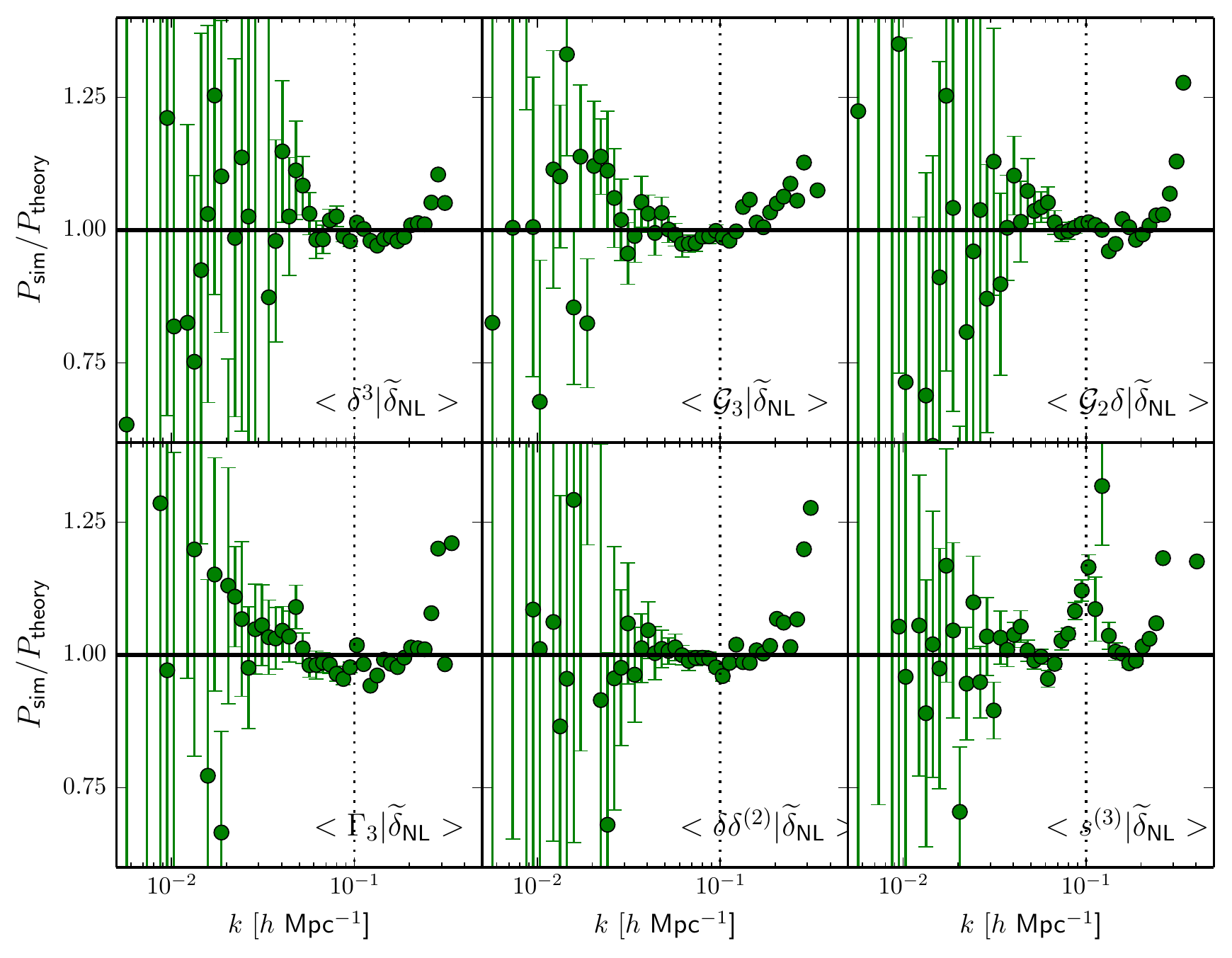}
\caption{
Ratio of the cross-correlations of cubic fields with the orthogonalized non-linear matter field as measured in simulations and predicted in perturbation theory. The cubic fields are smoothed with $R_{\text{f}} = 20 h^{-1}$ Mpc. As discussed in the text, these cross correlations are described by the two-loop irreducible diagram in PT. The vertical dotted line is drawn at $k = 0.1 h$ Mpc$^{-1}$ to separate the region of validity of the PT. For $k> 0.1 h$ Mpc$^{-1}$ PT results can not be trusted. To ensure convergence of PT, we have chosen the maximum wavenumber $k_{\text{max}} = 0.057 h$ Mpc$^{-1}$ for parameter estimation.}
\label{cubicratio}
\end{figure}

\item  Measuring bias parameters from large-scale, tree-level bispectrum and trispectrum is the cleanest way to avoid the degeneracies of the bias parameters. Therefore, we want to choose the maximum wavenumber $k_{\text{max}}$ in our analysis such that we are in the regime where PT is valid. To get an idea of the regime of validity of the tree-level trispectrum, we show in Fig.~\ref{cubicratio} the ratio of the cross-correlations of cubic fields with the non-linear matter field as measured in simulations and predicted by perturbation theory. We see that the data points start deviating from theory around wavenumber $k=0.1 h $Mpc$^{-1}$, which means that as we go to higher $k$-modes, loop corrections in the $T_\text{hmmm}$ trispectrum become important. We therefore make the conservative choice of $k_{\text{max}} = 0.057 h$ Mpc$^{-1}$ to ensure that we remain in the perturbative regime.


\item  To check that the orthogonalized cross-spectra of cubic fields obtained from simulations agree with a numerical evaluation of the perturbation theory integrals in Eq.~\eqref{D32loopIR}, we plot the irreducible parts of cubic cross-correlations in the seven by seven matrix plot in Fig.~\ref{7by7}. The solid lines are predictions of perturbation theory, whereas the data points with errorbars show simulation results. We see an excellent agreement between the simulations and the numerical two-loop integrals.

\end{itemize}

\subsection{Bias Constraints}
We are now ready to discuss our main results. We measured the bias parameters in Lagrangian and Eulerian space and compare our results with the co-evolution predictions described in Eq.~\eqref{predictions1} and Eq.~\eqref{predictions2}. We then discuss Eulerian and Lagrangian models with different number of parameters. 
\begin{figure}[h]
\begin{center}
\includegraphics[width=11cm]{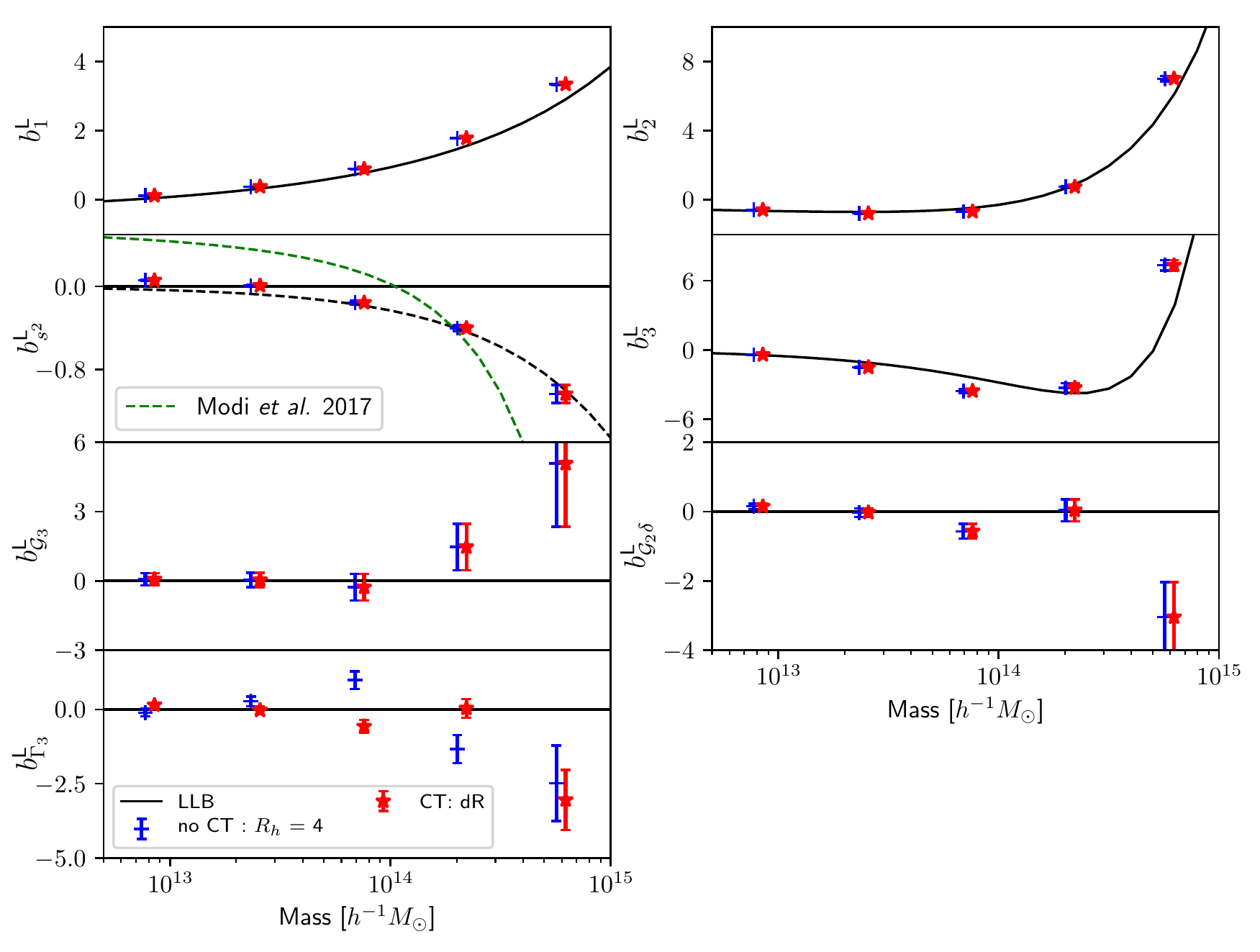}
\end{center}
\caption{Constraints on bias parameters from the protohalo statistics. The solid lines are the predictions of the local Lagrangian model calculated from the ST mass function. The dashed line is a fit to the observed non-zero Lagrangian tidal parameter given in Eq.~\eqref{Ltidal}. We are plotting two cases. The fit leading to the red points includes the counter term $\text{d}R$, which is just the Taylor expansion coefficient around our fiducial choice of smoothing scale $R_h$, while the blue points are without a counter term. The employed fiducial $R_\text{h}=4h^{-1} \text{Mpc}$ does not reflect the correct Lagrangian scale for all mass bins. Thus, we are more confident in the measurements with the counter term. \correc{We also overplot the fitting functions for $b_{s^2}$ given in Eq.(22) of \cite{Modi:2016dah} (rescaled to Lagrangian space, shown by the dashed-green line), which  contrary to our findings indicates a positive tidal bias for low masses and a stronger effect for large masses.}}
\label{proto}
\end{figure}

\subsubsection{Lagrangian bias from protohalos}
We obtained the protohalo density field in Lagrangian space by tracing back the constituent particles and assigning the Lagrangian center of mass to the grid. In Lagrangian space all the operators involving non-linear gravitational kernels vanish and thus we discard $\de^{(2)}$, $\de\de^{(2)}$ and $s^{(1)}s^{(2)}$ as shown in Eq.~\eqref{Lbias2a}. However, we still cross-correlate full basis of three quadratic and seven cubic operators with the protohalo density field in Lagrangian space and measure all bias parameters up to cubic order. The resulting bias measurements are shown in Fig.~\ref{proto}. We have detected a clear evidence of the presence of non-local Lagrangian tidal bias, the mass dependence of which is well captured by the fitting function
\begin{equation}
b_{s^2}^{\text{L}}(M) = -\frac{1}{2}\left(\frac{M}{4\times 10^{14}h^{-1}M_\odot}\right)^{0.8} 
\label{Ltidal}
\end{equation}
shown by the dotted black line in Fig.~\ref{proto}. This fitting function will be the basis of predictions of the LLB+$b_{s^2}^\text{L}$ model in the rest of this paper.
The measurements of the linear Lagrangian bias $b^{\text{L}}_{1}$ are strongly constrained by the halo-matter cross power spectrum and are in good agreement with the trends of the ST bias function, except for two highest mass bins that show slight deviation. Note that we include the ST bias predictions only as a reference to guide the eye rather than expecting perfect agreement.
Similarly, the measurements of non-linear local Lagrangian bias parameters $b_{2}^{\text{L}}$ and $b_{3}^{\text{L}}$ qualitatively agree with the predictions of the ST bias function. However, quantitatively we see deviations which are more obvious in the case of $b_{3}^{\text{L}}$. The theory lines for $b_{2}^{\text{L}}$ and $b_{3}^{\text{L}}$ are calculated from the second and third derivatives of the mass function. The detection of non-zero Lagrangian tidal bias clearly shows the failure of the spherical collapse model, partially explaining the disagreement of the measurements of local Lagrangian bias parameters with the theory predictions. A previous attempt at measuring cubic local Lagrangian bias was presented in \cite{Modi:2016dah}, \correc{where in agreement with our results, evidence for negative Lagrangian tidal bias at the high mass end is found. At the quantitative level however, their measurements and in particular their fitting function indicate a larger effect than what we find here. In particular, we don't find any evidence for positive $b_{s^2}^\text{L}$ at the low-mass end. For reference, we overplot their fitting function in Fig.~\ref{proto}.}

Next, we consider the non-local cubic bias parameters. We do not find significant detection of the presence of $b_{\Gamma_3}^{\text{L}}$, $b_{\mathcal{G}_2\delta}^{\text{L}}$, and $b_{\mathcal{G}_3}^{\text{L}}$ for low masses. Mass bin V, however, shows some mild evidence for non-vanishing cubic non-local Lagrangian bias. We have to caution however, that the employed smoothing and cutoff scales might be insufficient to suppress the impact of derivative bias corrections for these high mass, large radius tracers (see for instance \cite{Baldauf:2014fza,Modi:2016dah} for the scale dependence of Lagrangian bias).

\correc{Fig.~\ref{proto} reveals statistically significant changes between the fits with and without the counterterm d$R$.  Naively, one might have expected that the halo scale dependence is insignificant due to the large fiducial smoothing scale $R_\text{f}$. However, as we discuss in detail in App.~\ref{app:UV}, there is a several percent level residual dependency of the correlators on the halo smoothing scale. This sensitivity is at the same order as the relative errors on some of the bias parameters and can thus induce significant parameter shifts. At the same time, the presence of the counterterm can actually account for some of the stochasticity in the data. Thus, the constraints including the counterterm may show smaller error bars despite the larger parameter set.}

\begin{figure}[t]
\begin{center}
\includegraphics[width=11cm]{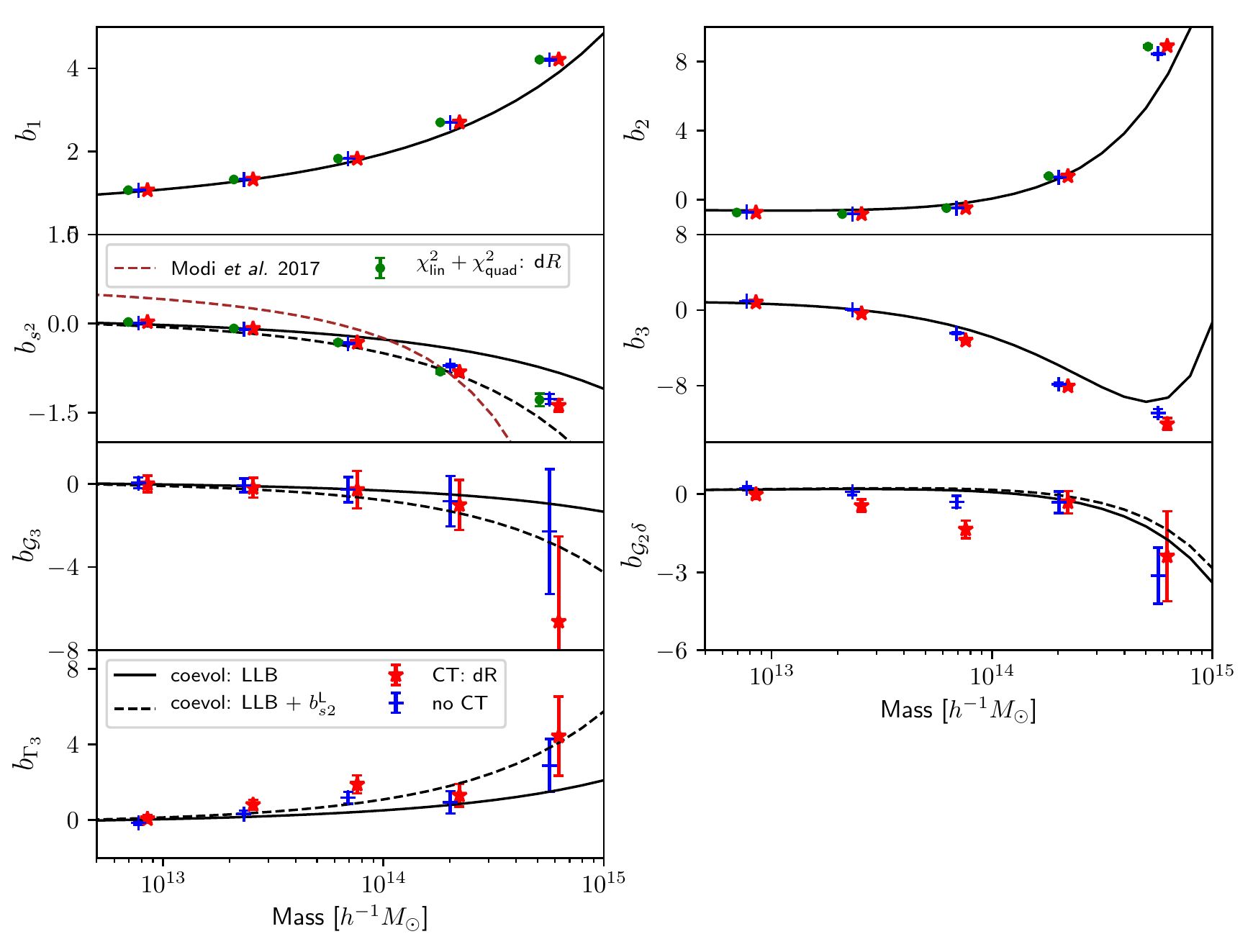}
\end{center}
\caption{Constraints on Eulerian bias parameters from seven and eight parameter fits to the late-time halo field. The red points depict the fits with an eight parameter model including the counterterm $\text{d}R$, whereas the green points show the results from a seven parameter model without the counterterm. The black dashed lines show the co-evolution prediction based on a local Lagrangian bias model and the black lines arise from a local Lagrangian model extended by a non-vanishing tidal term whose amplitude was fitted in Lagrangian space and is given by Eq.~\eqref{Ltidal}. Green points show constraints on linear and quadratic biases obtained from a fit to $\chi^2_\text{lin}+\chi_\text{quad}^2$. The constraints are in perfect agreement with the results from the full fits, which supports the consistency of our model and fitting procedure. \correc{The fitting function of \cite{Modi:2016dah} for $b_{s^2}$ is shown by the brown dashed curve.}}
\label{cubicmass}
\end{figure}

\subsubsection{Eulerian bias from the late-time halo field}
We now turn to the constraints on Eulerian bias parameters. In Fig.~\ref{cubicmass} we show the bias constraints for five mass bins obtained from seven and eight parameter fits to the late-time halo field. The solid lines are the predictions of the co-evolution of the local Lagrangian bias model, whereas the dashed lines are the predictions of the co-evolution of the local Lagrangian bias model extended by a non-local Lagrangian tidal term (LLB+$b_{s^2}^\text{L}$). We have plotted the constraints with and without the counter term $\text{d}R$. The measurements of the local Eulerian bias parameters $b_{1}^{\text{E}}$, $b_{2}^{\text{E}}$ and $b_{3}^{\text{E}}$ are following the trends of the ST bias function, with slight deviations towards the high mass end. As we noted before, we don't expect perfect agreement with this particular bias function. Our measurements of the tidal bias $b_{s^2}$ fall below the prediction based on co-evolution of the local Lagrangian bias model. The reason for this is the presence of the initial Lagrangian tidal field discussed in the previous section. The measurements show a preference for the predictions of the LLB+$b_{s^2}^\text{L}$ model. To check the consistency of our model, we have also performed fits to the propagator and quadratic field correlators ($\chi^2_\text{lin}+\chi^2_\text{quad}$) using only $b_1^\text{E}$, $b_2^\text{E}$, $b_{s^2}^\text{E}$ and $\text{d}R$. We find that the constraints are in good agreement with the ones obtained from the full eight parameter fits to linear, quadratic and cubic statistics.
\begin{figure}[t]
\begin{center}\includegraphics[width=11cm]{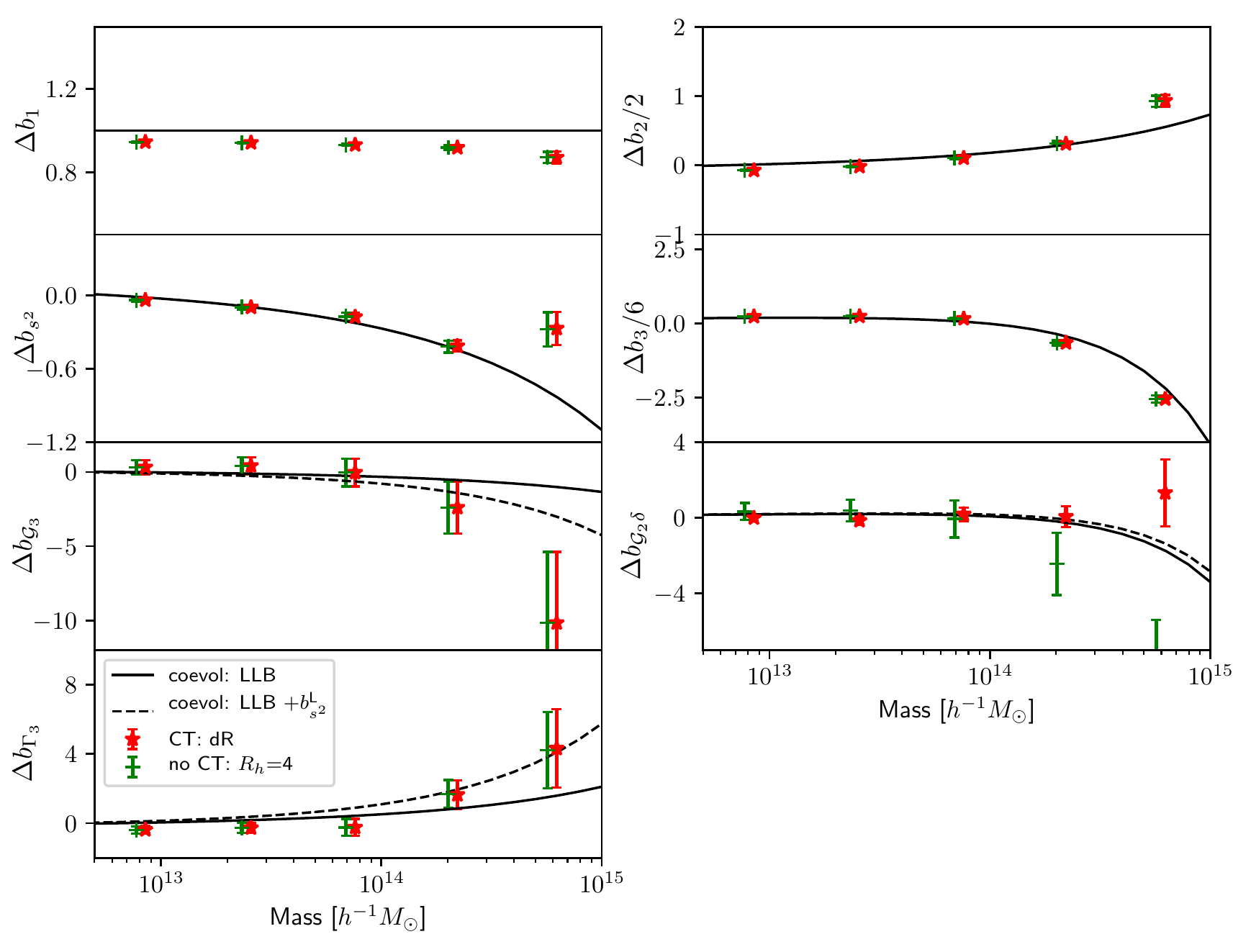}
\end{center}
\caption{Co-evolution Check: Difference of Eulerian and Lagrangian constraints on the local and non-local cubic bias parameters. We overplot the predictions of the local Lagrangian bias model (shown by solid black curves) and the Lagrangian model with an initial tidal field (LLB+$b_{s^2}^\text{L}$ shown by black dotted lines). At low mass the bias generally shows the trends of local Lagrangian bias, but at the hight mass end there are deviations especially for $b_{\Gamma_{3}}$ and $b_{\mathcal{G}_3}$.}
\label{cubicmass2}
\end{figure}

\begin{figure}[h]
\centering
\begin{subfigure}{.33\textwidth}
  \centering
\includegraphics[width=5cm]{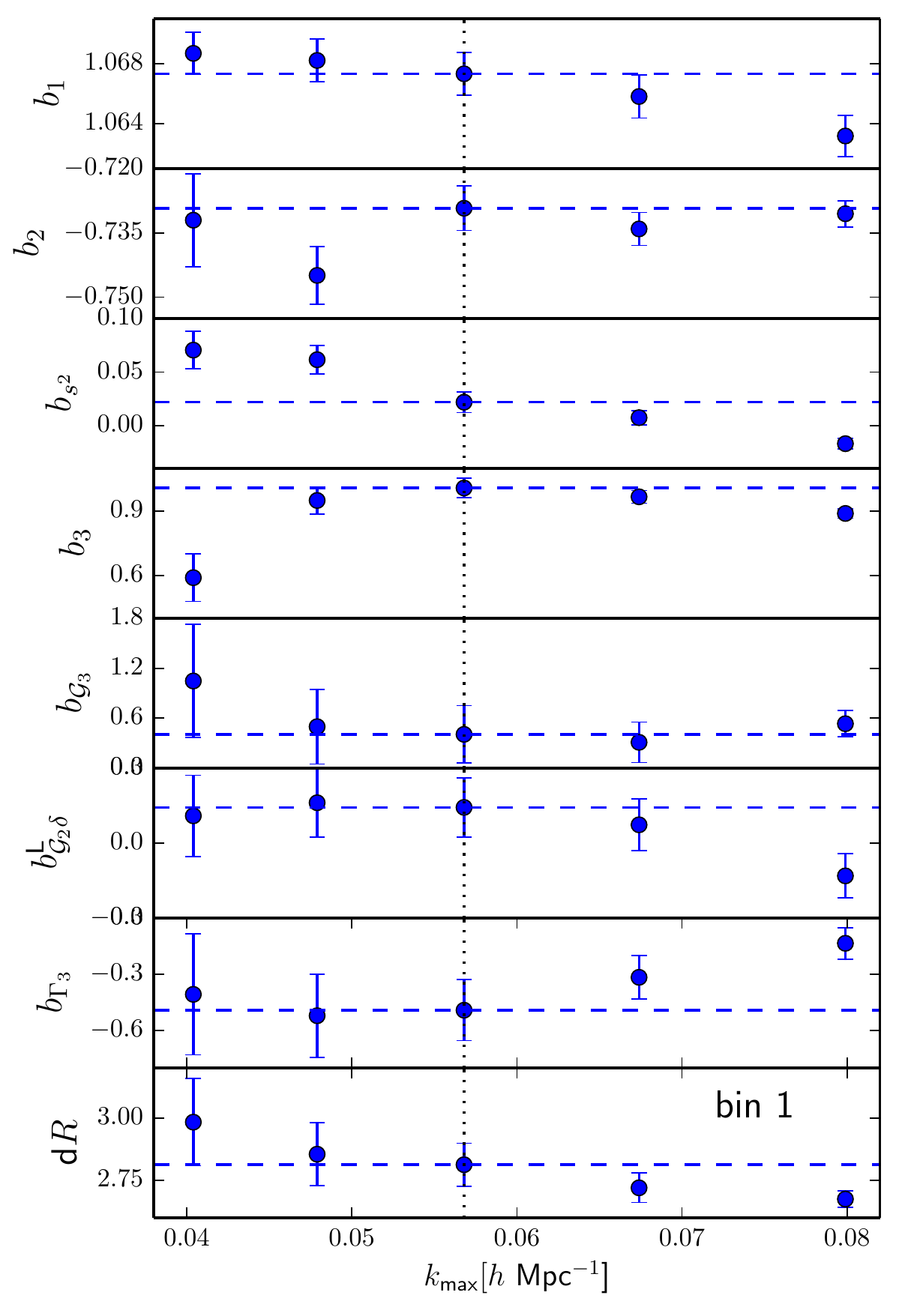}
  \caption{bin I}
  \label{rhplots:a}
\end{subfigure}
\begin{subfigure}{.33\textwidth}
  \centering
\includegraphics[width=5cm]{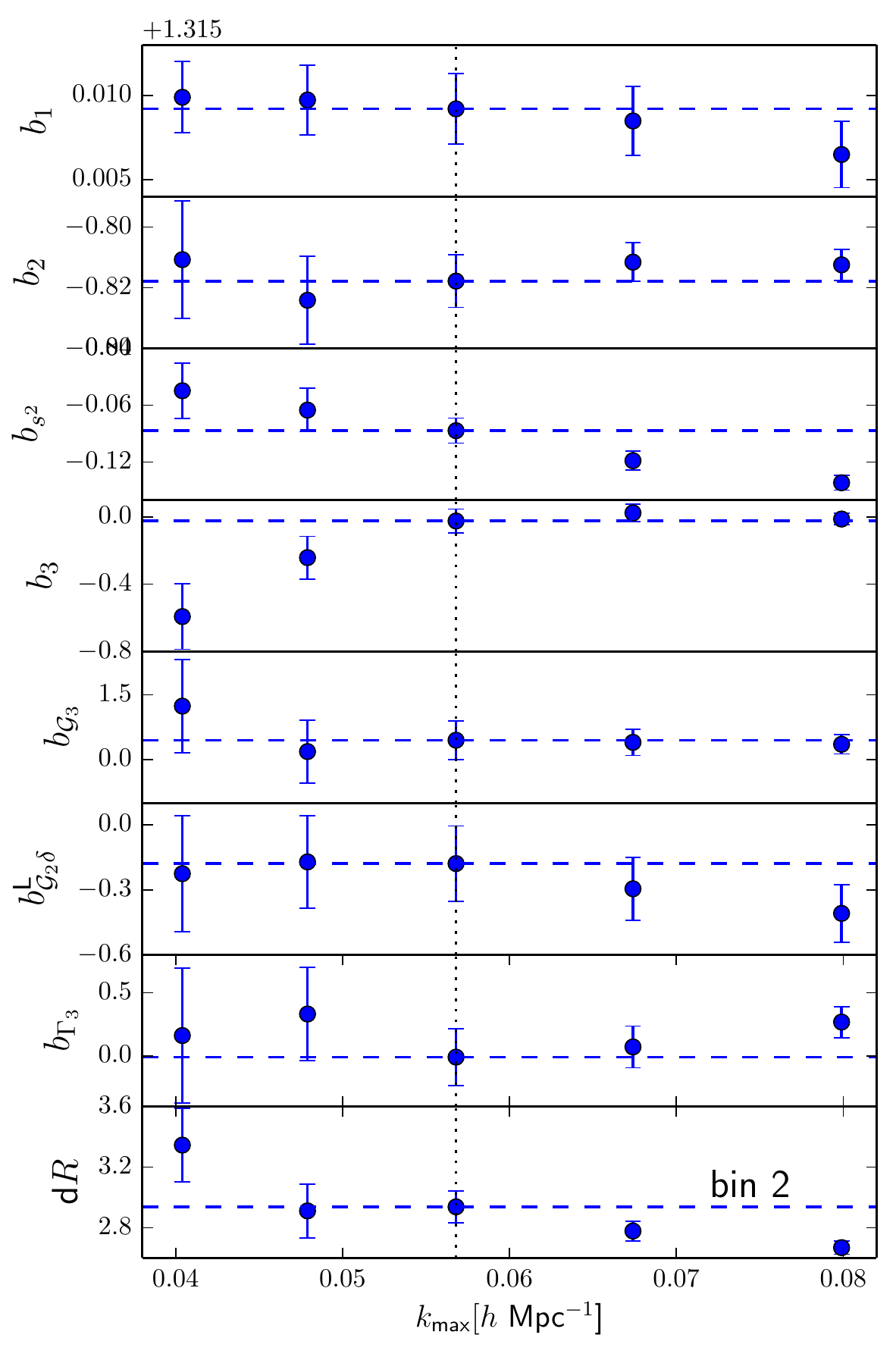}
  \caption{bin II}
  \label{rhplots:b}
\end{subfigure}%
\begin{subfigure}{.33\textwidth}
  \centering
\includegraphics[width=5cm]{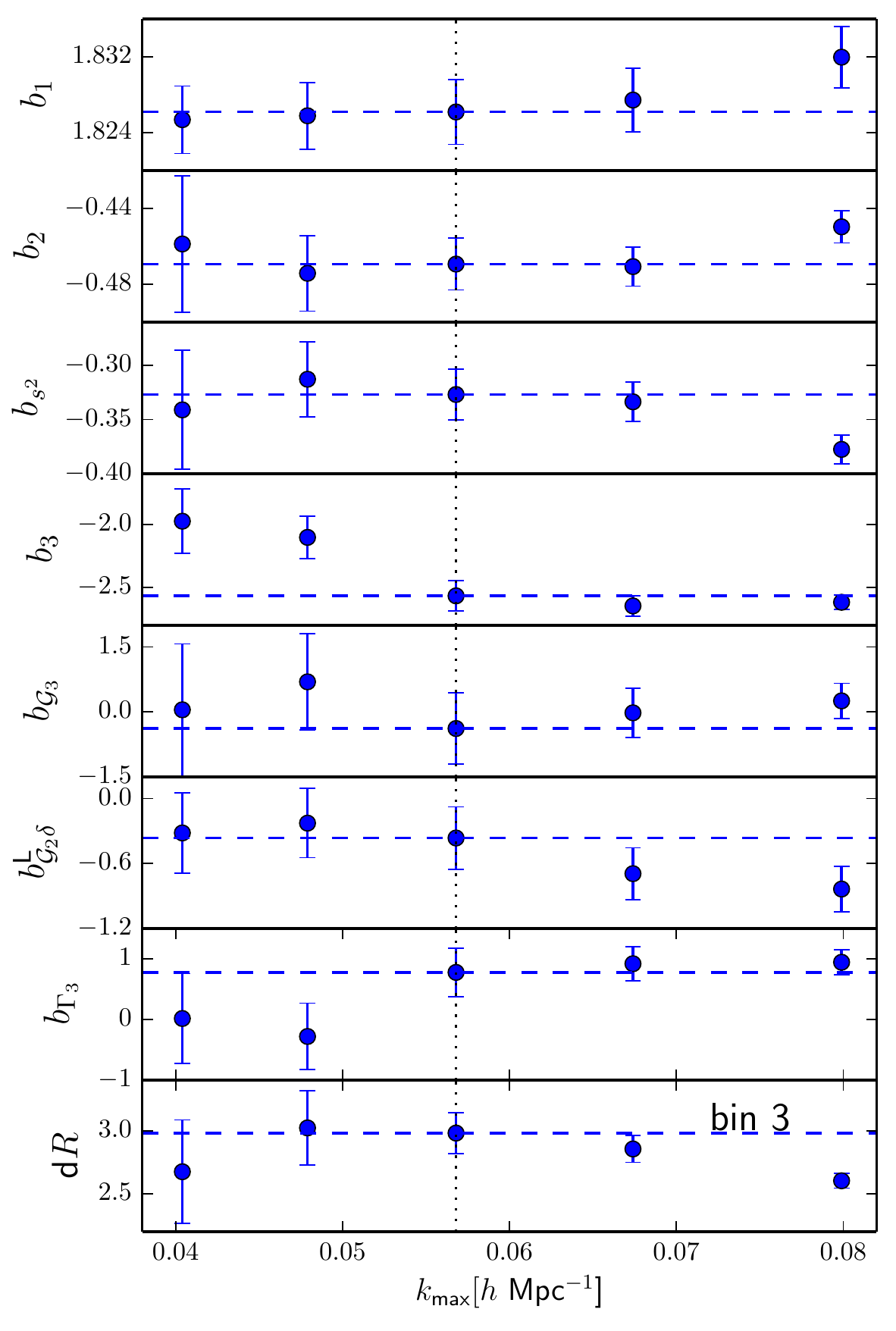}
  \caption{bin III}
  \label{rhplots:b}
\end{subfigure}
\begin{subfigure}{.33\textwidth}
  \centering
\includegraphics[width=5cm]{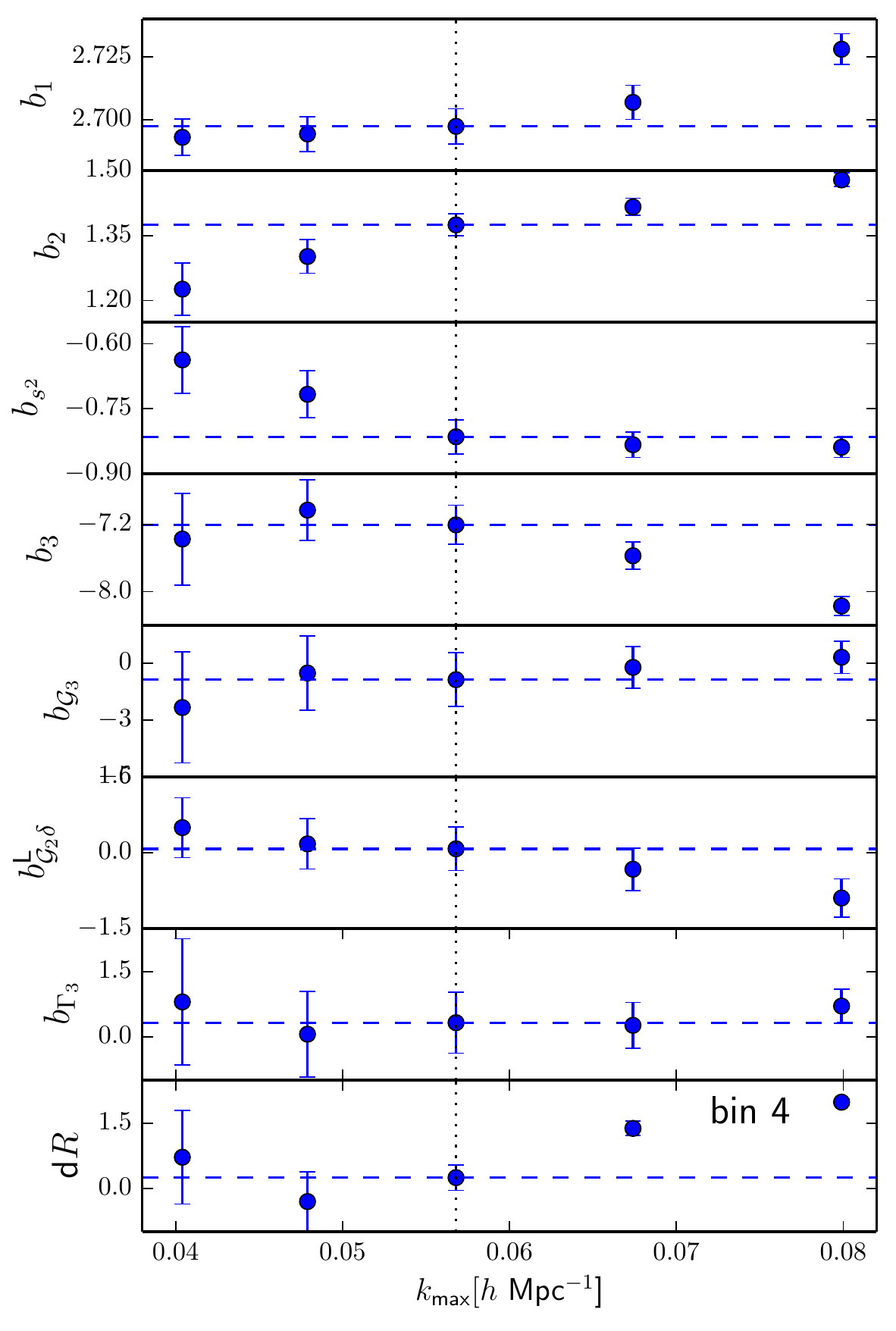}
  \caption{bin IV}
  \label{rhplots:a}
\end{subfigure}%
\begin{subfigure}{.33\textwidth}
  \centering
\includegraphics[width=5cm]{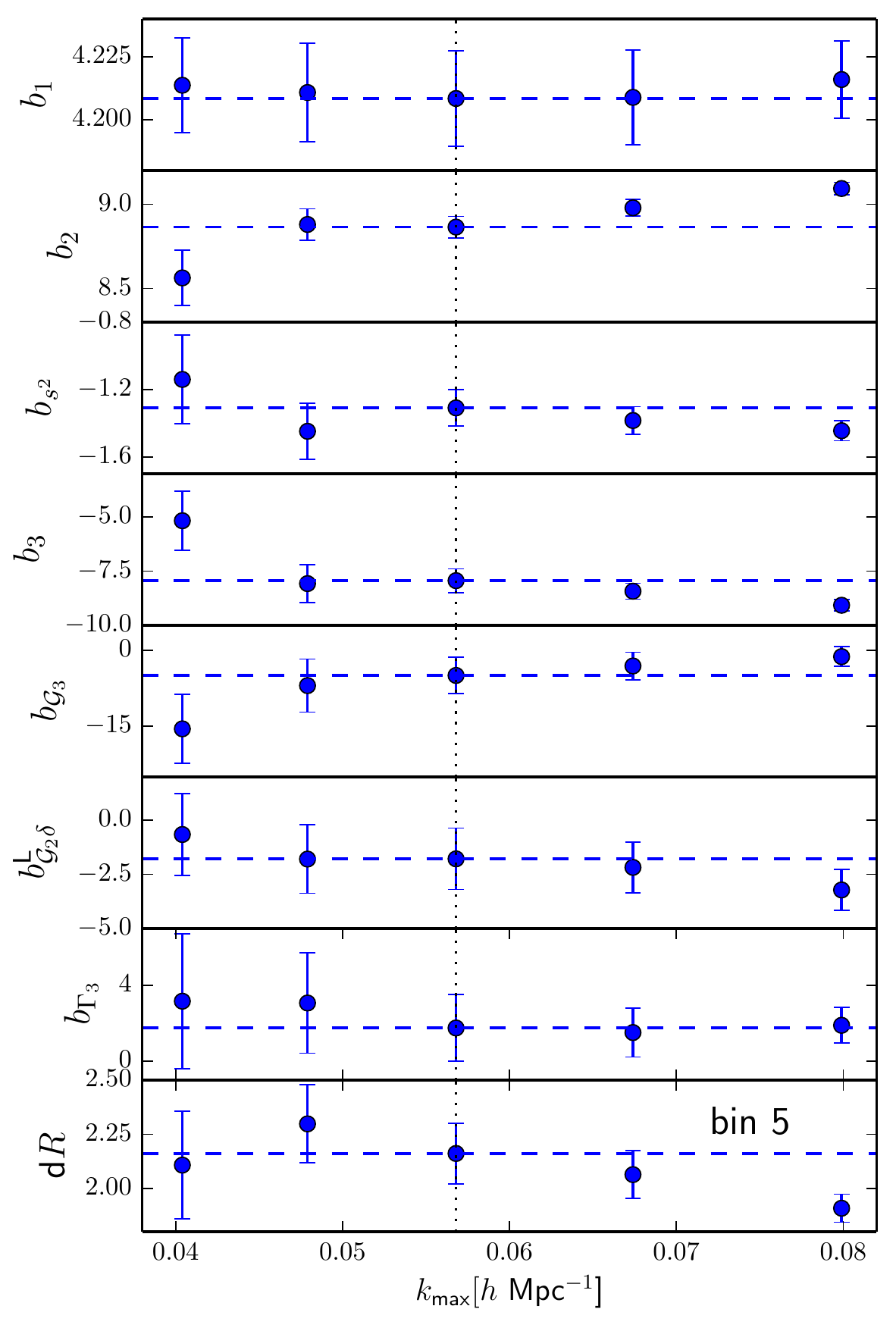}
  \caption{bin V}
  \label{rhplots:b}
\end{subfigure}
\caption{Measurements of the Eulerian bias parameter as a function of maximum wavenumber $k_{\text{max}}$. The horizontal blue-dashed lines are the best-fit values evaluated at $k_{\text{max}} = 0.057 h$ Mpc$^{-1}$ (vertical dotted line).}
\label{rhplots}
\end{figure}
Even though we are fitting for bias measurements on large scales, as ensured by the cutoffs $k_{\text{max}}=0.057 h$ Mpc$^{-1}$ and $R_\text{f}=20h^{-1}$ Mpc, to avoid corrections from non-linear modes, the higher mass bins can already be affected by higher derivative corrections. Going beyond the (integrated) tree-level trispectrum requires additional bias parameters and the inclusion of higher derivative bias operators. In fact, it has been shown that in the framework of EFTofLSS including higher derivative bias in the model improves the model performance for massive halos \cite{Fujita:2016dne}. In the EFTofLSS, the halo density is written in terms of the expansion in $(k/k_{\text{NL}})$ and $(k/k_{\text{M}})$, where $k_{\text{NL}}$ is the non-linear scale of the theory and $k_{\text{M}}$ corresponds to the scale of the derivative bias. For massive halos $k_{\text{M}}$ decreases and therefore derivative corrections become more important compared to low mass halos.

\begin{figure}[h]
\begin{center}
\includegraphics[width=\textwidth]{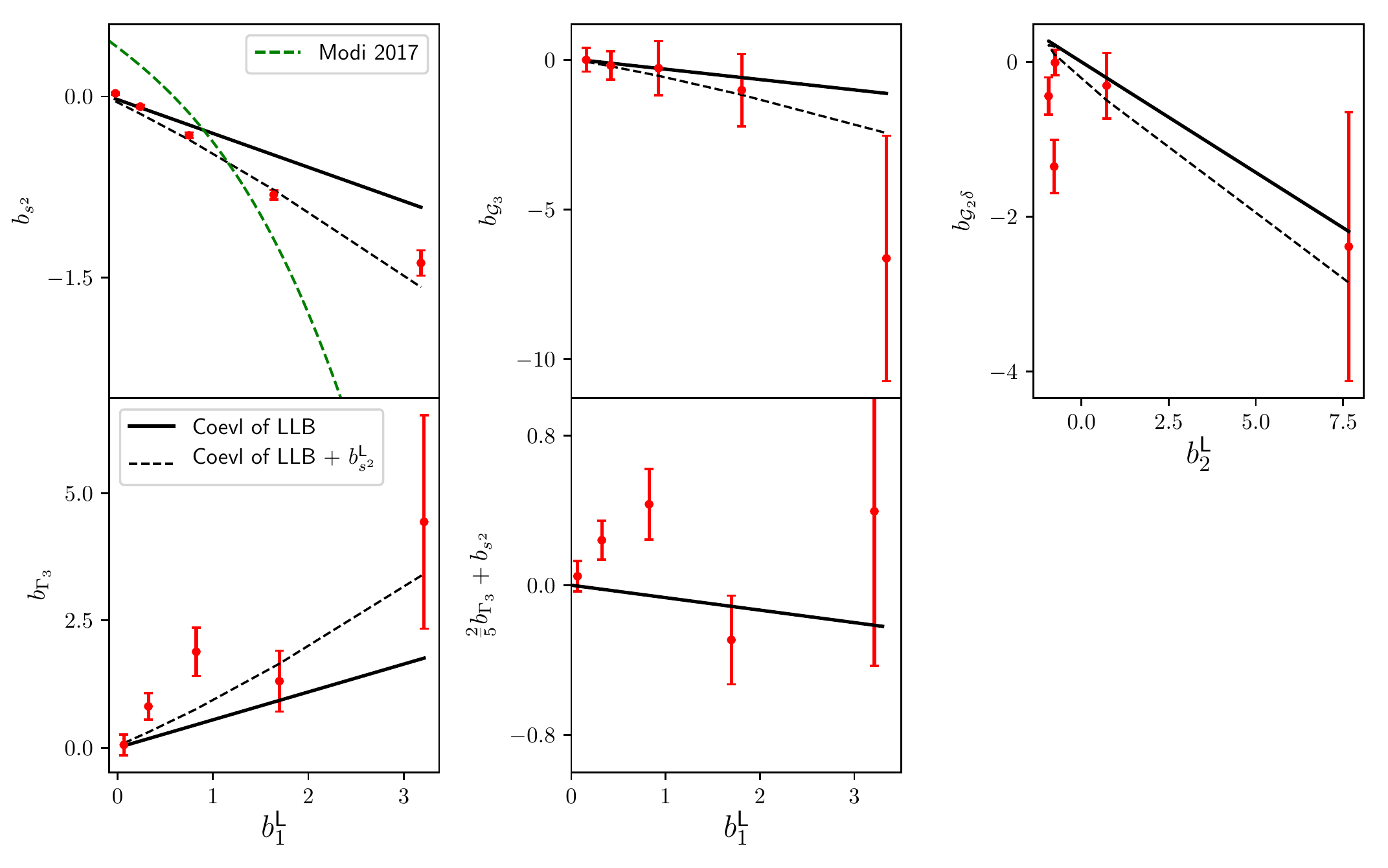}
\end{center}
\caption{Non-local bias constraints plotted against the linear Lagrangian bias and quadratic local Lagrangian bias. As stated in Eq.~\eqref{predictions1} the non-local bias parameters are predicted to follow a linear relation with the Lagrangian bias parameters shown as the black line. The data points show a preference for the model in which the LLB is extended by a Lagrangian tidal tensor contribution leading to the predictions in Eq.~\eqref{predictions2} (dashed curve). \correc{ We also overplot the fitting function  for $b_{s^2}$ given in Eq.~(22) of \cite{Modi:2016dah} (shown by the dashed-green line). Our measurements for $b_{s^2}$ are clearly not consistent with their fitting function. }}
\label{coevol}
\end{figure}
Coming back to bias measurements, we detect the presence of the non-local cubic bias in the late-time halo field at a significant level. The measurements for $b_{\mathcal{G}_3}$,$b_{\mathcal{G}_2\delta}$, and $b_{\Gamma_3}$, however, do not follow the predictions of the co-evolution of LLB; rather, in general, they are in slightly better agreement with the predictions of the co-evolution of LLB with initial Lagrangian tidal bias. 

In Fig.~\ref{cubicmass2} we highlight the dynamical contribution to the bias parameters by showing the difference of the initial and late-time measurements, and comparing them to the co-evolution predictions of the LLB and LLB+$b_{s^2}^{\text{L}}$. For $b_{\Gamma_3}$ the measurements follow the trend of the latter, except for a small deviation for mass bin III. For $b_{\mathcal{G}_2\delta}$ and $b_{\mathcal{G}_3}$ we see that the lowest three mass bins are in good agreement with the predictions of LLB+$b_{s^2}^{\text{L}}$, whereas the highest two mass bins clearly disagree from the predictions of both LLB and LLB+$b_{s^2}^{\text{L}}$ models. 

In Fig.~\ref{coevol} we plot $b_{s^2}$, $b_{\Gamma_3}$, and $b_{\mathcal{G}_3}$ against the linear bias and $b_{\mathcal{G}_2 \delta}$ against the non-linear local quadratic bias. In addition, we have also plotted the combination $2/5 b_{\Gamma_3} + b_{s^2} = -1/15 b^{\text{L}}_{1}$ that appears in the predictions for the halo-matter power spectrum at one-loop (see Sec.~\ref{sec:phm} below). We see that except for the fifth mass bin, the measurements of $b_{\mathcal{G}_3}$ are in good agreement with the predictions of both LLB and LLB+$b_{s^2}^{\text{L}}$. We see that $b_{\Gamma_3}$ is increasing with the linear bias but quantitatively mass bins II and III are clearly in disagreement with the co-evolution predictions. Unfortunately the errorbars are huge and affect the predictions for the one-loop halo-matter cross power spectra which we discuss below in Sec.~\ref{sec:phm}. These measurements are the best we can obtain from the cubic field method given our ensemble of simulations. 

Finally, we also show the bias measurements as a function of the cutoff wavenumber $k_{\text{max}}$ in Fig.~\ref{rhplots}. As one increases the maximum  $k$-mode, non-linear modes start affecting the measurements and should be taken care of by including appropriate loop corrections in the model. The measurements are fairly consistent on large scales up to our fiducial $k_{\text{max}}$.

\correc{During the final stages of this study \cite{Lazeyras:2017hxw} presented a similar study of cubic non-local bias. These authors use the correlation of cubic operators with the halo field without orthogonalization but remove the matter non-linearities from the halo field. This leaves closed loops in the bias operators, which we remove due to their strong UV-sensitivity (as discussed in Appendix \ref{app:UV}).
Their analysis goes to higher wavenumbers and subtracts a subset of odd correlators. They marginalize over residual $k^2$ dependencies for each of the cubic bias parameters to capture higher derivative and higher-order perturbative corrections, while we aim to account for these effects by fitting to $\text{d}R$.
Qualitatively we agree with their finding that the Eulerian non-local bias parameters are in tension with the predications based on the LLB model. 
Both approaches show the potential of the cubic field approach and future high-precision implementations should aim to combine the respective advantages of the two methods.}

\subsubsection{Constraints on Lagrangian bias parameters from different models}

Given that the final halo field shows reasonable agreement with the LLB+$b_{s^2}^{\text{L}}$ model, we consider a direct fit of the final halo field using the template in Eq.~\eqref{ourpred3s2}, i.e. linking the amplitude of the final cubic operators to the local Lagrangian bias and the Lagrangian tidal tensor bias. The free parameters in this fit are thus $\{b_1^\text{L},b_2^\text{L},b_3^\text{L},b_{s^2}^\text{L}\}$ and $\text{d}R$. We perform this same fit on the protohalos as well.

\begin{figure}[h]
\begin{center}
\includegraphics[width=11cm]{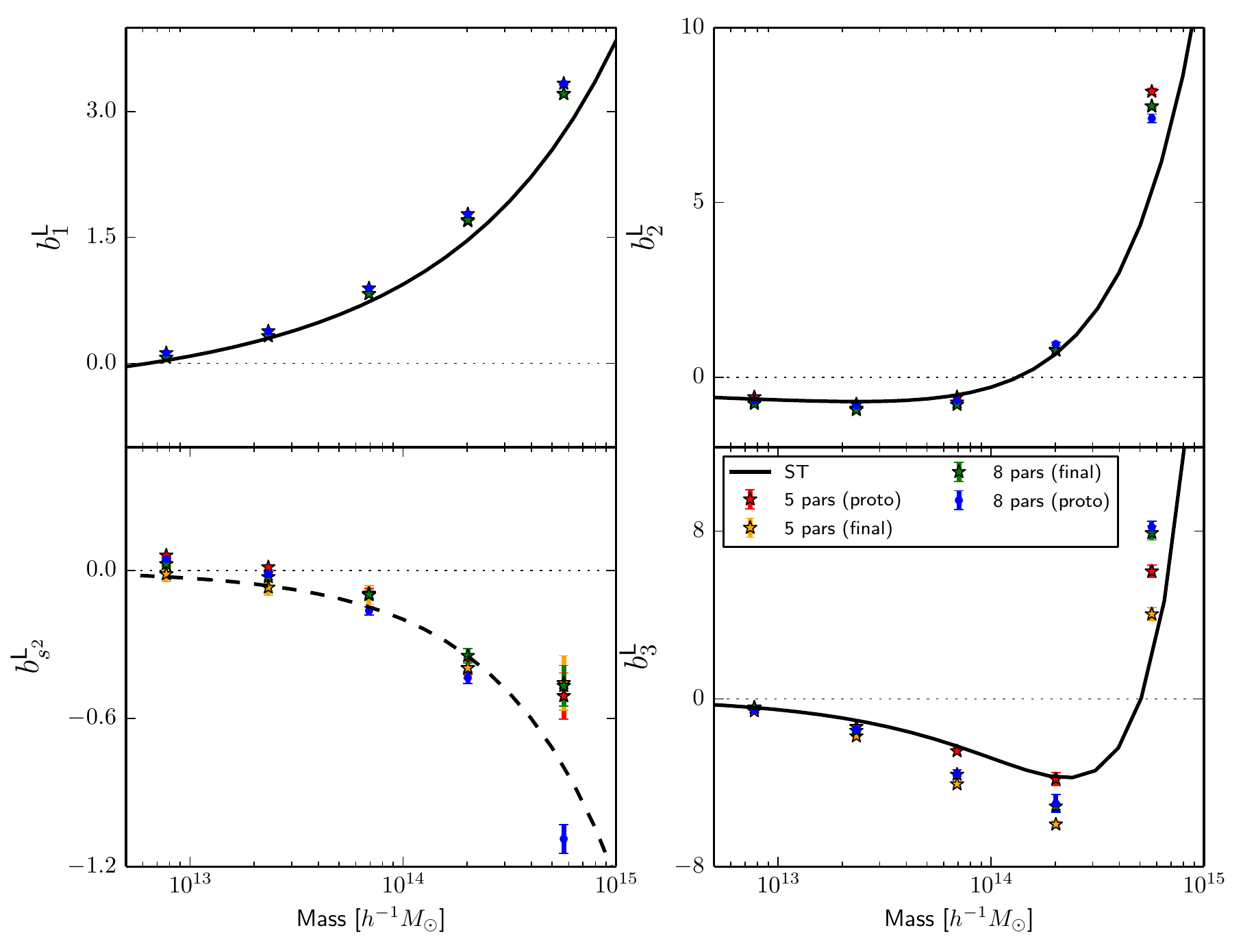}
\end{center}
\caption{Direct five parameter measurements of the Lagrangian bias parameters using the template Eq.~\eqref{ourpred3s2} for the final halo field
and reconstruction of the Lagrangian bias parameters from the full eight parameter fits of the final field described above.
We also show direct measurements of the Lagrangian bias parameters from a five and eight parameter fit to the protohalo field.
The solid lines are predictions of the ST bias function. The dashed curve for $b_{s^2}^{\text{L}}$ is our best-fit defined in Eq. \eqref{Ltidal}. }
\label{lagbias}
\end{figure}

We show the results of this study in Fig.~\ref{lagbias}, where we also show the Lagrangian bias parameters reconstructed from the eight parameter fits discussed above. In general we see a consistent picture, where all of the Lagrangian bias parameters obtained from the four different fitting procedures follow the same trend. There is some tension for the local cubic bias $b_3^\text{L}$, which is probably due to large parameter degeneracies in the protohalo fits. This might be partially due to us neglecting explicit $k^2$ bias contributions in the protohalo field as for instance predicted by the peak model \cite{Baldauf:2014fza,Modi:2016dah}.

Just for an example, Fig.~\ref{cubicmass} we plot marginalized posteriors of Lagrangian bias constraints for mass bin III obtained from the late-time halo field. We can see some mild degeneracies between the counterterm \text{d}R and the cubic local and quadratic non-local Lagrangian bias. These degeneracies are more severe in the constraints obtained from the protohalos.


\begin{figure}[h]
\begin{center}
\includegraphics[width=14cm]{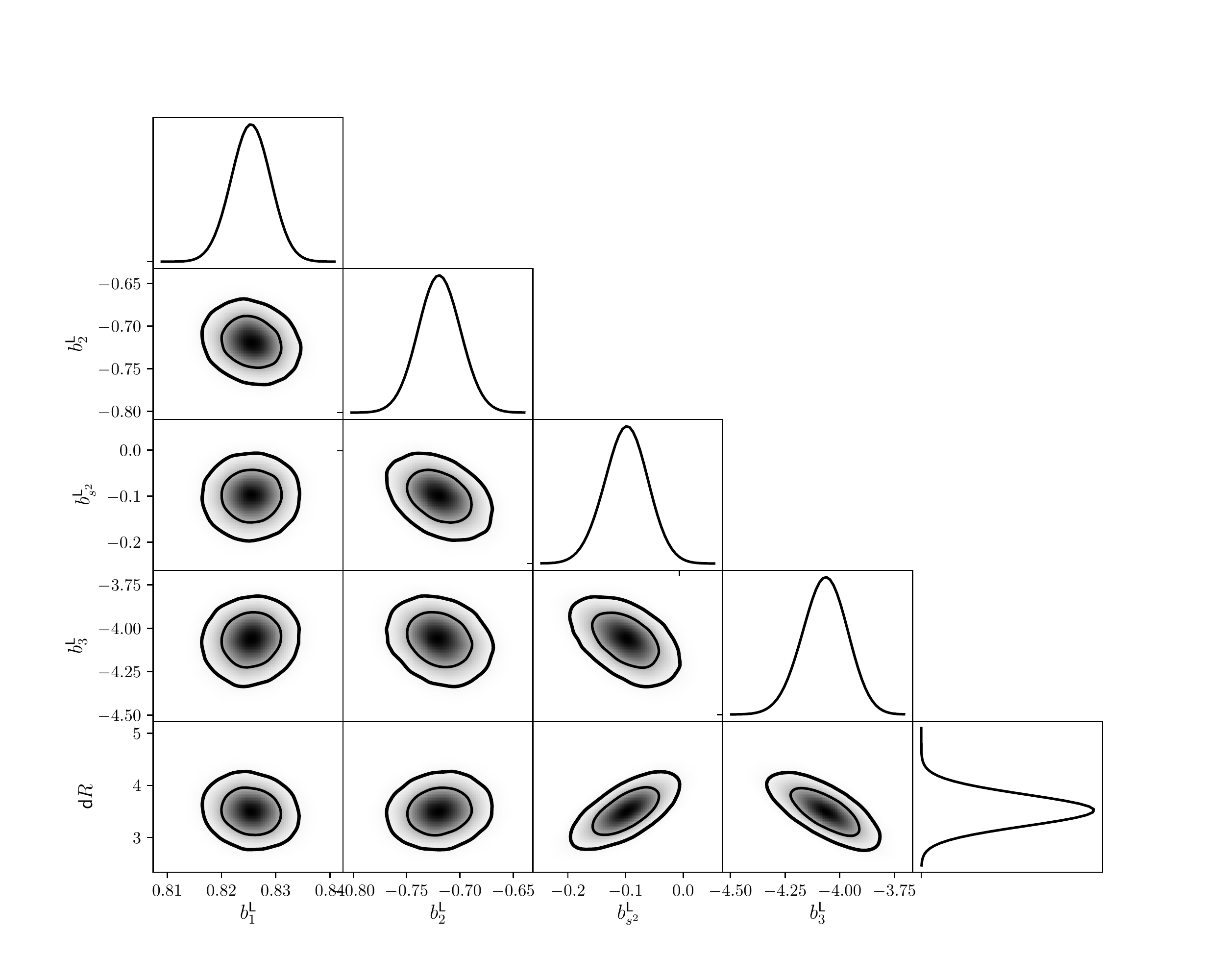}
\end{center}
\caption{Marginalized posteriors of Lagrangian bias constraints for mass bin III obtained from the late-time halo field. \correc{The counterterm d$R$ is shown in units of $1\ h^{-1}\text{Mpc}$. The dark and light regions represent 64.1 $\%$ and 95.4 $\%$ confidence regions respectively.} We clearly see detection of Lagrangian tidal bias and a deviation from the fiducial smoothing scale $R_{\text{h}}=4 h^{-1}$ Mpc. There are some mild degeneracies between the counterterm $\text{d}R$ and the local cubic and quadratic tidal tensor bias.}
\label{cubicmass}
\end{figure}

\begin{table}[h]
\small
\begin{center}
\resizebox{12cm}{!}{
\begin{tabular}{ |c|c|c|c|c|c|} 
 \hline
Models & Bin & Bin 2& Bin 3& Bin 4& Bin 5 \\
\hline
Eulerian 8 pars (with CT) &  2.782 & 1.914 & 1.080 & 1.006 & 1.297 \\
Eulerian 7 pars (without CT) &    2.417 & 1.708 & 1.120 & 1.144 & 1.405 \\
Lagrangian 8 pars (with CT) &   1.975 & 1.303 & 1.168 & 1.373 & 1.454 \\
Lagrangian 7 pars (without CT) &  2.680 & 1.636 & 1.162 & 1.263 & 1.267 \\
\hline
Lagrangian 5 pars (initial) & 
 1.800 & 1.296 & 1.267 & 1.370 & 1.497 \\
Lagrangian 5 pars (final) &  1.216 & 1.203 & 1.162 & 1.023 & 1.190 \\
 \hline
\end{tabular}}
 \label{tab1}
  \caption{Overview of reduced $\chi^2$ models considered. We have studied six different models which are summarized in the table. First, we note that the Eulerian and Lagrangian models with the counterterm $\text{d}R$ are statistically preferable compare to the ones without the counterterm. Second, we note that for both Eulerian and Lagrangian models, for mass bin I and II, five parameter fits are statistically preferable. However, for mass bins III, IV, and V the full model with eight parameters gives a lower reduced $\chi^2$ and is therefore preferable. This implies that low mass halos are in a better agreement with the co-evolution predictions of LLB+$b_{s^2}$.}
\end{center}
\end{table}


\subsection{Application: One-loop halo-matter power spectrum}
\label{sec:phm}
We are now ready to check the halo-matter cross spectrum and halo-propagator predictions. 
The halo-matter cross spectrum $P_{\text{hm}}(k)$ and the halo-propagator $P_{\text{hG}}(k)$ are defined through the two-point function in Fourier space as:
\begin{eqnarray}
\langle\de_{\text{h}}(\bk)\de_{\text{i}}(\bk')\rangle = (2\pi)^3\de^{(3)}_{\text{D}}(\bk+\bk')P_{\text{hi}}(k)\, ,
\end{eqnarray}
where $i=$m or G correspond to the non-linear and linear density field respectively. Up to one-loop in PT and at leading order in derivatives $P_{\text{hm}}$ and $P_{\text{hG}}$ are given by the following expressions \cite{McDonald:2009dh,Assassi:2014fva}
\begin{equation}
\ba
P_{\text{hm}}(k) &= b_{1}\Big(P_{\text{lin}}(k) + 2 P_{13}(k) + P_{22}(k) \Big) + \Big(b_{s^2}+ \frac{2}{5} b_{\Gamma_3} \Big) \mathcal{F}(q) - b_{\nabla^2\de}k^2P_{\text{lin}}(k)\\
&+\frac{1}{2}b_2\mathcal{I}_{\de^{(2)}\de^2}(k)+b_{s^2}\mathcal{I}_{\de^{(2)}s^2}(k)\\
P_{\text{hG}}(k) &= b_{1}\Big(P_{\text{lin}}(k) + P_{13}(k)\Big) + \Big(b_{s^2}+ \frac{2}{5} b_{\Gamma_3} \Big) \mathcal{F}(q) - \widetilde{b}_{\nabla^2\de}k^2P_{\text{lin}}(k)
\ea
\label{phm}
\end{equation}
where $\mathcal{F}(k)$, $\mathcal{I}_{\de^{(2)}\de^2}(k)$, and $\mathcal{I}_{\de^{(2)}s^2}(k)$ are defined as
\begin{eqnarray}
\mathcal{F}(k) = 4P_{\text{lin}}(k)\int_{\bq}\left(S_2(\bq,\bk-\bq)F_{2}(\bk,-\bq)-
\frac{34}{63}\right)P_{\text{lin}}(q)\, ,\\
\mathcal{I}_{\de^{(2)}\de^2}(k) = 2\int_{\bq}F_2(\bk-\bq,\bq)P_{\text{lin}}(q)P_{\text{lin}}(|\bk-\bq|)\, ,\\
\mathcal{I}_{\de^{(2)}s^2}(k) = 2\int_{\bq}F_2(\bk-\bq,\bq)S_2(\bk-\bq,\bq)P_{\text{lin}}(q)P_{\text{lin}}(|\bk-\bq|)\, .
\end{eqnarray}
Note that the above expressions do not contain a smoothing scale. For explicit expressions of $P_{13}$ and $P_{22}$ see \cite{Bernardeau:2001qr}. Note that naively, there would have been contributions proportional to $\sigma^2 P$ in the halo-matter power spectrum and propagator, which would renormalize the linear bias prefactor of the leading $P_\text{lin}$ contribution. This would cause the large scale limit to deviate  from $b_1 P_\text{lin}$ \cite{McDonald:2006mx}. However, the propagation of the 
$
-b^{\text{L}}_2/2\sigma^2 -b^{\text{L}}_3/2\sigma^2\de_{\text{G}}(\bq)  -2/3b^{\text{L}}_{s^2}\sigma^2
$
contributions to Eq.~\eqref{Lbias2a} leads to a
$(-b^{\text{L}}_2/2\sigma^2 -b^{\text{L}}_3/2\sigma^2-2/3b^{\text{L}}_{s^2}\sigma^2)P$
contribution to the power spectrum, which exactly cancels these renormalizing terms.

Having fixed the bias parameters from our measurements described earlier in this Section, we are only left with the $k^2$ term. The importance of $k^2$ corrections has been discussed in the literature in the context of peak model \cite{Desjacques:2008jj,Desjacques:2010gz,Baldauf:2014fza} or symmetry arguments \cite{McDonald:2009dh,Schmidt:2012ys}. Constraints on the $k^2$ bias or the leading derivative bias can be obtained by comparing Eqs.~\eqref{phm} with the simulation data. We quote our best-fit values for $b_{\nabla^2\de}$ and $\widetilde{b}_{\nabla^2\de}$ in Tab.~\ref{k2bias}. The large errorbars on the bias parameters $b_{s^2}+ \frac{2}{5} b_{\Gamma_3}$ do not allow for a significant detection of non-zero $k^2$ corrections for mass bins I and V. However, we get a significant detection of $k^2$ corrections for mass bins II, III and IV. \correc{The constraints obtained from the propagator and equal-time halo-matter power spectrum are consistent with each other. The amplitude of the parameter is non-monotonic with mass, which could be understood in the context of the peak model, where the Eulerian $k^2$-bias is given by sums of positive and negative contributions with different mass dependence \cite{Baldauf:2016aaw}.
The difference of $P_\text{hm}$ and $P_\text{hG}$ has its own residual $k^2$ correction $\tilde{b}_{\nabla^2\delta}-b_{\nabla^2\delta}=-c_\text{s}^2$, which is related to the EFT speed of sound in the matter power spectrum $P_{\text{mm}}(k) = P_{\text{lin}}(k) + 2 P_{13}(k) + P_{22}(k)- 2 c_\text{s}^2 k^2P_{\text{lin}}(k)$. As the difference $P_\text{hm}$ and $P_\text{hG}$ does not contain $\mathcal{F}$, it is less affected by the large error bars on $b_{s^2}+ \frac{2}{5} b_{\Gamma_3}$ and allows us to put tighter constraints on $c_s^2$ than on $\tilde{b}_{\nabla^2\delta}$ and $b_{\nabla^2\delta}$ individually. Within the error bars the results indeed agree with reported values in the literature \cite{Baldauf:2014qfa}. }

\begin{figure}[h]
\centering
\begin{subfigure}{.33\textwidth}
  \centering
\includegraphics[width=5cm]{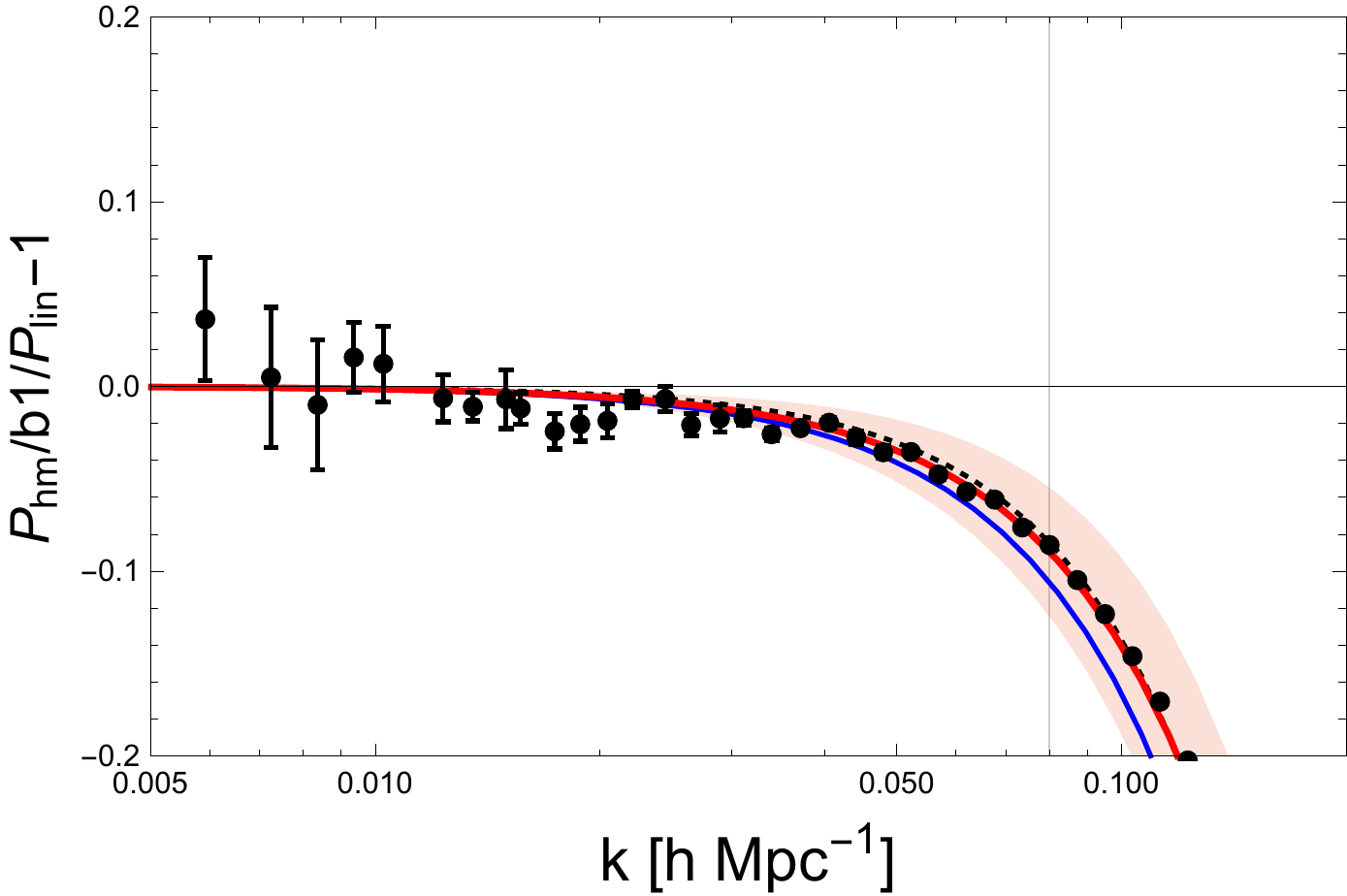}
  \caption{bin I}
\end{subfigure}
\begin{subfigure}{.33\textwidth}
  \centering
\includegraphics[width=5cm]{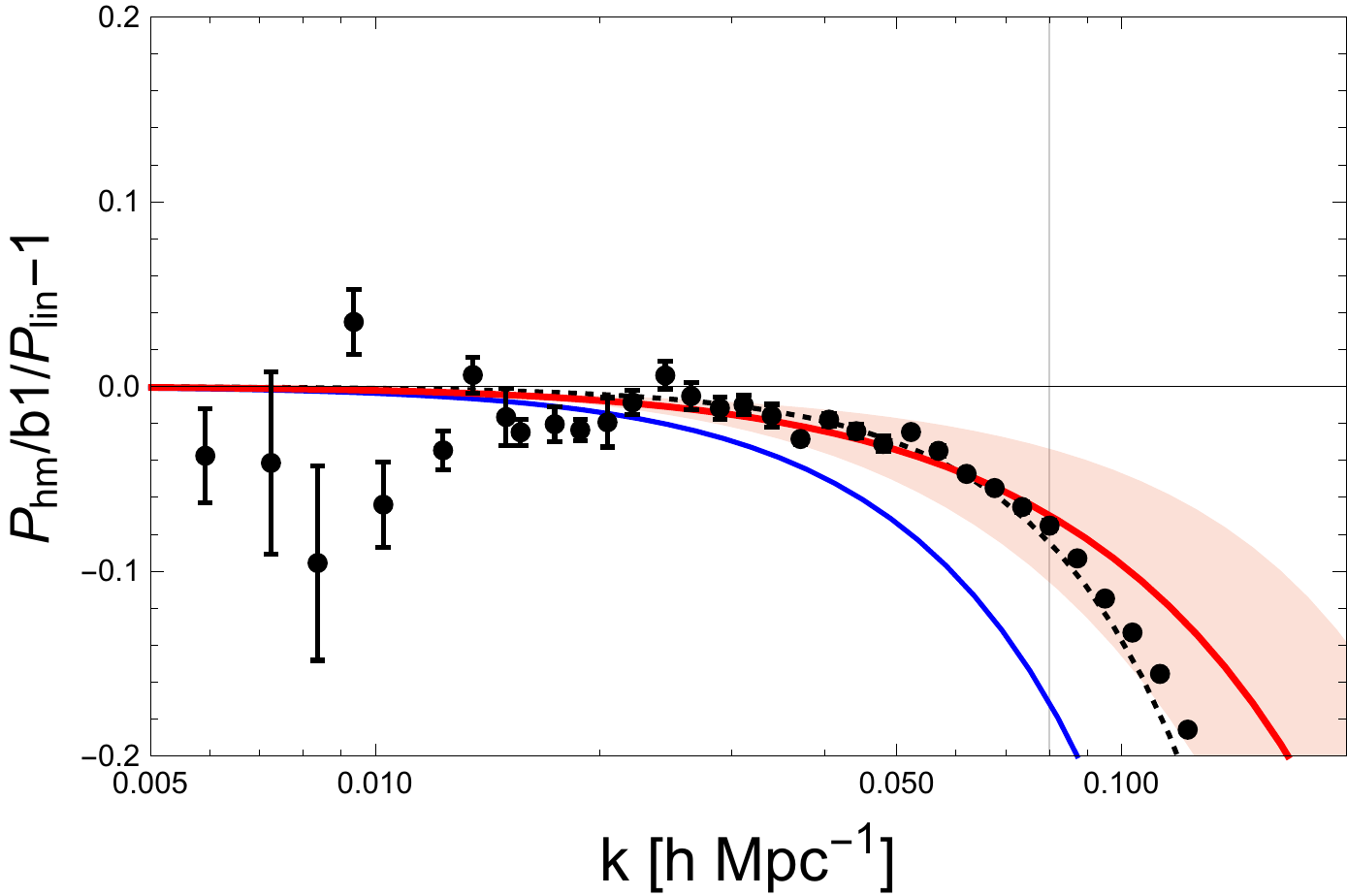}
  \caption{bin II}
\end{subfigure}%
\begin{subfigure}{.33\textwidth}
  \centering
\includegraphics[width=5cm]{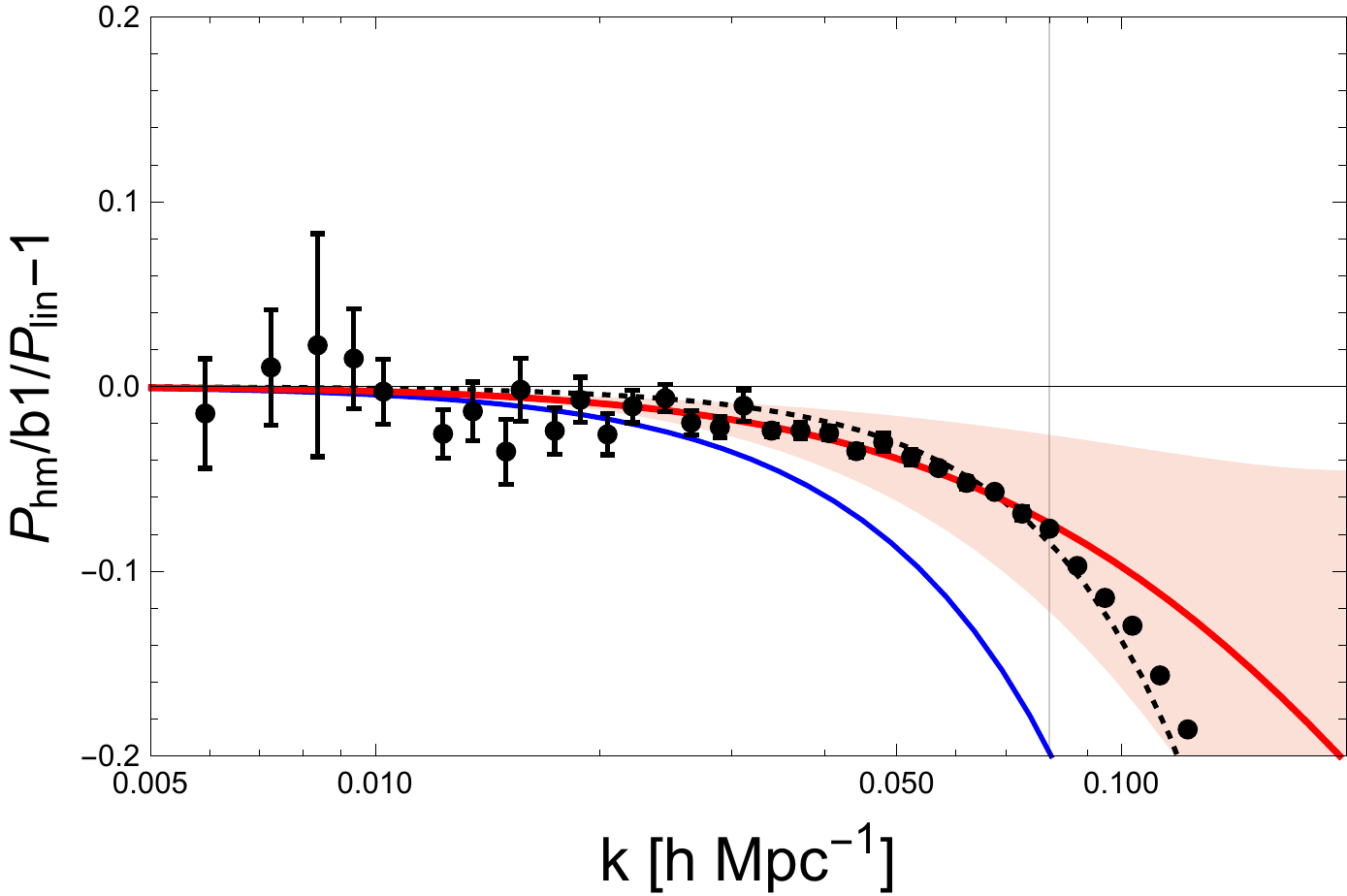}
  \caption{bin III}
\end{subfigure}
\begin{subfigure}{.33\textwidth}
  \centering
\includegraphics[width=5cm]{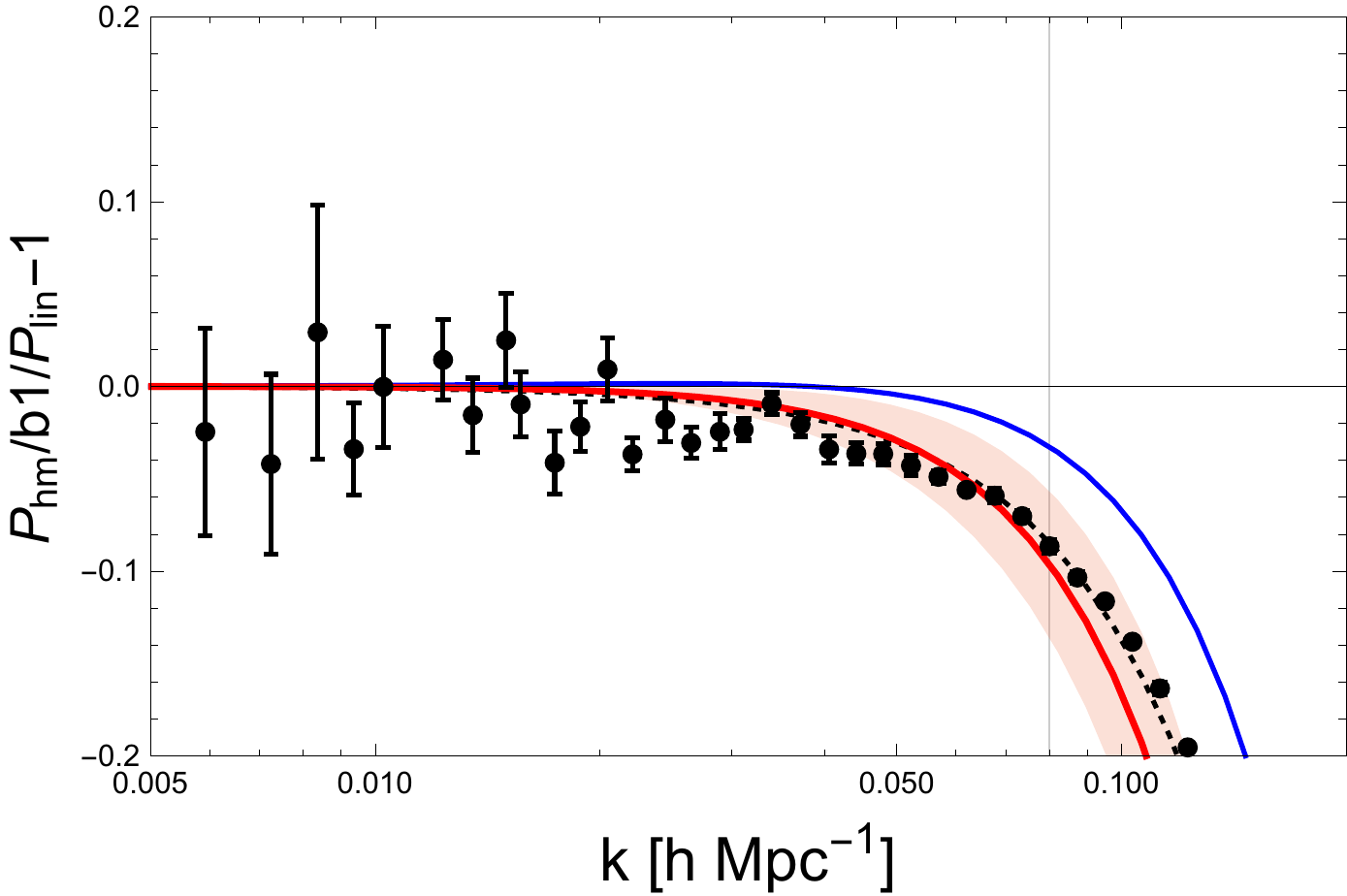}
  \caption{bin IV}
\end{subfigure}%
\begin{subfigure}{.33\textwidth}
  \centering
\includegraphics[width=5cm]{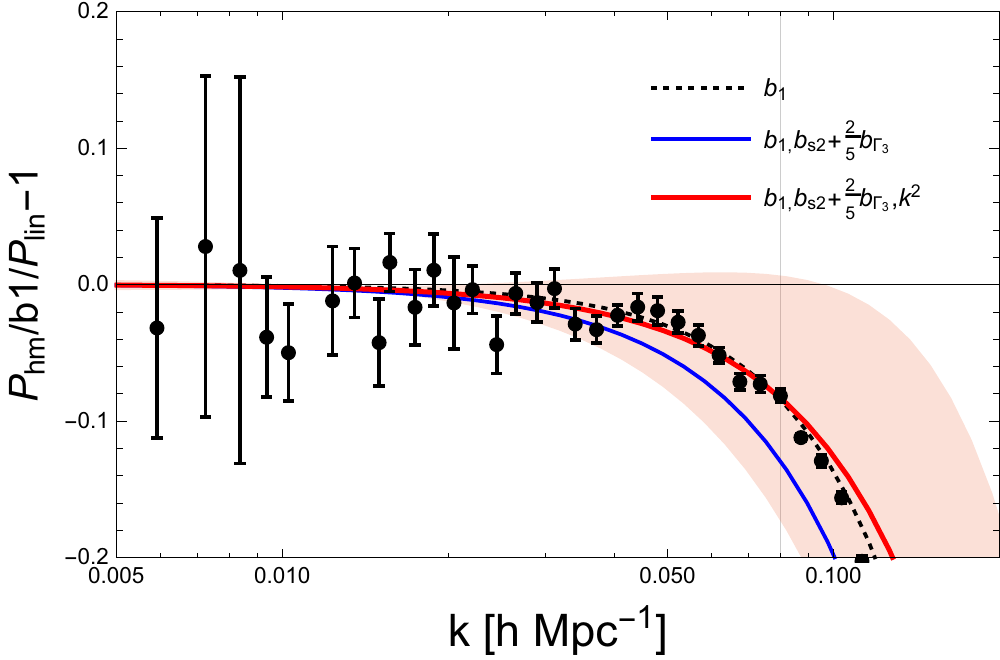}
  \caption{bin V}
\end{subfigure}
\caption{Cross correlation of the final halo field with the linear density field (propagator), normalized by the linear power spectrum. The red and blue lines show one-loop predictions with and without $k^2$ corrections. The shaded red region shows the effects of the bias errorbars on the predictions. }
\label{fig:prop}
\end{figure}

\begin{figure}[H]
\centering
\begin{subfigure}{.33\textwidth}
  \centering
\includegraphics[width=5cm]{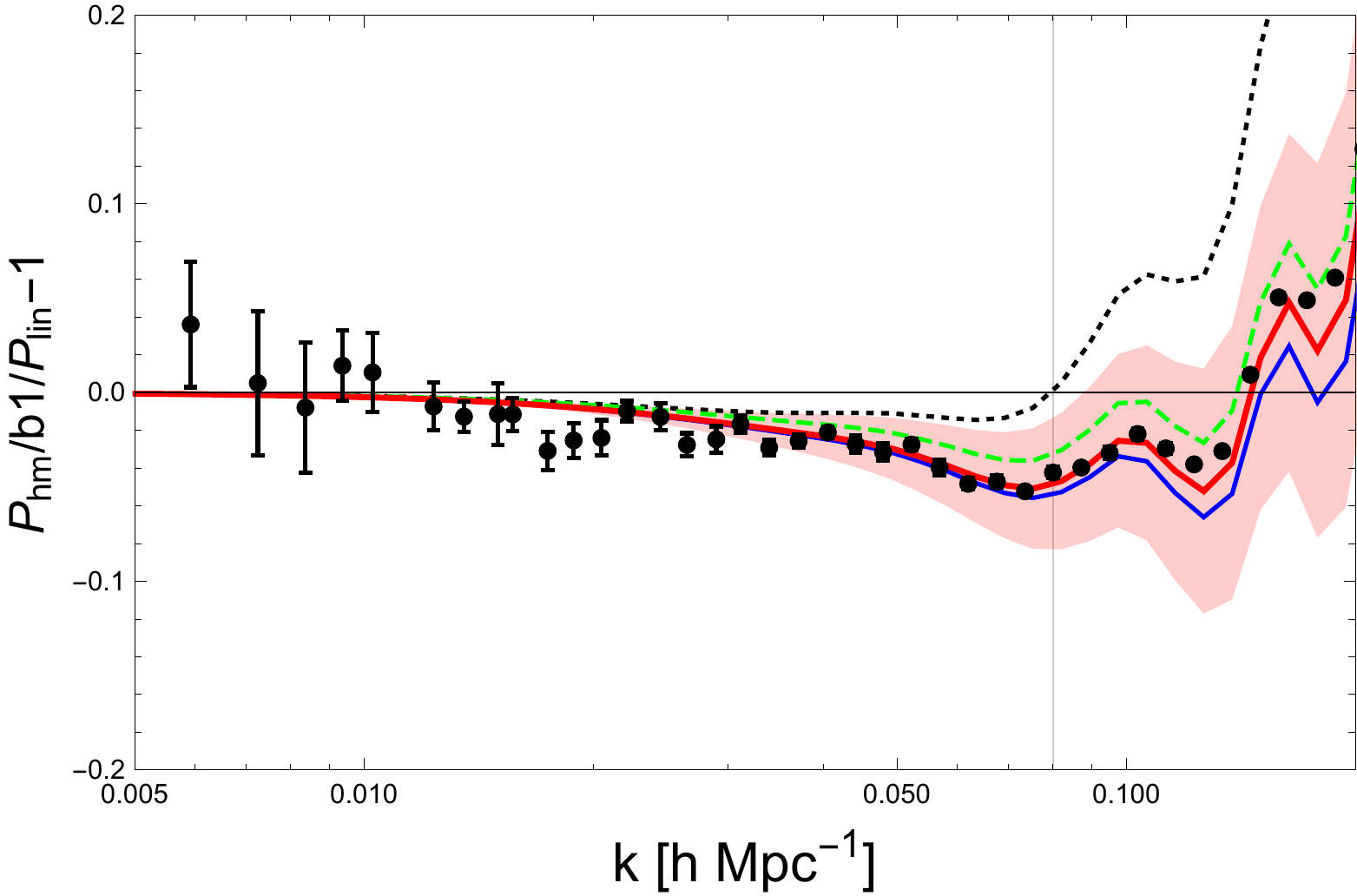}
  \caption{bin I}
\end{subfigure}
\begin{subfigure}{.33\textwidth}
  \centering
\includegraphics[width=5cm]{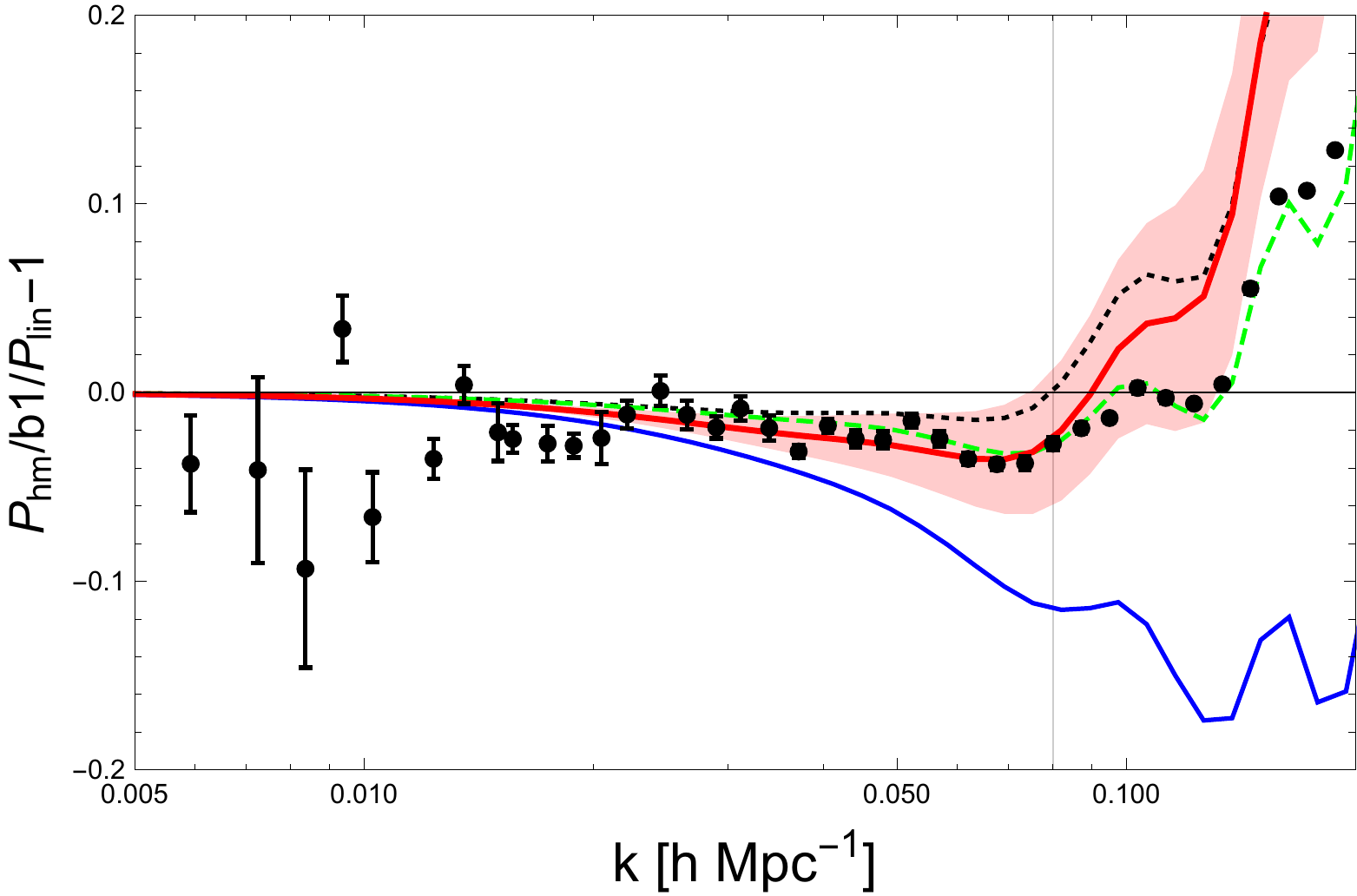}
  \caption{bin II}
\end{subfigure}%
\begin{subfigure}{.33\textwidth}
  \centering
\includegraphics[width=5cm]{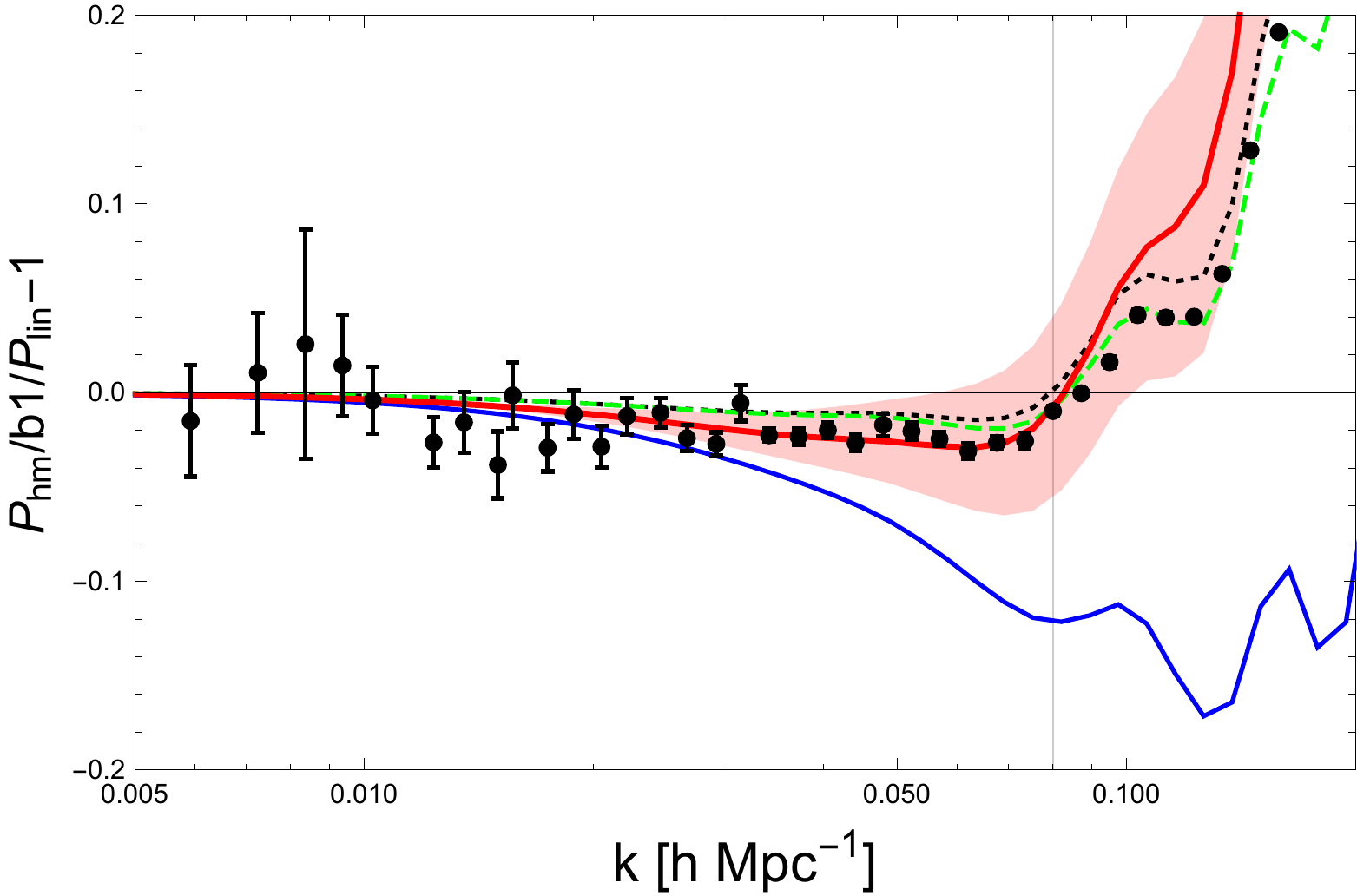}
  \caption{bin III}
\end{subfigure}
\begin{subfigure}{.33\textwidth}
  \centering
\includegraphics[width=5cm]{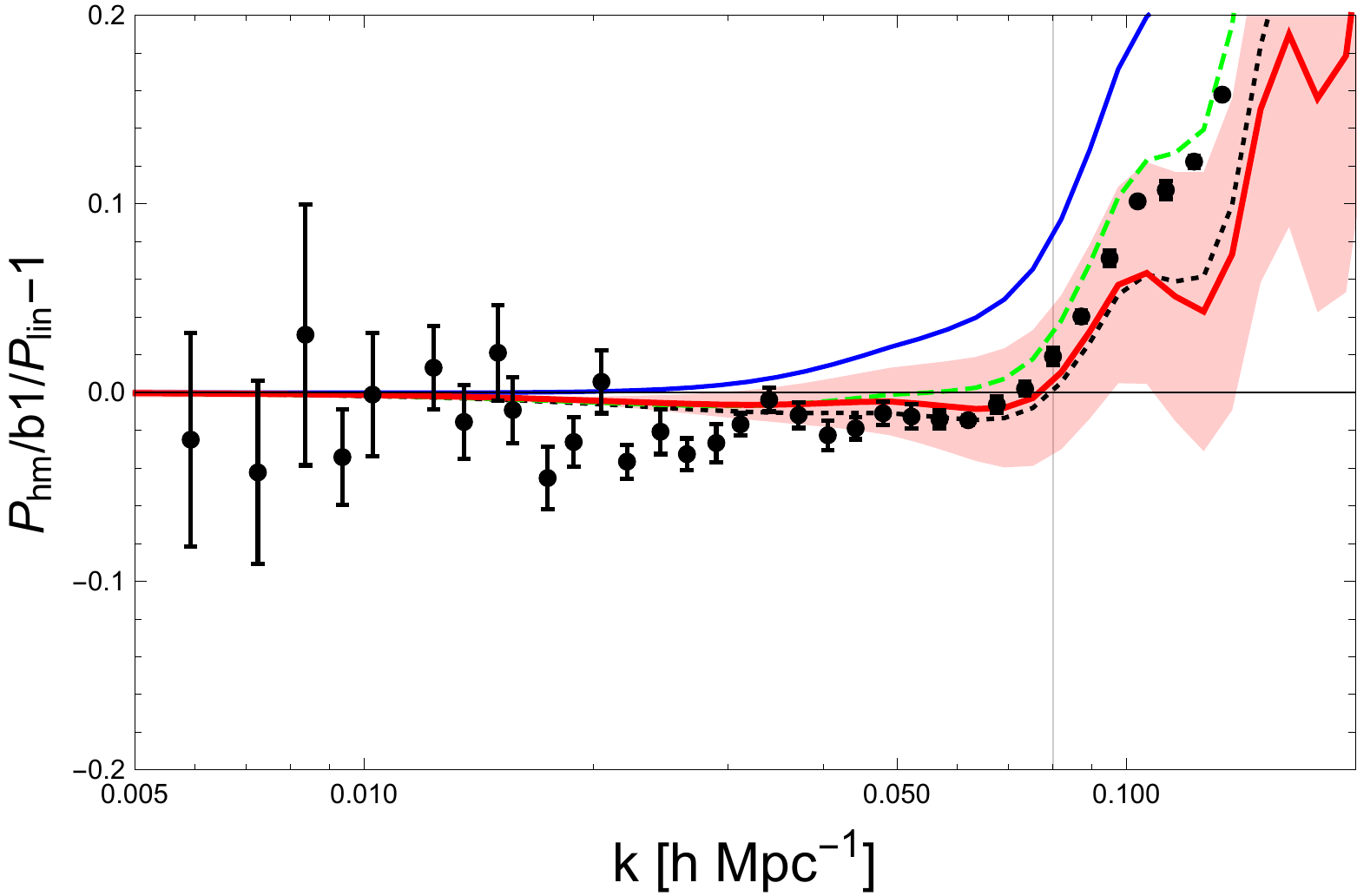}
  \caption{bin IV}
\end{subfigure}%
\begin{subfigure}{.33\textwidth}
  \centering
\includegraphics[width=5cm]{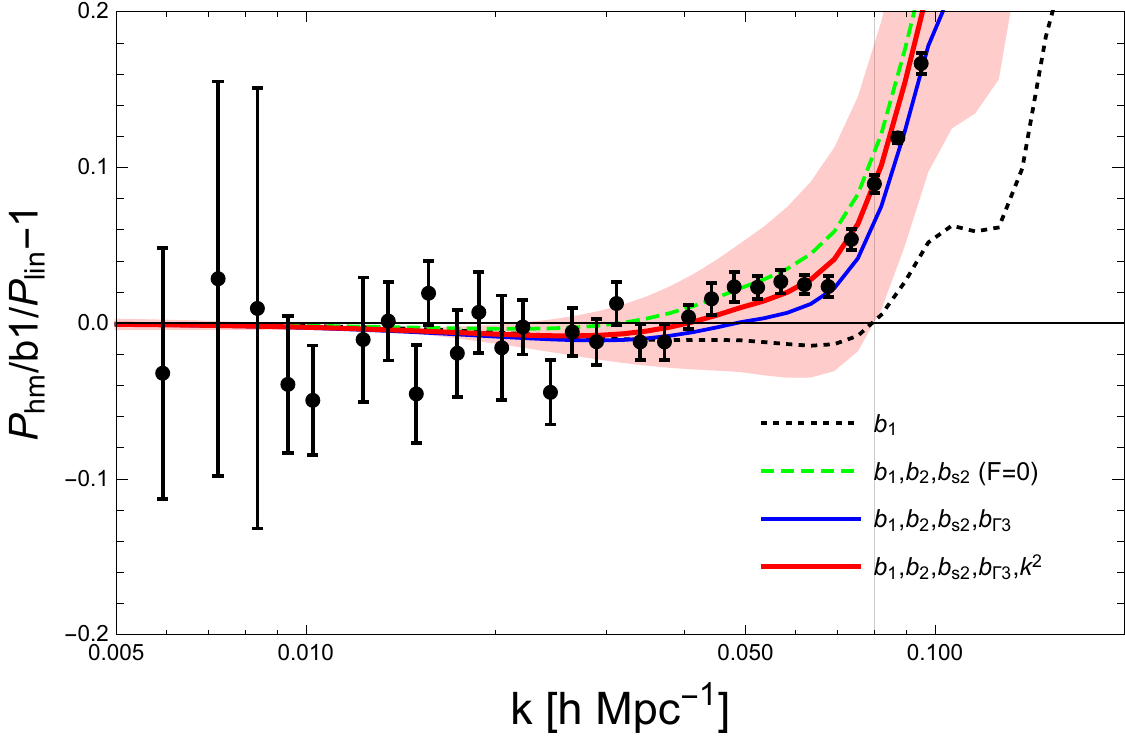}
  \caption{bin V}
\end{subfigure}
\caption{Ratio of halo-matter cross power spectrum and linear power spectrum. The red and blue lines show one loop predictions with and without the $k^2$ corrections. The shaded red region represents the uncertainty arising from the error on the cubic bias parameters (without the error on the $k^2$ term). We also show the predictions without quadratic and cubic bias reflected by the black dotted lines.}
\label{fig:phm}
\end{figure}

\begin{table}[H]
\small
\begin{center}
\begin{tabular}{|c|c|c|c|c|}
\hline
Mass Bin & $ \widetilde{b}_{\nabla^2\delta}$& $\Delta \widetilde{b}_{\nabla^2\delta}$ &$b_{\nabla^2\delta}$& $\Delta b_{\nabla^2\delta}$ \\  [0.2ex] 
\hline
 \text{I}& -2.74 & 6.52&  -0.92 & 6.59 \\
 \text{III} & -20.94 & 8.52&-18.65 & 8.58 \\
 \text{III} &-35.21 & 15.37 & -32.18 & 15.34 \\
 \text{IV} & 26.74 & 19.39& 32.39 & 19.44\\
 \text{V} & -30.35 & 66.35 & -16.34 & 67.12\\
\hline
\end{tabular}
\end{center}
\caption{Best-fit values for the $k^2$ bias coefficients for five mass bins obtained from the one-loop halo-matter statistics after fixing all other bias parameters. The maximum wavenumber used is $k_{\text{max}}=0.08h$ Mpc$^{-1}$. The quoted values are in units of $h^{-2}$ Mpc$^{2}$, i.e. inverse length-squared. \correc{The error bars are dominated by the uncertainty of $b_{s^2}+2/5 b_{\Gamma_3}$.}}
\label{k2bias}
\end{table}
In Figs.~\ref{fig:prop} and~\ref{fig:phm} we show our predictions for the one-loop halo propagator and halo-matter power spectrum. The predictions with and without $k^2$ are represented by solid blue and red lines respectively. In addition, the shaded green region shows the uncertainty arising from the error bars on the bias measurements (without the error on the $k^{2}$ correction). We see that after adding the $k^{2}$ corrections the theory agrees with the data up to wavenumber $k = 0.1h $ Mpc$^{{-1}}$. However, precision is highly affected by the large error bars on the combination of cubic parameters $b_{s^2}+ \frac{2}{5} b_{\Gamma_3} $.

\newpage
\section{Summary \& Conclusions}
\label{conclusion}

In this paper we have studied the measurement of bias parameters beyond leading order from cross-correlations of quadratic and cubic bias operators with halo fields in a suite of $N$-body simulations.


We summarize our results as follows:
\begin{itemize}
\item We find that we can model the final halo distribution with seven bias parameters (one linear, two second order and four third order bias parameters) with one additional parameter that accounts for the halo scale. We find clear evidence for non-zero quadratic and cubic non-local bias operators. The amplitude of the detected non-local bias deviates from the predictions of the evolved local Lagrangian model both at quadratic and cubic level.

\item The distribution of protohalos in Lagrangian space in turn shows evidence for the existence of a Lagrangian quadratic tidal bias contribution, i.e., a deviation from the local Lagrangian bias model. The presence of such a term indicates that the collapse threshold for halo formation depends not only on the density but also on the shear \cite{Castorina:2016tuc,Modi:2016dah}, and that the strength of this dependence increases with mass.

\item We have not detected any cubic non-local terms in Lagrangian space for low mass bins; however, for the highest mass bin V we find some evidence for the presence of these terms. We would like to emphasize again that our smoothing and cutoff scales might be insufficient to suppress the derivative bias corrections for high mass bins. Should this hint for the existence of cubic Lagrangian bias be confirmed, the modelling of collapse thresholds for halo formation would need to be extended to cubic fields.

\item The non-detection of cubic Lagrangian bias operators for low masses motivated us to consider the consequences of a Lagrangian bias model with a quadratic tidal component but no non-local cubic operators. The Lagrangian tidal bias contributes to both the quadratic and cubic non-local bias operators in Eulerian space. We were able to fit the final distribution with the simple five parameter model that contains Lagrangian local biases up to third order and a tidal Lagrangian bias.

\item We see some mild degeneracies between the counterterm $\text{d}R$ and the cubic local and quadratic non-local bias terms in Lagrangian space, which might be because of neglecting $k^2$ terms in the protohalo field.

\item Given the importance of the $k^2$ term, we constrained it from the one-loop halo-matter cross spectra statistics for five mass bins after having fixed the other bias parameters from our measurements. The constraints are given in Tab.~\ref{k2bias}. Because of large errorbars on the cubic bias parameters entering in these statistics, we have not detected the presence of non-zero $k^2$ term for \correc{some of the mass} bins. \correc{We find that the constraints from the propagator and equal time statistic are consistent and that their difference agrees with previous measurements of the dark matter speed of sound in the EFT framework.}

\item We plot the predictions of one-loop halo propagator and halo-matter cross spectrum in Fig.~\ref{fig:prop} and Fig.~\ref{fig:phm} respectively. Our predictions agree with the $N$-body simulation data up to $k=0.1h$ Mpc$^{-1}$. However, to make more precise \newcorrec{predictions}, one has to reduce the errorbars on the combination of the cubic bias $2/5 b_{\Gamma_3} + b_{s^2}$.
\end{itemize}

As we hinted in Sec.~\ref{sec:bitri}, it might be interesting to consider the bispectrum of quadratic field, linear field and halo field as an alternative means to extract bias information from the trispectrum. Furthermore, the strong filtering or derivative corrections in the protohalo statistics \cite{Baldauf:2015aha,Modi:2016dah} should be accounted for more directly in order to improve the reliability of the constraints on Lagrangian bias parameters. An application of the presented method to actual observations will be complicated by the nonavailability of a Gaussian reference field. This problem could potentially be alleviated by using cross correlation of squared and cubed lensing fields with the galaxy field.

The method presented here allows for straightforward extensions to quartic statistics, which will be relevant for computations of the one-loop halo or galaxy bispectrum. Furthermore, straightforward extensions of this method should allow to constrain cubic primoridal non-Gaussiantity such as the $g_\text{NL}$ \cite{Ade:2015ava} local model.

\acknowledgments
While this paper was being finished the study \cite{Lazeyras:2017hxw} appeared which uses a similar approach to estimate cubic bias parameters. We would like to thank M. Schmittfull for inspiring discussions and A. Challinor and P. Shellard for insightful comments. MA. would also like to thanks O. Leicht for useful discussions and Safwan A. Khan for proofreading.  The numerical part of this work was performed using the DiRAC COSMOS  supercomputer and greatly benefited from the support of K. Kornet. MA is funded by the Cambridge Trust and HEC Pakistan. TB is funded by a Stephen Hawking Advanced Fellowship from the Centre for Theoretical Cosmology, DAMTP, University of Cambridge.


\appendix
\section{Basis}
The authors in \cite{Angulo:2015eqa,Fujita:2016dne} define a basis
of operators
\begin{equation}
\boldsymbol{\mathcal{B}}_\text{FMSVA}=\lbrace\mathbb{C}_{\delta,1}^{(3)},\mathbb{C}_{\delta,2}^{(3)},\mathbb{C}_{\delta,3}^{(3)},\mathbb{C}_{\delta^2,1}^{(3)},\mathbb{C}_{\delta^2,2}^{(3)},\mathbb{C}_{\delta^3,1}^{(3)}\mathbb{C}_{s^2,2}^{(3)}\rbrace
\end{equation}
Note that their basis is equivalent to our basis
\begin{equation}
\boldsymbol{\mathcal{B}}_\text{here}=\lbrace F_3, 1,\mathcal{G}_3+\frac19,\mathcal{G}_2\delta+\frac23,\Gamma_3+\frac{16}{63},\delta^{(1)} \delta^{(2)},s_{ij}^{(1)}s_{ji}^{(2)}\rbrace
\end{equation}
\begin{equation}
\boldsymbol{\mathcal{B}}_\text{here}=\mathcal{M}
\boldsymbol{\mathcal{B}}_\text{FMSVA}
\end{equation}
where 
\begin{equation}
\mathcal{M}=
\left(
\begin{array}{ccccccc}
 1 & 1 & 1 & 0 & 0 & 0 & 0 \\
 0 & 0 & 0 & 0 & 0 & 1 & 0 \\
 0 & 0 & -\frac{45}{4} & 0 & \frac{151}{16} & -\frac{613}{72}
   & \frac{3}{4} \\
 0 & 0 & 0 & 0 & \frac{7}{4} & -\frac{17}{6} & 0 \\
 0 & 0 & -9 & 0 & \frac{79}{12} & -\frac{661}{126} & 1 \\
 0 & 0 & 0 & 1 & 1 & 0 & 0 \\
 0 & \frac{7}{2} & 0 & -\frac{17}{6} & 0 & 0 & 1 \\
\end{array}
\right)\, .
\end{equation}

\section{UV-sensitivity and EFT counterterms}
\label{app:UV}

\subsection{$R_{\text{h}}$-dependence of quadratic and cubic correlations}
In this section we discuss the UV sensitivity of the correlations of quadratic fields with the quadratic bias operators and cubic fields with the cubic bias operators. As discussed in the main text, the quadratic field correlations are represented by a one-loop power spectrum diagram.  To show the UV sensitivity of these diagrams, we calculate them theoretically using $R_{\text{h}}=4 h^{-1}$Mpc and $R_{\text{h}}=6 h^{-1}$Mpc and take the ratio at a fixed wavenumber $k=0.017 h$ Mpc$^{-1}$. We show the results in Table~\ref{App:tab:quadUV}. We can see a change at the $5\%$ level at the chosen wavenumber.
\begin{table}[h]
\begin{center}
\begin{tabular}{c|ccc}
\hline
 & $F_2$ & $\de^2$ & $S_2$ \\  [0.2ex] 
\hline
  $\de^2$& 0.002 & 0.039 & 0.049 \\
  $-\Psi\cdot\nabla\delta$& 0.010 & 0.041 & 0.049 \\
  $S_2$ & 0.012 & 0.049 & 0.052 \\
\hline
\end{tabular}
\end{center}
\caption{Quadratic fields: Relative change in the amplitude of cross-correlations of quadratic fields with the quadratic bias operators at $k = 0.042h$ Mpc$^{-1}$ as we change the halo smoothing scale from  $R_{\text{h}}=4 h^{-1}$ Mpc to $R_{\text{h}}=6 h^{-1}$ Mpc.}
\label{App:tab:quadUV}
\end{table}

We then repeat the same exercise for cubic correlations. We will show that the two-loop irreducible diagrams are more UV-sensitive than two-loop irreducible diagrams. We show the results in Tables~\ref{App:tab:IR},~\ref{App:tab:R}, and ~\ref{App:tab:full} for irreducible diagrams, reducible diagrams and the total contribution, respectively. One can see in Table~\ref{App:tab:R} that most of the reducible two-loop diagrams of cubic correlations change by more than $80\%$ as we change the halo smoothing scale from $4 h^{-1}$Mpc to $6 h^{-1}$Mpc. On the other hand, the irreducible two-loop diagrams show a weaker change at the $5\%$ level. Note that in Tables~\ref{App:tab:IR}, ~\ref{App:tab:R}, ~\ref{App:tab:full}, and ~\ref{App:tab:quadUV} we use $k = 0.042h$ Mpc$^{-1}$.
\begin{table}[h]
\begin{center}
\begin{tabular}{c|ccccccc}
\hline
&$F_3$ & $\de^3$ & $\mathcal{G}_3$ & $\mathcal{G}_2\de$ & $\Gamma_3$ &$\de\de^{(2)}$& $s^{(3)}$\\
\hline
$F_3$&   0.023 & -0.143 & 0.026 & -0.067 & 0.013 & -0.009 & 0.012 \\
$\de^3$ & -0.143 & 0.085 & 0.098 & 0.073 & 0.048 & 0.086 & 0.112 \\
$\mathcal{G}_3$ &   0.026 & 0.099 & 0.053 & 0.045 & 0.075 & 0.058 & 0.042 \\
$\mathcal{G}_2\de$ & -0.067 & 0.073 & 0.045 & 0.078 & 0.046 & 0.083 & 0.104 \\
$\Gamma_3$ &  0.013 & 0.048 & 0.075 & 0.046 & 0.049 & 0.049 & 0.053 \\
$ \de\de^{(2)}$&   -0.009 & 0.086 & 0.058 & 0.083 & 0.050 & 0.086 & 0.086 \\
$s^{(3)}$&  0.012 & 0.112 & 0.042 & 0.104 & 0.053 & 0.086 & 0.108 \\
\hline
\end{tabular}
\end{center}
\caption{Irreducible: Relative change in the amplitude of irreducible diagrams of the cross-correlations of cubic fields at $k = 0.042h$ Mpc$^{-1}$ as we change the halo smoothing scale from  $R_{\text{h}}=4 h^{-1}$ Mpc and $R_{\text{h}}=6 h^{-1}$ Mpc.}
\label{App:tab:IR}
\end{table}

\begin{table}[H]
\begin{center}
\begin{tabular}{c|ccccccc}
\hline
&$F_3$ & $\de^3$ & $\mathcal{G}_3$ & $\mathcal{G}_2\de$ & $\Gamma_3$ &$\de\de^{(2)}$& $s^{(3)}$\\
\hline
$\de^3$&  0.119 & 0.820 & - & 0.813 & 0.196 & 0.825 & 0.913 \\
\hline
\end{tabular}
\end{center}
\caption{Reducible: Relative change in the amplitude of reducible diagrams of the cross-correlations of cubic fields at $k = 0.042h$ Mpc$^{-1}$ as we change the halo smoothing scale from  $R_{\text{h}}=4 h^{-1}$ Mpc to $R_{\text{h}}=6 h^{-1}$ Mpc.}
\label{App:tab:R}
\end{table}

\begin{table}[H]
\begin{center}
\begin{tabular}{c|ccccccc}
\hline
&$F_3$ & $\de^3$ & $\mathcal{G}_3$ & $\mathcal{G}_2\de$ & $\Gamma_3$ &$\de\de^{(2)}$& $s^{(3)}$\\
\hline
$F_3$&   0.048 & 0.815 & 0.026 & 0.812 & 0.242 & 0.867 & 0.950 \\
$\de^3$ &  0.129 & 0.771 & 0.099 & 0.749 & 0.165 & 0.787 & 0.886 \\
$\mathcal{G}_3$ &  0.026 & 0.099 & 0.053 & 0.045 & 0.075 & 0.058 & 0.042 \\
$\mathcal{G}_2\de$ & 0.128 & 0.750 & 0.045 & 0.721 & 0.164 & 0.773 & 0.872 \\
$\Gamma_3$ &  0.153 & 0.752 & 0.075 & 0.722 & 0.135 & 0.762 & 0.874 \\
$ \de\de^{(2)}$&  0.153 & 0.787 & 0.058 & 0.772 & 0.185 & 0.792 & 0.888 \\
$s^{(3)}$&  0.105 & 0.798 & 0.042 & 0.809 & 0.184 & 0.801 & 0.930 \\
\hline
\end{tabular}
\end{center}
\caption{Full theory:  Relative change in the amplitude full cross-correlations of cubic fields (reducible + irreducible) at  $k = 0.042h$ Mpc$^{-1}$ as we change the halo smoothing scale from  $R_{\text{h}}=4 h^{-1}$ Mpc to $R_{\text{h}}=6 h^{-1}$ Mpc.  }
\label{App:tab:full}
\end{table}

\subsection{ Quadratic EFT counterterms}

After showing that the quadratic and orthogonalised cubic correlations in our model do indeed show some dependency on the halo smoothing scale, we want to discuss possible EFT counterterms to remove these UV-sensitivitives. First, let us consider again the correlations of the quadratic fields $\mathcal{D}_2$ with quadratic bias operators $\mathcal{O}_2$:
\begin{equation}
\ba
\langle \mathcal{D}_2|\mathcal{O}_2\rangle=\int_{\bq} W_{\text{R}_h}(|\bk-\bq|)W_{\text{R}_h}(q)P_{\text{lin}}(q)P_{\text{lin}}(|\bk-\bq|)&\mathcal{K}_{\mathcal{D}_2}(\bk-\bq,\bq)\mathcal{K}_{\mathcal{O}_2}(\bk-\bq,\bq)\\
&W_{\text{R}_f}(|\bk-\bq|)W_{\text{R}_f}(q) ,
\ea
\end{equation}
where $\mathcal{D}_2\in \{\de^2,\Psi\cdot\nabla\de,s^2\}$,  $\mathcal{O}_2\in \{\de^{(2)},\de^2,s^2\}$, and $q=k r_1$. We write the low-$k$ limits of the described above in a matrix notation as
\begin{equation}
\lim_{q\rightarrow \infty}\langle \mathcal{D}^{i}_2|\mathcal{O}^{j}_2\rangle=\int_{\bq} W_{\text{R}_f}(q)^2 P_{\text{lin}}(q)^2\mathcal{M}_{\mathcal{D}_2\mathcal{O}_2}^{ij}(k,r_1; R_h)\, 
\label{B2mat}
\end{equation}
Eq. \eqref{B2mat} is a $3\times 3$ matrix of the cross-correlations of quadratic fields with quadratic bias operators. The matrix $\mathcal{M}_{\mathcal{D}_2\mathcal{O}_2}$ represents the UV limits of the product of two kernels in terms of halo smoothing scale 
\begin{equation}
\mathcal{M}_{D_2 B_2}=
\left(
\begin{array}{ccc}
 \frac{1}{21} k^2 R_h^2-\frac{q^4}{21 k^4} & 2-\frac{2 k^6 R_h^2}{q^4} & -\frac{4 k^6 R_h^2}{3 q^4}+\frac{4}{3}
   k^2 R_h^2-\frac{4 q^4}{3 k^4}+\frac{4}{3} \\
 \frac{q^4}{42 k^4}-\frac{1}{42} k^2 R_h^2 & 0 & \frac{4}{15} k^2 R_h^2-\frac{4 q^4}{15 k^4} \\
 \frac{2}{63} k^2 R_h^2-\frac{2 q^4}{63 k^4} & -\frac{4 k^6 R_h^2}{3 q^4}+\frac{4}{3} k^2 R_h^2-\frac{4 q^4}{3
   k^4}+\frac{4}{3} & -\frac{8 k^6 R_h^2}{9 q^4}+\frac{16}{9} k^2 R_h^2-\frac{16 q^4}{9 k^4}+\frac{8}{9} \\
\end{array}
\right)
\label{app:UVmatrix}
\end{equation}
This shows that the halo smoothing affects the low-$k$ limit of quadratic correlations and hence the measurements of bias parameters. This dependency should be removed by adding appropriate counterterm. At the leading order we can add a constant and a $k^{2}$ counter term. There are two ways to include these counter term: (1) power spectrum level and (2) at field level. We discuss both cases now.

\begin{enumerate}
\item \textbf{At the power spectrum level}:The counterterms at the power spectrum are constant terms $\alpha_0$ and the $k^2$, the coefficient of which is denoted by $\alpha_2$. These two counterterms take into account the effects of the smoothing. The final expression thus reads:
\begin{equation}
\ba
\langle \de^2|\de_{\text{h}}\rangle' =  b_1\langle \de^2|\de^{(2)}\rangle'+b_2\langle \de^2|\delta^2\rangle' +&b_{s^2}\langle \de^2|s^2\rangle'+ \alpha_1 + \beta_1 k^2
\ea
\end{equation}
\begin{equation}
\ba
\langle -\Psi\cdot\nabla\de|\de_{\text{h}}\rangle' =  b_1\langle- \Psi\cdot\nabla\de|\de^{(2)}\rangle'&+b_2\langle- \Psi\cdot\nabla\de|\delta^2\rangle'&+b_{s^2}\langle -\Psi\cdot\nabla\de|s^2\rangle'
+ \alpha_2 + \beta_2 k^2
\ea
\end{equation}
\begin{equation}
\ba
\langle s^2|\de_{\text{h}}\rangle' =  b_1\langle s^2|\de^{(2)}\rangle'+b_2\langle s^2|\delta^2\rangle' 
&+b_{s^2}\langle s^2|s^2(\bk')\rangle'+ \alpha_3 + \beta_3 k^2
\ea
\end{equation}
The functional form of $\alpha_i$ and $\beta_i$ (where $i=1,2$ and 3) in these statistics come from the large scale limit of the quadratic field kernels. We can easily define them from Eq.~\eqref{app:UVmatrix} as follows:
\begin{equation}
\ba
\alpha_1 &=\frac{1}{50} \left(2 b_{\text{s}^2}+3 b_2\right) \left(363360 R_h^2+145318897\right)\\
\beta_1 &=-\frac{1}{700} \left(28 b_{\text{s}^2}+b_1\right) \left(25103 R_h^2+9902453\right)
\ea
\end{equation}
\begin{equation}
\ba
\alpha_2 &= 0\\
\beta_2 &= \frac{\left(5 b_1-56 b_{\text{s}^2}\right) \left(25103 R_h^2+9902453\right)}{7000}
\ea
\end{equation}
\begin{equation}
\ba
\alpha_3 &=\frac{1}{75} \left(2 b_{\text{s}^2}+3 b_2\right) \left(363360 R_h^2+145318897\right)\\
\beta_3 &=-\frac{\left(56 b_{\text{s}^2}+b_1+42 b_2\right) \left(25103 R_h^2+9902453\right)}{1050}
\ea
\end{equation}

One can easily see that $\alpha_3 = \frac{2}{3}\alpha_1$ and $\alpha_2 = 0$ which eventually brings down the number of counterterms to four ($\alpha_1,\beta_1,\beta_2,\beta_3$). One disadvantage of defining the counterterms at the power spectrum rather than field level, is that it doesn't allow for cosmic variance cancellation.

\item \textbf{At the field level}: at the field level the EFT counterterms correspond to two derivative operators:
\begin{equation}
\de_{\text{h}} (\bx) = b_1 \de(\bx) + b_2\de^2(\bx) + b_{s^2}s^2(\bx) + \beta\nabla^2\de^2(\bx) + \alpha\de\nabla^2\de(\bx) + \dots
\end{equation}
On large scales, the cross correlation of $\de^2$ with the counterterms give
\begin{equation}
\lim_{k\rightarrow 0}\langle \de^2|\nabla^2\de^2\rangle = -k^2\int_{\bq}P_{\text{lin}}(q)^2W_{\text{R}_{\text{f}}}(q)^2W_{\text{R}_{\text{h}}}(q)^2 \Rightarrow k^2 \times \text{constant}
\end{equation}
\begin{equation}
\lim_{k\rightarrow 0}\langle \de^2|\de\nabla^2\de\rangle = -\int_{\bq}q^2P_{\text{lin}}(q)^2W_{\text{R}_{\text{f}}}(q)^2W_{\text{R}_{\text{h}}}(q)^2 \Rightarrow \text{constant}
\end{equation}
For the three quadratic statistics, there are total of six counterterms. However, as already shown above, two counterterms can be eliminated giving us final four counterterms. The advantage of including the EFT counterterms at the field level is that we can not only compare the magnitudes of Fourier components but also their phases. In other words, we can obtain the constraints by minimizing $\langle D_{2}[\de]|\de^{\text{sim}}_{\text{h}}-\de^{\text{model}}_{\text{h}} \rangle$. If we compute the terms in $\de^{\text{model}}_{\text{h}}$ with the same phase as $\de^{\text{sim}}_{\text{h}}$, the random fluctuations (from sampling initial conditions) will be canceled and the bias constraints will improved significantly.
\end{enumerate}

\subsection{Cubic EFT counterterms}

Exactly the same procedure can be applied to study the cubic EFT counter term. First, let us consider the UV limits of smoothed cubic kernels $\mathcal{O}^{R_{\text{h}}}_{3}(\bk,-\bq,\bq) =W_{\text{R}_{h}}(q)^{2}W_{\text{R}_{h}}(k)\mathcal{O}_{3}(\bk,-\bq,\bq) $, which appear in the reducible diagrams :

\begin{equation}
\ba
&\lim_{k \rightarrow 0} F^{R_{\text{h}}}_3(\bk,-\bq,\bq) \approx \left(\frac{61 R_h^2}{1890}-\frac{61}{1890
   q^2}\right)+\frac{37 k^4 R_h^2}{3780 q^2}+k^2 \\
&\lim_{k \rightarrow 0} \de^{3,\text{R}_{f}}(\bk,-\bq,\bq) \approx -\frac{1}{2} k^2 R_h^2-q^2 R_h^2+1\\
&\lim_{k \rightarrow 0} \mathcal{G}^{R_{\text{h}}}_3(\bk,-\bq,\bq) \approx 0\\
&\lim_{k \rightarrow 0} \de\mathcal{G}^{R_{\text{h}}}_2(\bk,-\bq,\bq) \approx \frac{2}{9} k^2
   R_h^2+\frac{4}{9} q^2 R_h^2-\frac{4}{9}\\
&\lim_{k \rightarrow 0} \Gamma^{R_{\text{h}}}_3(\bk,-\bq,\bq) \approx \frac{32 \text{k}^4 R_h^2}{2205 q^2}+k^2
   \left(\frac{64 R_h^2}{315}-\frac{64}{315 q^2}\right)\\
&\lim_{k \rightarrow 0} \de F^{R_{\text{h}}}_2(\bk,-\bq,\bq) \approx -\frac{17}{63} k^2 R_h^2-\frac{34}{63}
   q^2 R_h^2+\frac{34}{63}\\
&\lim_{k \rightarrow 0} S_2F^{R_{\text{h}}}_2(\bk,-\bq,\bq) \approx k^2 \left(\frac{2
   R_h^2}{27}-\frac{16}{63 q^2}\right)+\frac{8 k^4 R_h^2}{441 q^2}-\frac{68}{189} q^2 R_h^2+\frac{68}{189}
\ea
\end{equation}

The UV limits of smoothed cubic irreducible kernels $\lim_{k \rightarrow 0} \mathcal{O}^{R_{\text{h}}}_3(\bk-\bq-\bp,\bq,\bp)=\lim_{k \rightarrow 0} W_{\text{R}_{h}}(|\bk-\bp-\bq|)W_{\text{R}_{h}}(p)W_{\text{R}_{h}}(q)\mathcal{O}_3(\bk-\bq-\bp,\bq,\bp)$ contain many terms and it is therefore not convenient to write down the full expressions here. However, we refer to \cite{Baldauf:2015aha} where one can find a good discussion about the UV limits of the two-loop power spectrum integrals and gravitational kernels in detail. In addition, to get an intuition of the low-$k$ behaviour of the orthogonalized cubic correlations we refer to Fig~\ref{7by7}.

Similar to quadratic statistics, the UV limits of cubic kernels require the inclusion of two counter terms (a constant term $\alpha$ and a $k^2$ term $\beta$) for each cubic statistic. At the field level, these two counter terms correspond to two higher derivative bias operators, that is $\de^2\nabla^2\de$ and $\nabla^2\de^3$ for constant and $k^2$ counterterms respectively. To summarise, we need at least two EFT counterterms for each quadratic and cubic statistics which means that to obtain consistent halo bias constraints up to cubic order from three quadratic and seven cubic statistics one is required to include $\mathcal{O}(15)$-$\mathcal{O}(20)$ EFT counterterms. The large number of the EFT counterterms for bias measurements is the main motivation for us to use the Taylor expansion method described in the main text.

\bibliographystyle{JHEP}
\bibliography{test}


\end{document}